\pdfoutput=1
\documentclass[american,english]{svjour3}
\usepackage[T1]{fontenc}
\usepackage[latin9]{inputenc}
\usepackage{color}
\usepackage{array}
\usepackage{float}
\usepackage{url}
\usepackage{multirow}
\usepackage{graphicx}
\usepackage{subscript}

\makeatletter

\providecommand{\tabularnewline}{\\}

\@ifundefined{showcaptionsetup}{}{%
 \PassOptionsToPackage{caption=false}{subfig}}
\usepackage{subfig}
\makeatother

\usepackage{babel}
\begin{document}

\title{Photoabsorption in Sodium Clusters: First Principles Configuration
Interaction Calculations}

\author{Pradip Kumar Priya, Deepak Kumar Rai, and Alok Shukla}

\institute{Department of Physics, Indian Institute of Technology Bombay, Powai,
Mumbai 400076, INDIA\\
e-mails: eccpradip@gmail.com, deepakrai@phy.iitb.ac.in, shukla@phy.iitb.ac.in}
\maketitle
\begin{abstract}
We present systematic and comprehensive correlated-electron calculations
of the linear photoabsorption spectra of small neutral closed- and
open-shell sodium clusters (Na$_{n}$, n=2-6), as well as closed-shell
cation clusters (Na$_{n}$$^{+}$, n=3, 5). We have employed the configuration
interaction (CI) methodology at the full CI (FCI) and quadruple CI
(QCI) levels to compute the ground, and the low-lying excited states
of the clusters. For most clusters, besides the minimum energy structures,
we also consider their energetically close isomers. The photoabsorption
spectra were computed under the electric-dipole approximation, employing
the dipole-matrix elements connecting the ground state with the excited
states of each isomer. Our calculations were tested rigorously for
convergence with respect to the basis set, as well as with respect
to the size of the active orbital space employed in the CI calculations.
These calculations reveal that as far as electron-correlation effects
are concerned, core excitations play an important role in determining
the optimized ground state geometries of various clusters, thereby
requiring all-electron correlated calculations. But, when it comes
to low-lying optical excitations, only valence electron correlation
effects play an important role, and excellent agreement with the experimental
results is obtained within the frozen-core approximation. \textcolor{black}{For
the case of Na$_{6}$, the largest cluster studied in this work, we
also discuss the possibility of occurrence of plasmonic resonance
in the optical absorption spectrum.}
\end{abstract}

\section{Introduction}

\label{sec:introd}

Clusters of atoms and molecules are considered to be strong candidates
for developing future nanotechnological devices ranging from electron-transport
based devices such as switches and transistors, to opto-electronic
devices such as light-emitting diodes.\cite{yoshiyuki} However, regardless
of the nature of the application, in order to harness full potential
of these atomic assemblies, it is important to achieve a deep theoretical
understanding their electronic structure.\cite{alonso,jellinek,Heer,Ekarat}
Clusters can be seen as an intermediate state of matter lying somewhere
between the atomic state, and the bulk state. Therefore, they can
provide us important insights into the evolution of the matter from
the microscopic to the macroscopic length scales. Of particular interest
are issues such as the nature of chemical bonding, low-lying excited
states, and the geometrical structure of the clusters as their size
grows. One would intuitively expect that clusters whose bulk states
are insulating, or semiconducting, will exhibit localized electrons.
Likewise, electrons in clusters of atoms whose bulk form is metallic,
will be expected to be itinerant. Indeed, in cluster science, an important
direction of research is to understand as to what are the signatures
of metallicity in metal atom clusters.\cite{alonso,jellinek,Heer,Ekarat,bowlan}If
the electrons in the metal atom clusters are indeed metallic, can
one use simple jellium model or some variant thereof, to describe
their physics?\cite{brack,Heer} It is well known that the nature
of low-lying optical excitation in bulk metals is plasmonic, that
is collective. But what happens when we consider a metal atom cluster:
is the nature of low-lying excited states for such systems molecular
or collective?\cite{brack,Heer,pollack}

Alkali atoms have a rather simple electronic structure, with a single
valence electron outside the closed shell. In the bulk form, alkali
metals are considered to be a protypical examples of free-electron
gas formed by the one electron contributed by each atom. Therefore,
alkali atom clusters, will be the simplest systems on which the issue
of metallicity in finite sized systems can be probed. Perhaps because
of this simplicity that a number of theoretical and experimental probes
of alkali atom clusters have been performed over the years. Early
description of the electronic structure of clusters was performed
in terms of the free-electron gas based jellium model, perhaps due
to its general nature, and inherent simplicity.\cite{clemenger,beck,wekardt}
Jellium model was also employed to describe the photoabsorption in
neutral and charged sodium clusters.\cite{photoabs} First principles
methodologies were also used several years back to compute the optimized
geometries, binding energies, and ionization potentials of small alkali
metal clusters.\cite{koutecky,jena,martin,boustani,kotec,rao,bona,bonacic,na-cation-kout-jcp-96}
Rao and Jena used first principles Hatree-Fock (HF) and configuration-interaction
(CI) methodology to study the electronic structure, geometry, and
magnetic properties of lithium based homo-atomic and hetero-atomic
clusters.\cite{jena,rao} Martins \emph{et al}. performed calculations
employing first principles density-functional theory (DFT) within
the local-spin-density approximation (LSDA) to study the ground state
geometries and electronic structure of small sodium clusters.\cite{martin}
First principles multi-reference configuration interaction (MRCI)
approach was employed by Kouteck\'y and coworkers to study the low-lying
excited states and photoresponse of a number of lithium and sodium
clusters.\cite{boustani,kotec,bona,na4-fantucci,bonacic,koutecky}
R\"othlisberger and Andreoni\cite{rothlisberger} used a first-principles
molecular dynamics based approach to compute the structure and electronic
properties of sodium clusters ranging from Na\textsubscript{2} to
Na\textsubscript{20 }. Employing an \emph{ab initio} DFT based all
electron approach, Blundell \emph{et al}.\cite{blundell-polariz}
reported calculations of static polarizabilities of several Na clusters.
 Balb\'{a}s performed time-dependent density functional theory (TDDFT)
based calculations of the photoabsorption spectra of relatively large
sodium clusters Na\textsubscript{20} and Na\textsubscript{40 }.\cite{torres-na-theory}
First principles approaches based upon HF, DFT, and second-order M{\o}ller-Plesset
perturbation theory (MP2) were employed by Solov'yov \emph{et al}.\cite{solovyov}
to determine the geometry and other ground state properties of a number
of sodium clusters. DFT based calculations on the electronic structure
and geometry of the ground state of several neutral and cationic clusters
of sodium were performed by K\"ummel and \emph{et al}.\cite{kmmel}
An \emph{ab initio} time-dependent local-density approximation (TDLDA)
based methodology was adopted by Pacheco and Martins\cite{li-na-photo-theory}
to study the photoresponse of lithium and sodium clusters. First principles
TDLDA was also employed by Moseler \emph{et al}.\cite{moseler} to
compute the optical absorption spectra of cationic clusters of sodium.
Guti\'{e}rrez \emph{et al}. employed a first principles DFT based
approach to compute the transport properties of small sodium clusters.
Rubio and coworkers computed the optical absorption spectra of small
sodium and silicon clusters using TDDFT based approaches.\cite{rubio,marques}TDDFT
was also employed by Joswig \emph{et al}.\cite{na-theory-photo-joswig}
to study the photoabsorption spectra of a large number of sodium clusters.
Wang \emph{et al}.\cite{na-plane-plasmon-theory} employed a TDDFT
based approach to study plasmonic excitation in planar clusters of
sodium. Recently, Pal \emph{et al}.\cite{pal,bethe-salpeter-na-cluster}
employed a first principles linear-response theory based Bethe-Salpeter
equation approach to compute the optical absorption spectra of a number
of closed-shell neutral and cationic sodium clusters.

Experimental measurements of photoabsorption processes in alkali metal
clusters has been performed for a long time. Early experiments of
photoabsorption in sodium and potassium dimers were performed by Fredrickson
and Watson.\cite{watson} For the sodium trimer, optical absorption
measurements have been reported by Herrmann \emph{et al}.,\cite{hermann}
Wang \emph{et al}.,\cite{wang} and Broyer \emph{et al}.\cite{broyer}
As far as the spectroscopy of Na\textsubscript{4} cluster is concerned,
Wang \emph{et al}. have again performed the measurements.\cite{wang,pollack}
In a separate paper, the same group also reported the photoabsorption
measurements in Na\textsubscript{4}, Na\textsubscript{5}, and Na\textsubscript{7 }clusters.\cite{dahlseid}
Schmidt and Haberland\cite{schmidt} measured the photoabsorption
cross-sections of sodium mono-cationic clusters ranging from Na$_{3}{}^{+}$
to Na$^{+}{}_{64}$. Several measurements of photoabsorption cross-sections
and static polarizabilities of sodium clusters have also been reported
by de Heer and coworkers over the years.\cite{saunders,Heer2,selby,photoabs2,bowlan} 

As reviewed above, most of the calculations of the electronic structure
and related properties of alkali metal clusters have been performed
either using the jellium model, or, more recently, using the DFT based
methodologies. Except for the early works of Rao and Jena,\cite{jena,rao}
and those of Kouteck\'y and coworkers,\cite{boustani,kotec,bona,na4-fantucci,bonacic,koutecky,na-cation-kout-jcp-96}
we are not aware of studies of the low-lying excited states and optical
properties of alkali metal clusters which employ wave function based
methodologies such as the CI approach, or the coupled-cluster (CC)
method. While DFT based methodologies are certainly computationally
very efficient because they can, in principle, include electron-correlation
effects at the mean-field level in the Kohn-Sham equations. However,
the exact form of the exchange-correlation functional, which is supposed
to incorporate these effects, is unknown. Therefore, the choice of
the exchange-correlation functional (XCF) is a matter of speculation,
and as a result a huge array of such functionals are available. It
is a well-known fact that results of calculations depend critically
on the choice of the functional, and sometimes, one has to use different
functionals to describe different properties of the same system. Some
functionals provide a good description of the ground state geometry,
while, others provide a better description of the cohesive energies.
As far as the description of excited states of extended systems is
concerned, at present the method of choice is TDDFT, in which a number
of strides have been made recent times.\cite{tddft-review-casida}
Nevertheless, TDDFT also suffers from the same problems as DFT, \emph{i.e.},
the results obtained are highly functional dependent. It also appears
to be most effective in describing low-lying singly-excited states,
while at higher energies, its accuracy appears to be questionable.
For all these reasons we believe that it is very important to benchmark
excited state results by performing accurate electron-correlated first-principles
calculations using the Born-Oppenheimer Hamiltonian, and wave-function-based
methodologies. Because the Hamiltonian, unlike the functional is well-known,
the only problem which wave-function-based methods experience are
related to the truncation of the many-particle wave function. However,
by systematically expanding the wave-function in terms of the many-particle
basis, this problem can be curtailed. It is with this philosophy in
mind that we have undertaken systematic large-scale first-principles
CI calculations of sodium clusters ranging from Na$_{2}$ to Na$_{6}$,
to compute their linear photoabsorption spectra, and carefully compare
our results to the spectroscopic data where available. In addition
to the neutral clusters, calculations have also been performed for
the closed-shell cationic clusters Na$_{3}{}^{+}$ and Na$_{5}{}^{+}$.
For several clusters, in addition to the global minima, we have also
considered various possible low-lying geometries, with the intention
of probing structure-property relationship which can serve as fingerprint
for detection of cluster geometries using photo-absorption experiments.
The computational methodology we have employed for the purpose is
the frozen-core configuration interaction (CI) method at the full-CI
or quadruple-CI level, which has been extensively used in our group
in earlier studies of the optical properties of a number of conjugated
polymers,\cite{mrsd_prb_02,Shukla-TPA,Shukla-THG,mrsd_prb_05,mrsd_prb_07,himanshu,himanshu-triplet}
and small boron and aluminum clusters.\cite{sahu,shinde,shinde-al}
We also compare our results to the experimental ones where available,
and find excellent agreement. 

\section{Theoretical Background}

\label{sec:theory}

In this work we have adopted a first-principles wave function based
methodology in which the molecular orbitals (MOs) are expanded as
a linear combination of Cartesian Gaussian basis functions. Calculations
are initiated by performing either the restricted Hartree-Fock (RHF),
or the restricted open-shell Hartree-Fock (ROHF) calculations depending
upon whether the cluster under consideration is electronically closed
shell, or open-shell type, so as to obtain the MO basis set. Subsequently,
correlated calculations are performed using the configuration-interaction
(CI) methodology, and the approach is employed at various approximation
levels such as singles-CI (SCI), full-CI (FCI), or quadruple-CI (QCI).
For performing the CI calculations, we employed a configuration state
function (CSF) basis, in which the CSFs were eigenfunctions of both
the total spin ($S^{2}$), as well as the $z$-component of the spin
($S_{z}$). Furthermore, wherever possible, point group symmetries
at the level of $D_{2h}$ and its subgroups were also used to reduce
the computational effort. We have employed this wave-function-based
first principles approach in several of our recent works on clusters,\cite{shinde,shinde-b6,shinde-al}and
for performing these calculations we have used the MELD program package.\cite{meld}

As far as geometries of various isomers are concerned, in some cases
they were obtained manually, by performing electron-correlated calculations
for different geometries, while in other cases, automated geometry
optimization at various levels of theory, as implemented in the Psi4
computer program,\cite{psi4} was employed.\cite{emsl_bas1,emsl_bas2}
For a few isomers, we used the geometries reported by other authors,
in previous papers.

Once the ground state geometry of an isomer is decided, ground and
excited state calculations are performed on it using the CI methodology
described in our previous works.\cite{mrsd_prb_02,Shukla-TPA,Shukla-THG,mrsd_prb_05,mrsd_prb_07,sahu,himanshu,shinde,shinde-al,himanshu-triplet}
The level of the CI approach adopted in the calculations is based
upon the size of the isomer, as also the number of active electrons
and orbitals. For the smaller isomers frozen-core FCI method was used,
while for the larger ones, the frozen-core QCI approach was employed.
For the purpose of benchmarking the frozen-core approximation, we
employed the SCI method both at the all-electron, as well as frozen-core
levels. Using the many-body wave functions obtained from the CI calculations,
optical absorption spectrum was computed in the electric-dipole approximation,
assuming a Lorentzian line shape. 

\section{Calculations and Results}

\label{sec:results}

In this section, first we shall discuss the convergence of our calculations
with respect to various approximations. Thereafter, the results of
our calculations for various clusters will be presented with comprehensive
discussion.

\subsection{Convergence of Calculations}

Here, we carefully examine the convergence of the calculated absorption
spectra with respect to the size and quality of the basis set, along
with various truncation schemes in the CI calculations. 

\subsubsection{Choice of the basis set}

\begin{figure}[h]
\includegraphics[width=8cm]{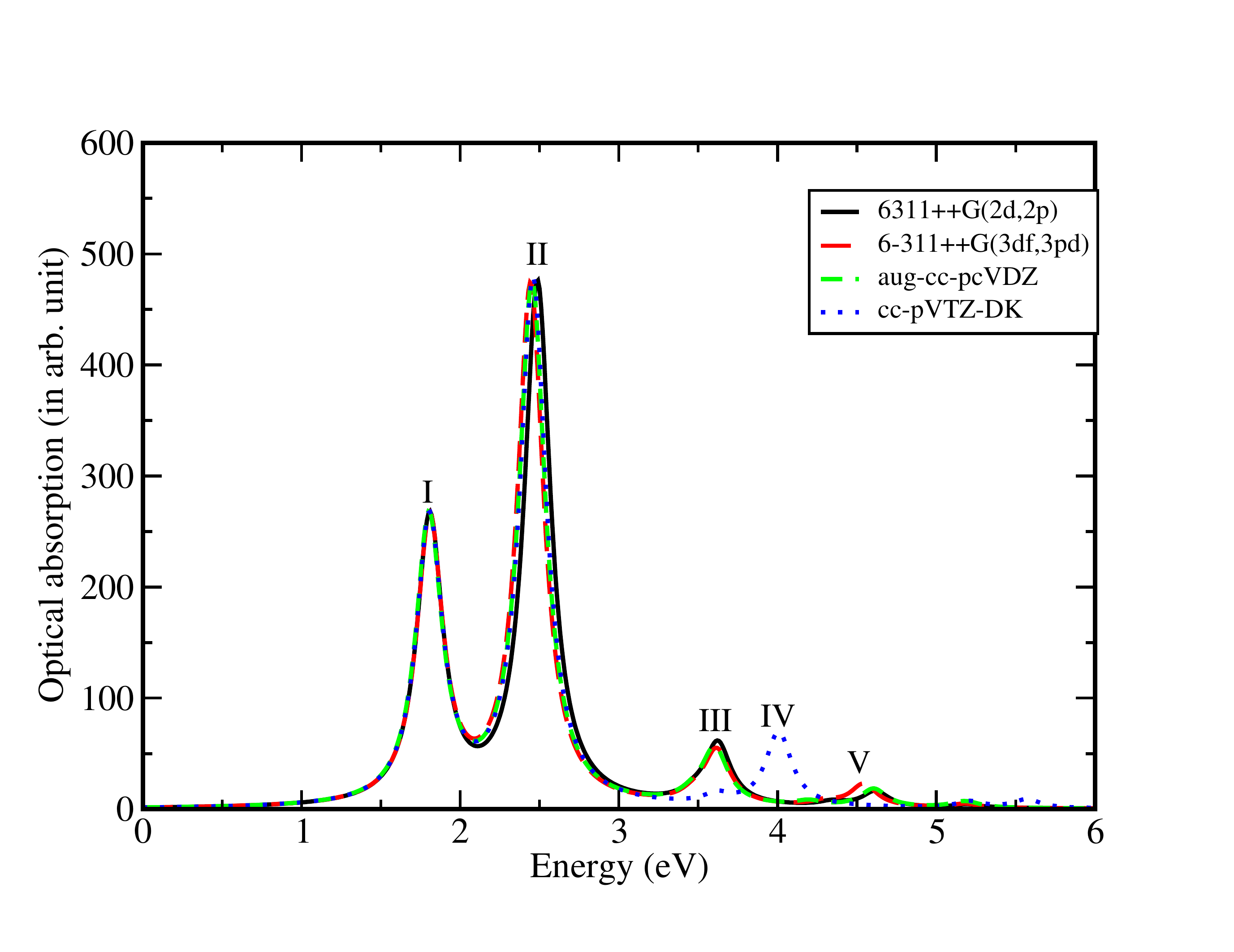}

\caption{ Optical absorption in Na$_{2}$ calculated using various Gaussian
contracted basis sets. }

\label{basis_dep}
\end{figure}

Because finite basis set is employed in electronic structure calculations,
it is but natural that the results will depend upon the quality and
the size of the basis set employed. Therefore, before comparing our
results with the experiments, we want to explore the dependence of
photo-absorption spectra on the basis set. It is obvious from the
metallic nature of bulk sodium that its 3s valence electron exhibits
delocalized behavior, and, therefore, the basis set used for calculations
should have diffuse Gaussian basis functions. We calculated the absorption
spectra of Na$_{2}$ employing basis sets 6-311++G(3df,3pd), 6-311++G(2d,2p),
and also correlation-consistent basis sets aug-cc-pcVDZ and cc-pVTZ-DK,
all of which have been taken from the EMSL Gaussian Basis Set Library.
\cite{emsl_bas1,emsl_bas2} Upon examining the calculated spectra
presented in Fig. \ref{basis_dep}, the following trends emerge: the
spectra computed by augmented basis sets 6-311++G(3df,3pd), 6-311++G(2d,2p)
and aug-cc-pcVDZ are in very good agreement with each other in the
energy range examined, while the one computed by the set cc-pVTZ-DK
is a bit blue shifted as far as higher energy peaks are concerned.
Considering good agreement among three basis sets, we choose to perform
calculations for all the isomers (unless, indicated otherwise) of
sodium clusters using the basis set 6-311++G(3df,3pd). 

\subsubsection{Orbital Truncation Schemes}

It is a well-known fact that the computational effort in a CI calculation
performed using $N$ MOs scales as $\approx N^{6}$, which can become
quite unfavorable for large values of $N$. For first principles calculations
utilizing large basis sets, generally $N$ tends to be large, and,
therefore, it is very important to reduce the number of MOs used in
the CI calculations. The number of occupied MOs is reduced by employing
the so-called ``frozen-core approximation'' discussed earlier, while
the number of unoccupied (virtual) states are truncated by ignoring
high-energy MOs lying far from the Fermi level.

\begin{figure}[h]
\includegraphics[width=8cm]{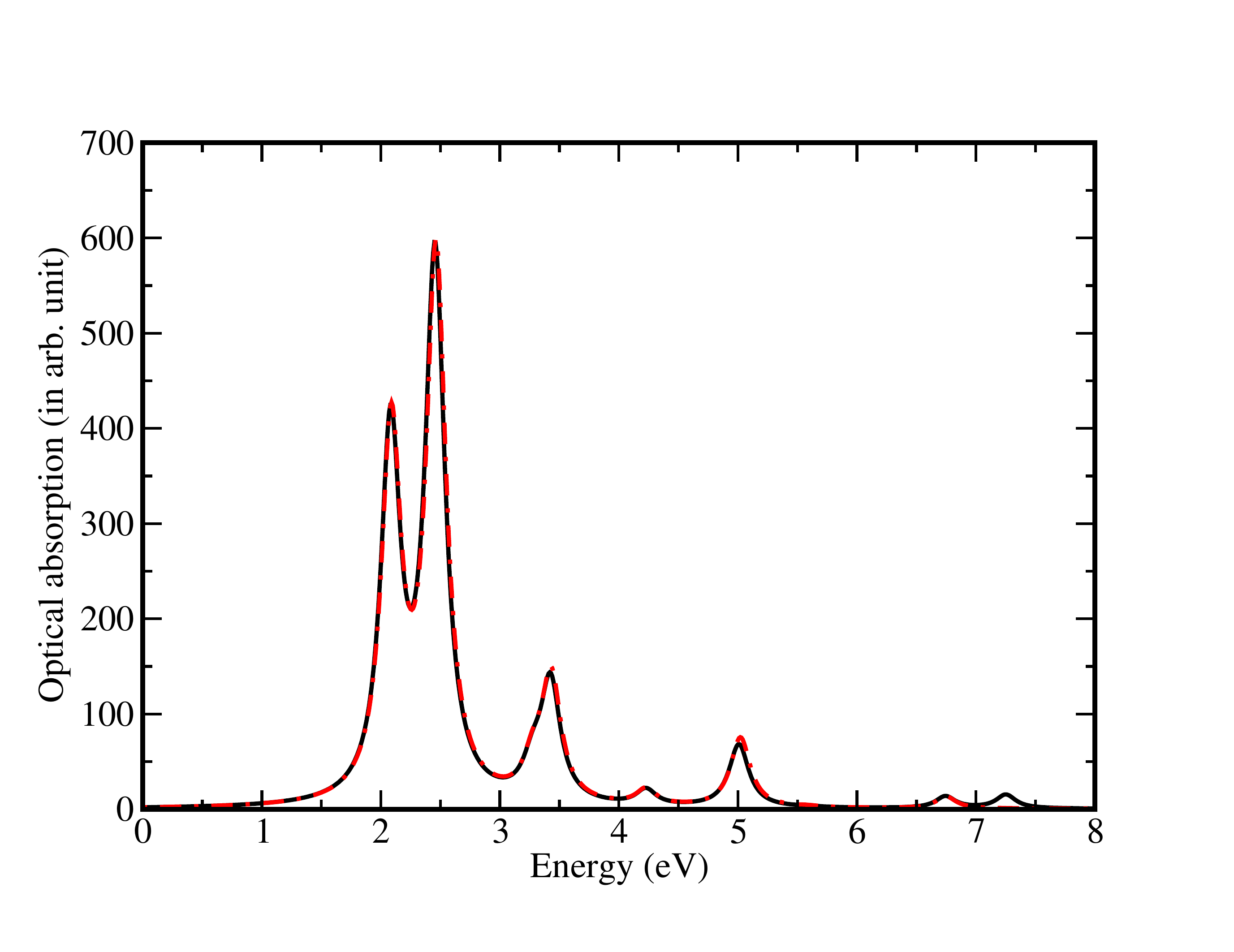}

\caption{(Color online) Comparison of the SCI level optical absorption spectrum
of Na$_{2}$ computed at the all-electron level ( in black), with
that obtained using the frozen-core approximation (in red). Hardly
any difference is visible between the two spectra.}

\label{na2_core_freez}
\end{figure}

\begin{figure}[h]
\includegraphics[width=8cm]{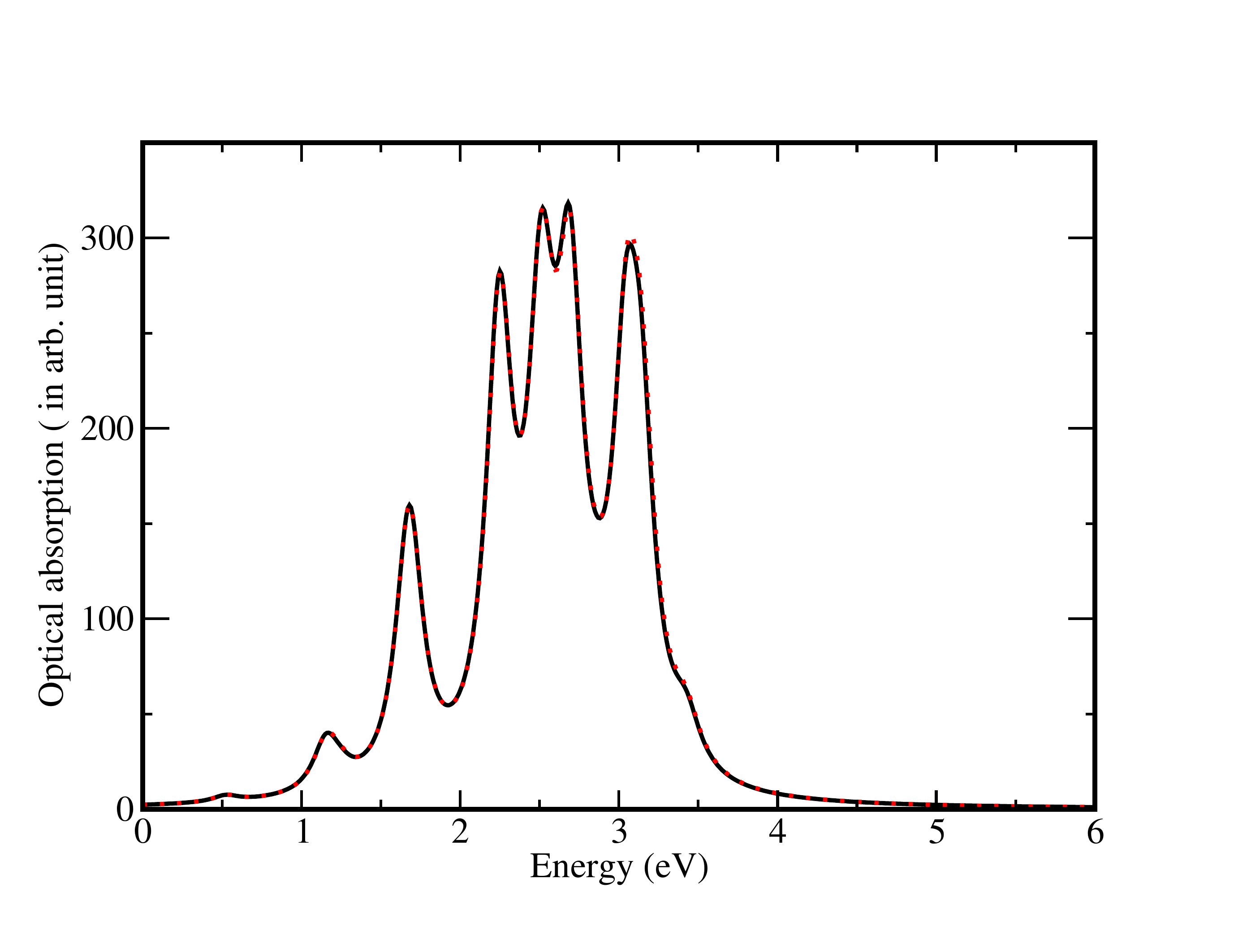}

\caption{(Color online) Same comparison as shown in Fig. 2, but for the case
of Na$_{3}$. }

\label{na3_core_freez}
\end{figure}

\begin{figure}[h]
\includegraphics[width=8cm]{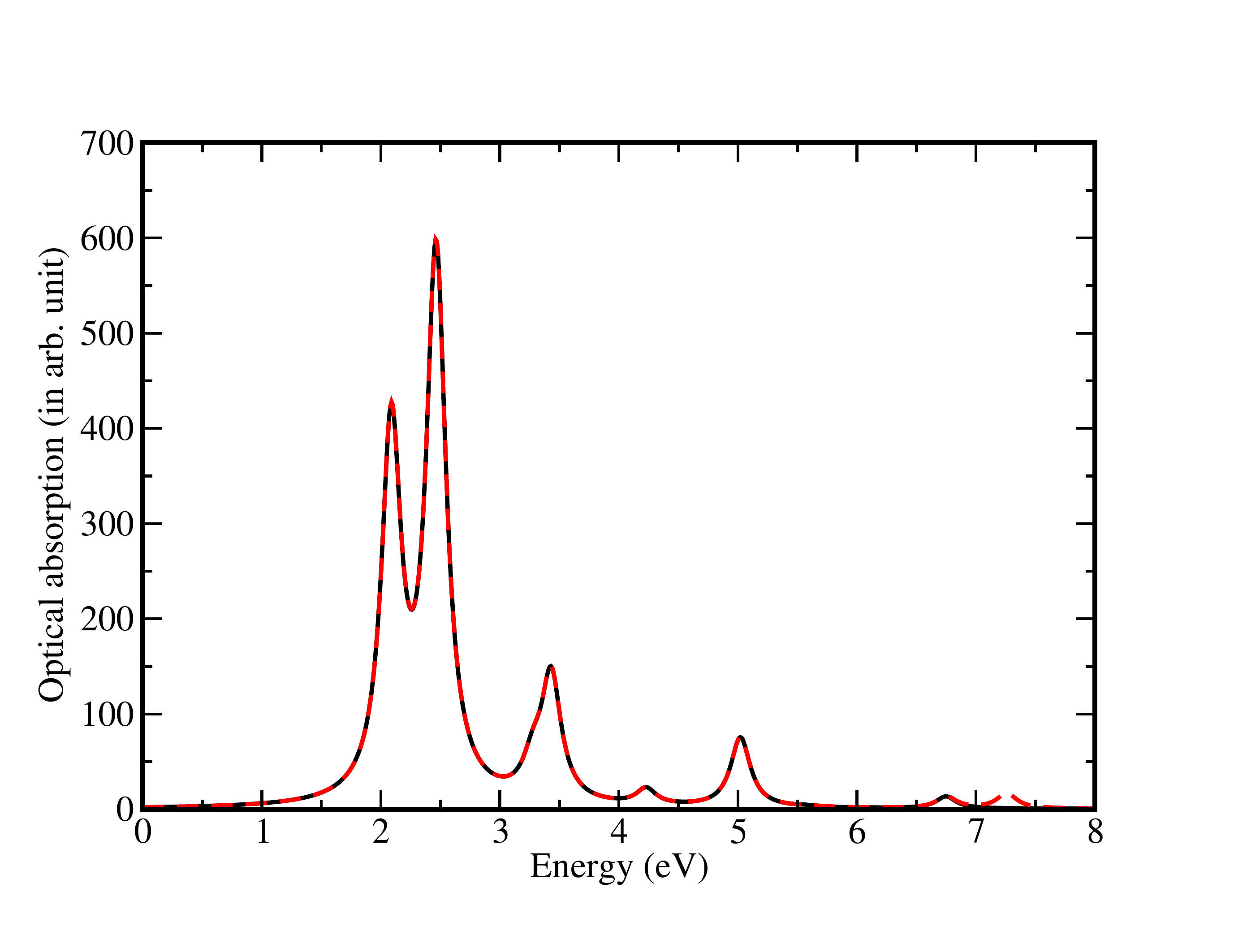}

\caption{(Color online) A comparison of all-electron SCI optical absorption
spectrum of Na$_{2}$ (in black), with that obtained by removing the
virtual orbitals with energies larger than 1.0 Hartree (in red), from
the list of active orbitals. The differences between two sets of calculations
are insignificant. }

\label{na2_active}
\end{figure}

\begin{figure}[h]
\includegraphics[width=10cm]{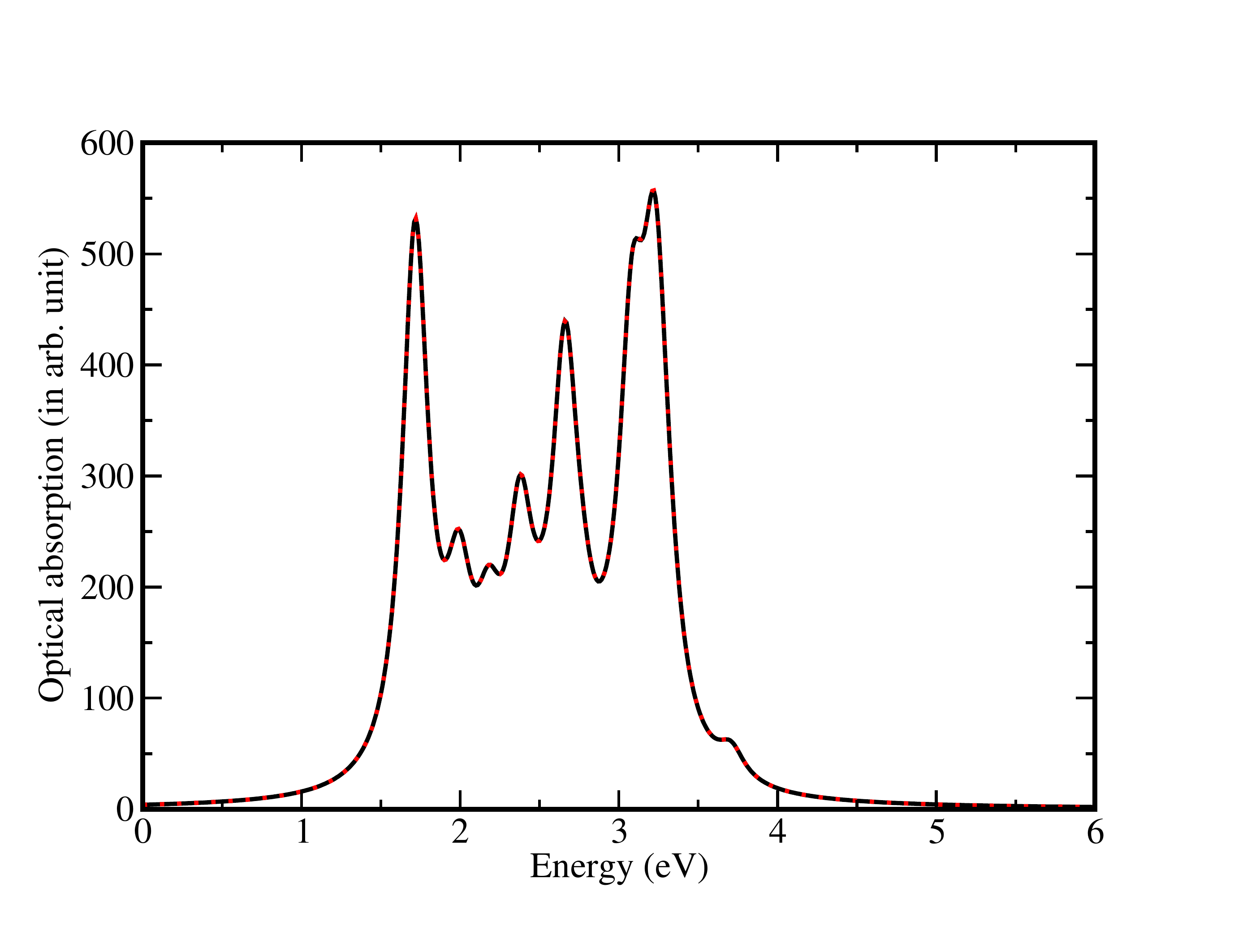}

\caption{(Color online) Same comparison as shown in Fig. 4, but for the case
of Na$_{4}$. \label{na4_freeze_act}}
\end{figure}

The influence of freezing the core orbitals on the optical absorption
spectrum of Na$_{2}$ cluster is displayed in Fig. \ref{na2_core_freez},
while for Na$_{3}$ the results are presented in Fig. \ref{na3_core_freez}.
It is obvious from the figures that it virtually makes no difference
to the computed spectra when the core orbitals are frozen. The effect
of removing the high-energy virtual orbitals on the absorption spectrum
of clusters Na$_{2}$ and Na$_{4}$ are examined in Figs. \ref{na2_active}
and \ref{na4_freeze_act}, respectively. From these figures it is
obvious that if the virtual orbitals above the energy of 1 Hartree
are removed from the list of active orbitals, the absorption spectrum
stays unchanged. This is understandable on physical grounds because
we are interested in computed absorption spectrum in the energy range
much lower than 1 Hartree. Therefore, in rest of the calculations,
all the core orbitals were frozen, and virtual orbitals with energies
larger than this cutoff were not considered. 

\subsubsection{Size of the CI expansion}

In CI calculations, greater numerical accuracy demands the inclusion
of a rather large number of configurations which can make the calculations
numerically intractable. However, when it comes to the computation
of optical properties, our focus is on calculating accurate energy
differences between the ground state, and the excited states, rather
than their absolute energies, which is possible even with moderately
large CI expansions. In Table \ref{tab_ref} we present the average
total number of configurations (N$_{total}$) for different isomers,
with the average computed across different irreducible representations
in these symmetry adapted calculations. The fact that these calculations
employ more than three million configurations for the largest cluster
considered (Na$_{6}$), makes us confident that our results are fairly
accurate.

Before we discuss the absorption spectrum for each isomer, we present
the ground state energies along with the relative energies of each
isomer in Table \ref{tab_ener}, from which it can be easily deduced
as to which isomer of a given cluster corresponds to the minimum energy
ground state geometry. Furthermore, symmetry of the ground state wave
function of each isomer is also listed in the same table. 

\begin{table}[h]

\caption{Average number of total configurations (N$_{total}$) involved in
frozen-core CI calculations of optical absorption spectra of various
isomers of sodium clusters. }

\begin{tabular}{|c|c|c|c|}
\hline 
Cluster  & Isomer  & CI method & N$_{total}$ \tabularnewline
\hline 
\hline 
Na$_{2}$ & Linear ($D_{\infty h}$) & FCI & 7056\tabularnewline
\hline 
 &  &  & \tabularnewline
\hline 
Na$_{3}{}^{+}$ & Triangular ($D_{3h}$) & FCI & 992250\tabularnewline
\hline 
 &  &  & \tabularnewline
\hline 
Na$_{3}$ & Triangular ( $C_{2v}$) & FCI & 992250\tabularnewline
\hline 
 & Linear ($D_{\infty h}$) & FCI & 992250\tabularnewline
\hline 
 & Equilateral ($D_{3h}$) & FCI & 992250\tabularnewline
\hline 
 &  &  & \tabularnewline
\hline 
Na$_{4}$ & Rhombus ($D_{2h}$ ) & FCI & 725903\tabularnewline
\hline 
 & Square ($D_{4h}$) & FCI & 1153510\tabularnewline
\hline 
 &  &  & \tabularnewline
\hline 
Na$_{5}{}^{+}$ & $D_{2d}$  & FCI & 406456\tabularnewline
\hline 
 & $D_{2h}$  & FCI & 762300\tabularnewline
\hline 
 & $D_{3h}$ & FCI & 442176\tabularnewline
\hline 
 & $C_{2v}$ & FCI & 192660\tabularnewline
\hline 
 &  &  & \tabularnewline
\hline 
Na$_{5}$ & Planar ( $C_{2v}$) & QCI & 3349740\tabularnewline
\hline 
 & Bipyramid ( $C_{2v}$) & QCI & 1463592\tabularnewline
\hline 
 &  &  & \tabularnewline
\hline 
Na$_{6}$ & Planar ($D_{3h}$) & QCI & 3463230\tabularnewline
\hline 
 & $C_{5v}$  & QCI & 3463230\tabularnewline
\hline 
 & $D_{4h}$  & QCI & 3162995\tabularnewline
\hline 
\end{tabular}\label{tab_ref}

\end{table}

\begin{table}[h]
\caption{Ground state (GS) energies (in Hartree) and relative energies (in
eV) corresponding to the geometries of various isomers depicted in
Fig. \ref{fig:Ground-state-geometries}, and the symmetry of their
wave function. For all the cases except Na$_{6}$ isomers, geometries
were optimized using an all-electron CCSD(T) approach, as implemented
in the Psi4 program,\cite{psi4} employing the cc-PVTZ basis set.\cite{emsl_bas2}
For Na$_{6}$ isomers, CCSD method\cite{psi4} was utilized for geometry
optimization, coupled with the cc-pvdz basis set.\cite{emsl_bas2}
In a few cases, previously published geometries, cited in the text,
were used.}

\begin{tabular}{|c|c|c|c|c|}
\hline 
Cluster & Isomer & Symmetry & GS energy  & Relative energy \tabularnewline
\hline 
 &  &  & (Ha) & (eV)\tabularnewline
\hline 
\hline 
Na$_{2}$ & Linear ($D_{\infty h}$) & $^{1}\Sigma_{g}$ & -323.7694 & 0.0000\tabularnewline
\hline 
 &  &  &  & \tabularnewline
\hline 
Na$_{3}{}^{+}$ & Triangular ($D_{3h}$) & \textcolor{black}{$^{1}A'_{1}$} & -485.5130 & 0.0000\tabularnewline
\hline 
 &  &  &  & \tabularnewline
\hline 
Na$_{3}$ & Triangular ( $C_{2v}$) & $^{2}B_{2}$ & -485.6519 & 0.0000\tabularnewline
\hline 
 & Equilateral ($D_{3h}$) & $^{2}A'_{1}$ & -485.6490 & 0.07891\tabularnewline
\hline 
 & Linear ($D_{\infty h}$) & $^{2}\Sigma_{u}$ & -485.6474 & 0.1225\tabularnewline
\hline 
 &  &  &  & \tabularnewline
\hline 
Na$_{4}$ & Rhombus ($D_{2h}$ ) & $^{1}A_{g}$  & -647.5553 & 0.0000\tabularnewline
\hline 
 & Square ($D_{4h}$) & $^{1}A_{1g}$ & -647.5360 & 0.5252\tabularnewline
\hline 
 &  &  &  & \tabularnewline
\hline 
Na$_{5}{}^{+}$ & $D_{2d}$  & $^{1}A_{1}$ & -809.3119 & 0.0000\tabularnewline
\hline 
 & $D_{2h}$  & $^{1}A_{g}$  & -809.3117 & 0.0054\tabularnewline
\hline 
 & $D_{3h}$ & \textcolor{black}{$^{1}A'_{1}$} & -809.3071 & 0.1306\tabularnewline
\hline 
 & $C_{2v}$ & $^{1}A_{1}$ & -809.2882 & 0.6449\tabularnewline
\hline 
 &  &  &  & \tabularnewline
\hline 
Na$_{5}$ & Planar ( $C_{2v}$) & $^{2}A_{1}$ & -809.4505 & 0.0000\tabularnewline
\hline 
 & Bipyramid ( $C_{2v}$) & $^{2}B_{1}$ & -809.4463 & 0.1143\tabularnewline
\hline 
 &  &  &  & \tabularnewline
\hline 
Na$_{6}$ & $C_{5v}$  & $^{1}A_{1}$ & -971.2509 & 0.0000\tabularnewline
\hline 
 & Planar ($D_{3h}$) & \textcolor{black}{$^{1}A'_{1}$} & -971.2476 & 0.0898\tabularnewline
\hline 
 & $D_{4h}$  & $^{1}A_{1g}$ & -971.2441 & 0.1850\tabularnewline
\hline 
\end{tabular}\label{tab_ener}
\end{table}

\subsection{Calculated Absorption Spectra of Various Clusters}

\begin{figure}[h]
\subfloat[Na$_{2}$, $\mbox{D}_{\infty h}$]{\includegraphics[scale=0.18]{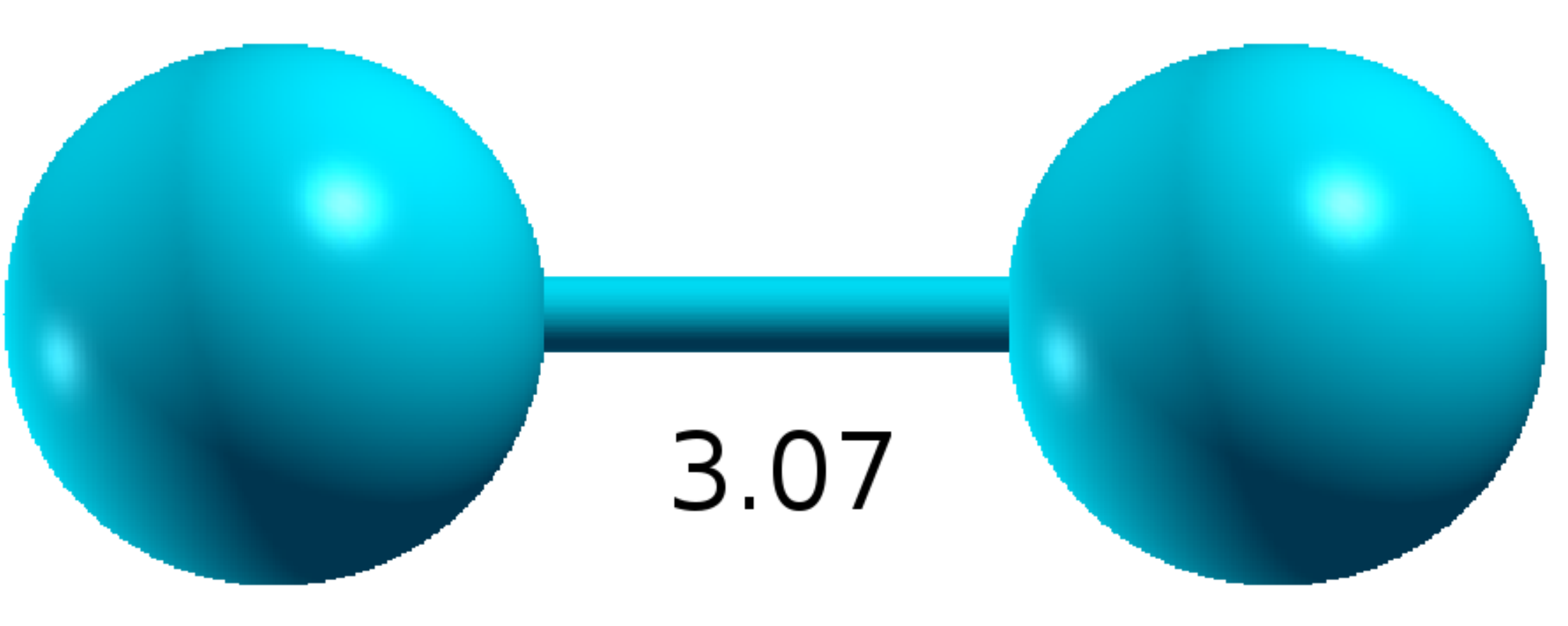}

}\subfloat[\textcolor{black}{Na$_{3}^{+}$, }$\mbox{D}_{3h}$]{\includegraphics[scale=0.2]{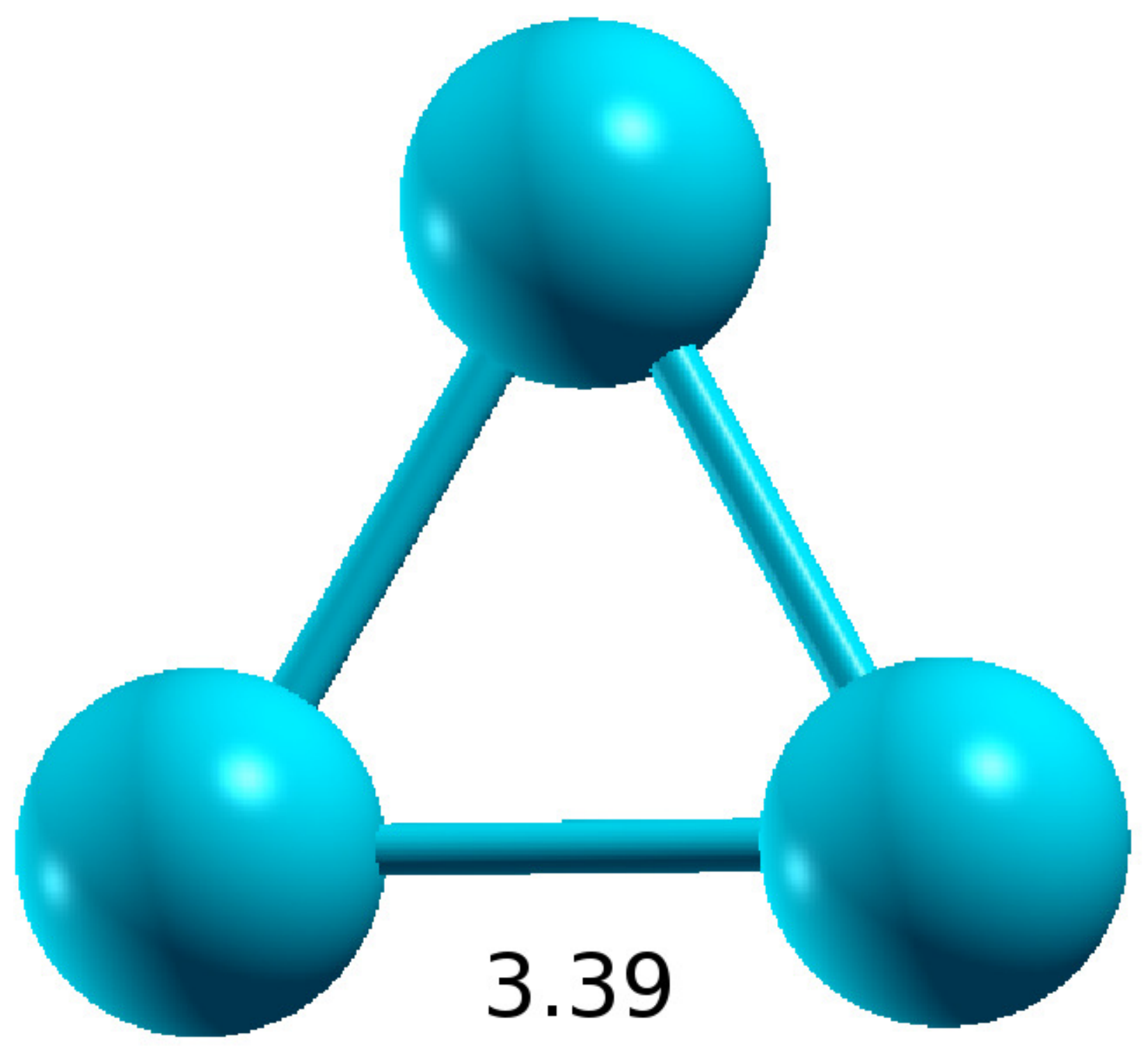}

}\subfloat[\textcolor{black}{Na$_{3}$, }$\mbox{C}_{2v}$]{\includegraphics[scale=0.2]{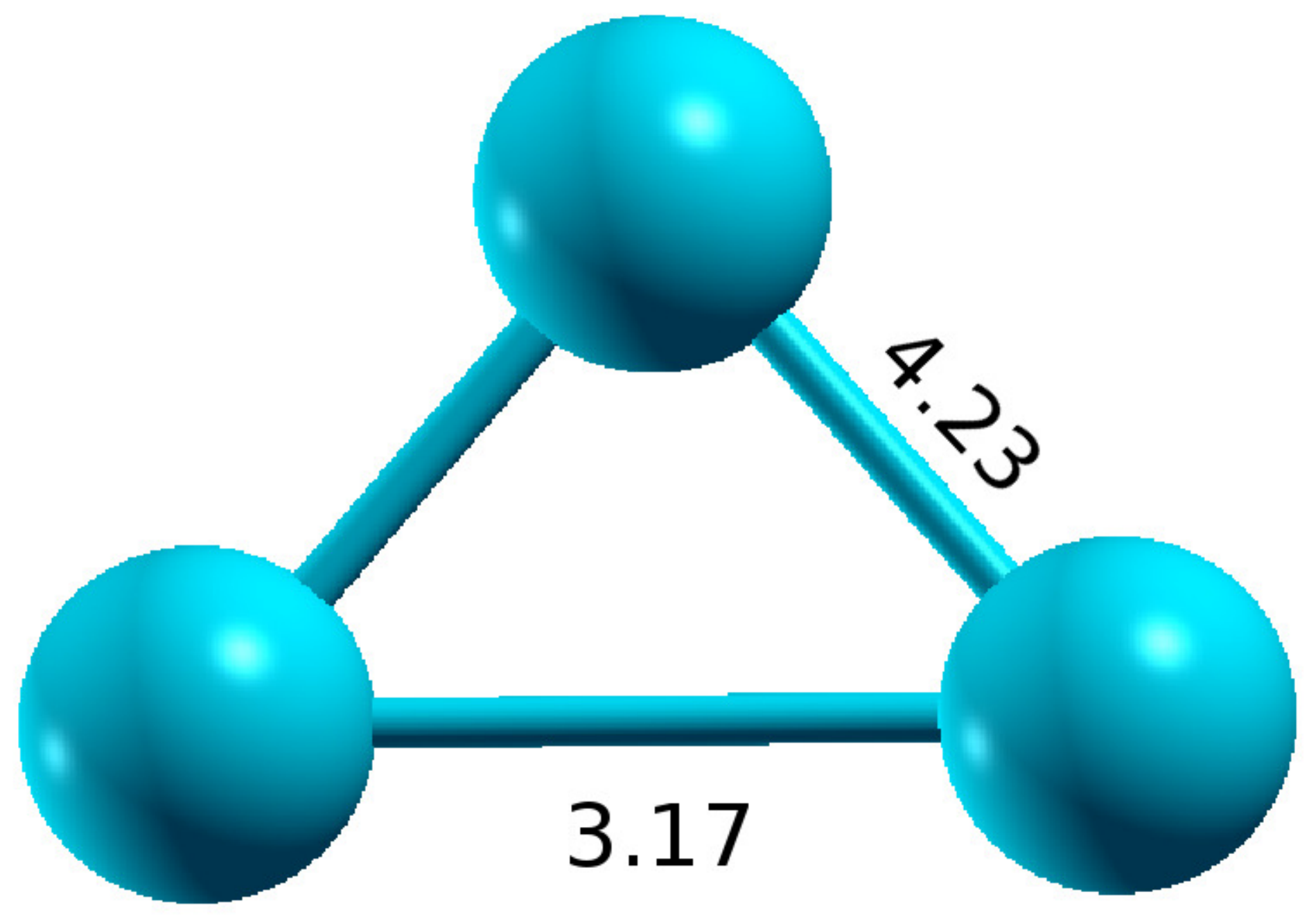}

}\subfloat[\textcolor{black}{Na$_{3}$, }$\mbox{D}_{\infty h}$]{\includegraphics[scale=0.22]{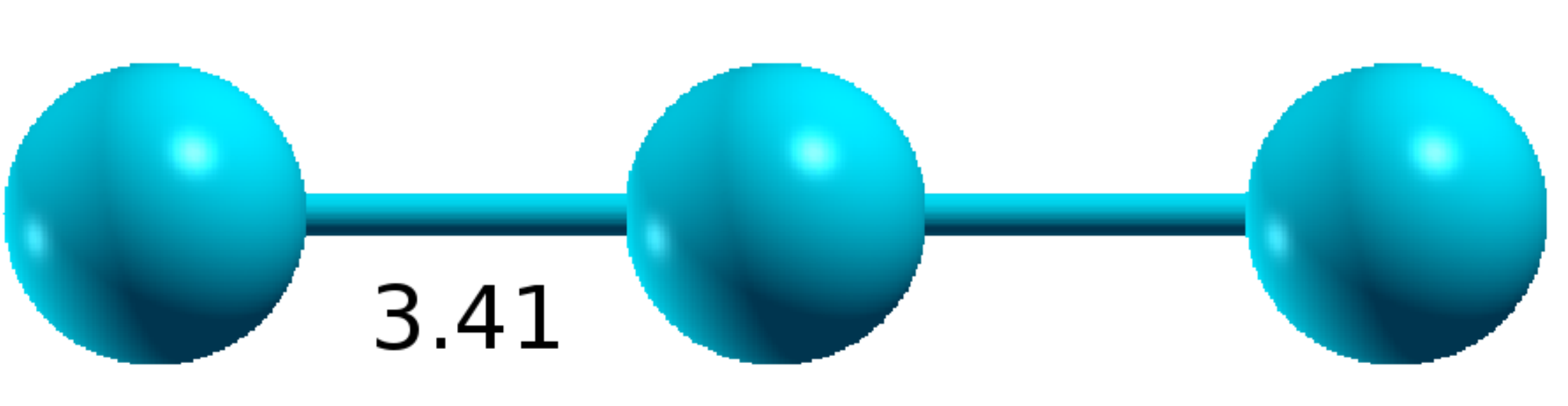}

}

\subfloat[\textcolor{black}{Na$_{3}$, }$\mbox{D}_{3h}$]{\includegraphics[scale=0.2]{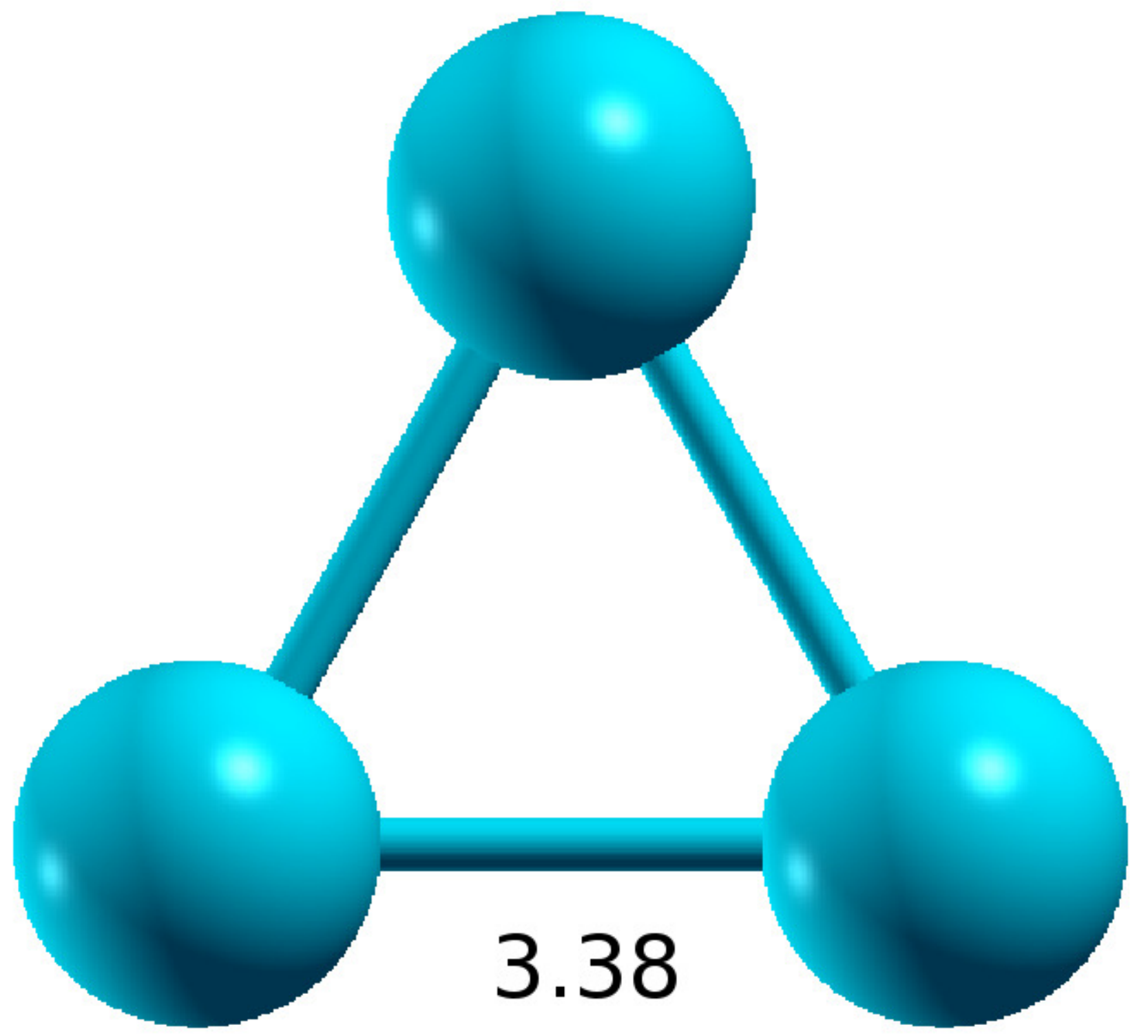}

}\subfloat[\textcolor{black}{Na$_{4}$, }$\mbox{D}_{2h}$]{\includegraphics[scale=0.22]{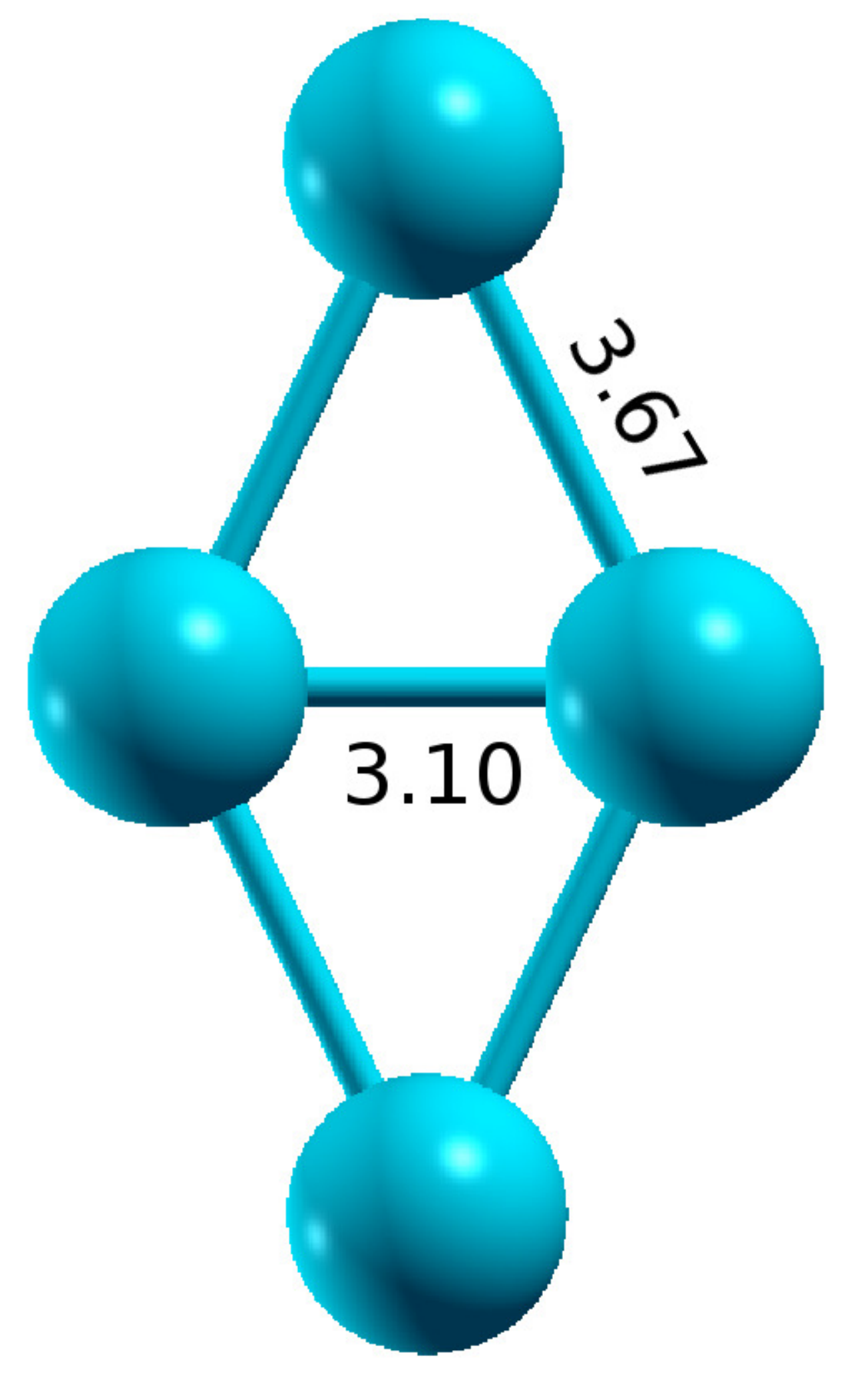}

}\subfloat[\textcolor{black}{Na$_{4}$, }$\mbox{D}_{4h}$]{\includegraphics[scale=0.18]{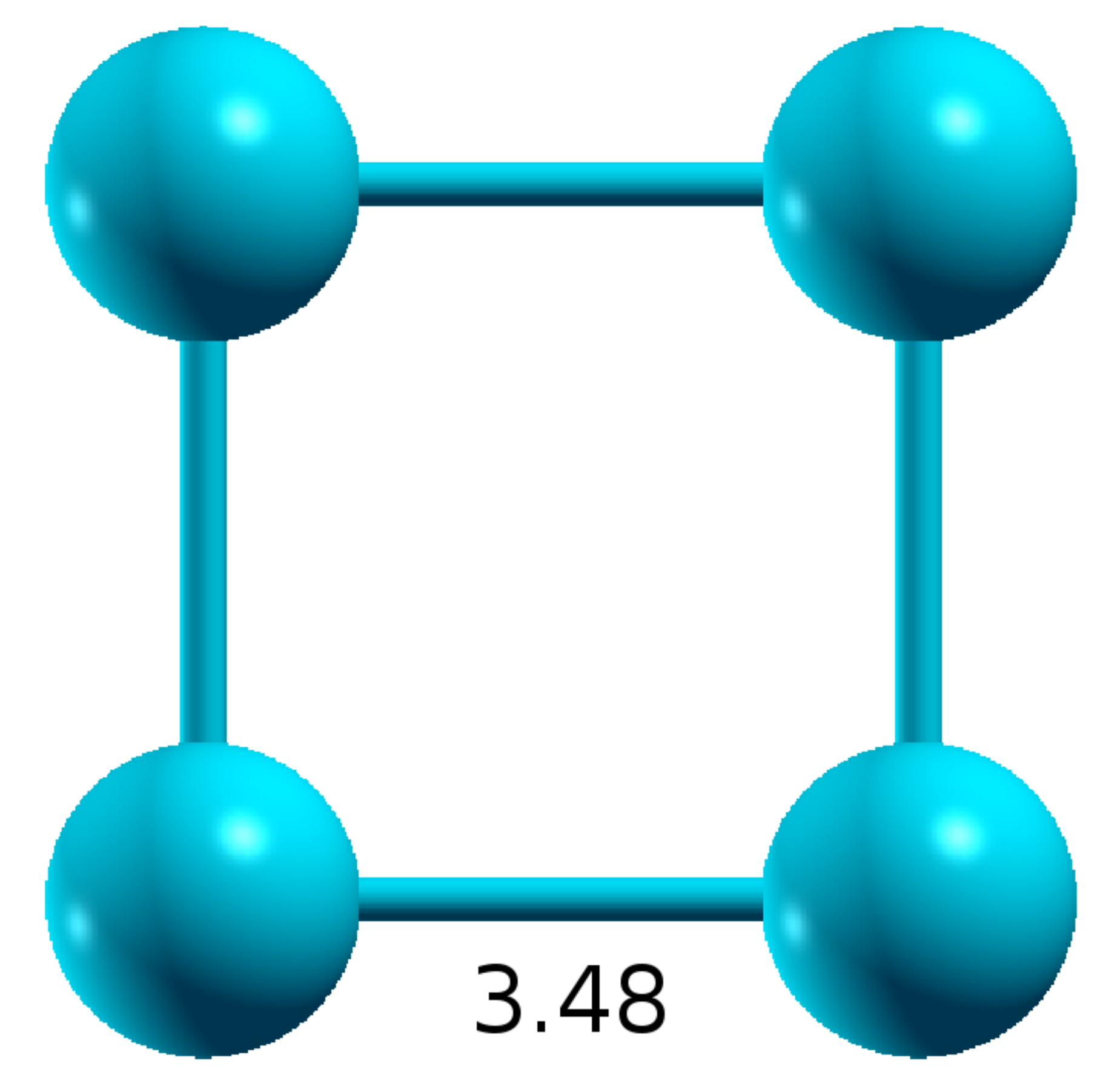}

}\subfloat[\textcolor{black}{Na$_{5}^{+}$, }$\mbox{D}_{2d}$]{\includegraphics[scale=0.3]{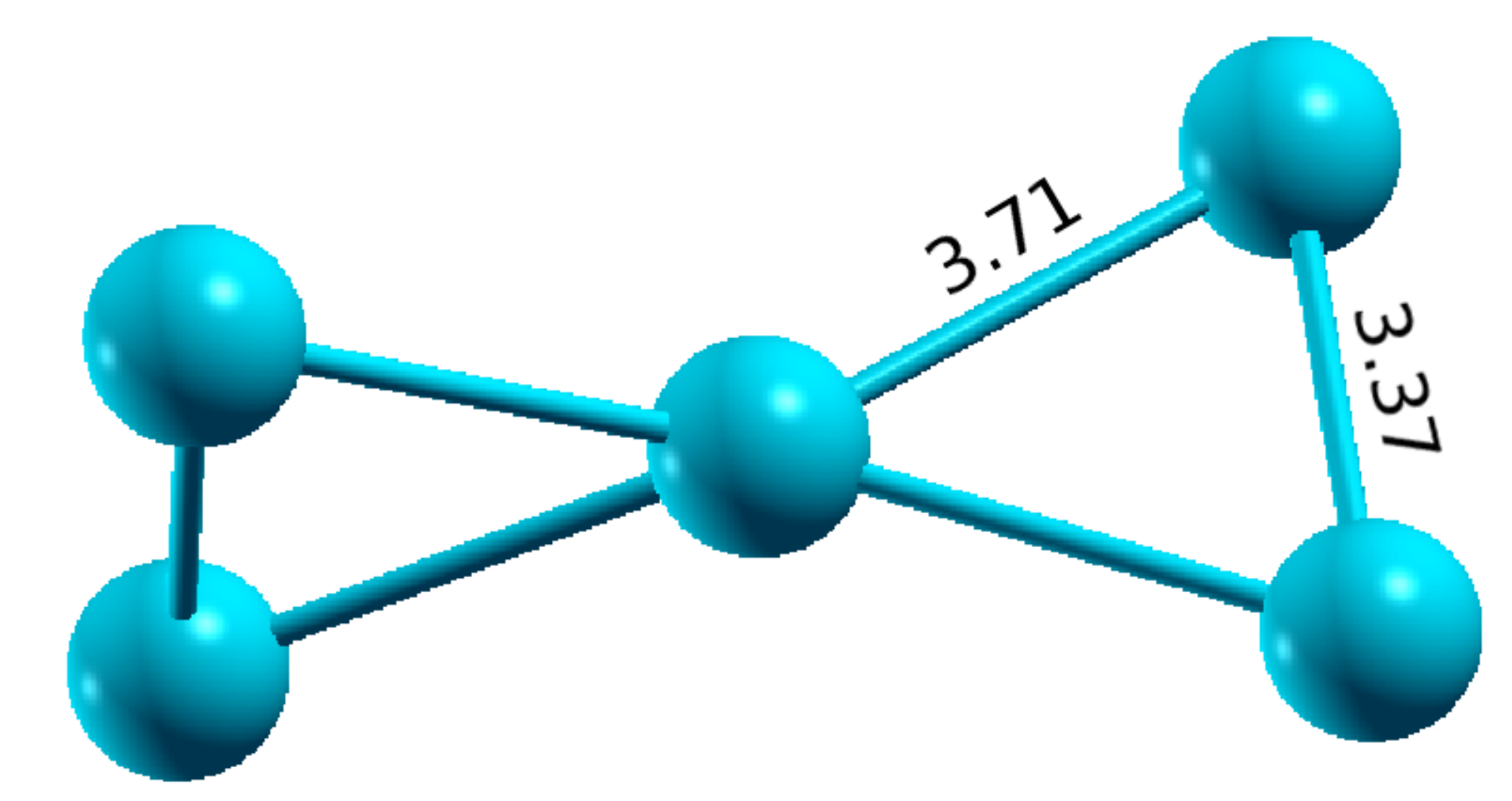}}

\subfloat[\textcolor{black}{Na$_{5}^{+}$, }$\mbox{D}_{2h}$]{\includegraphics[scale=0.25]{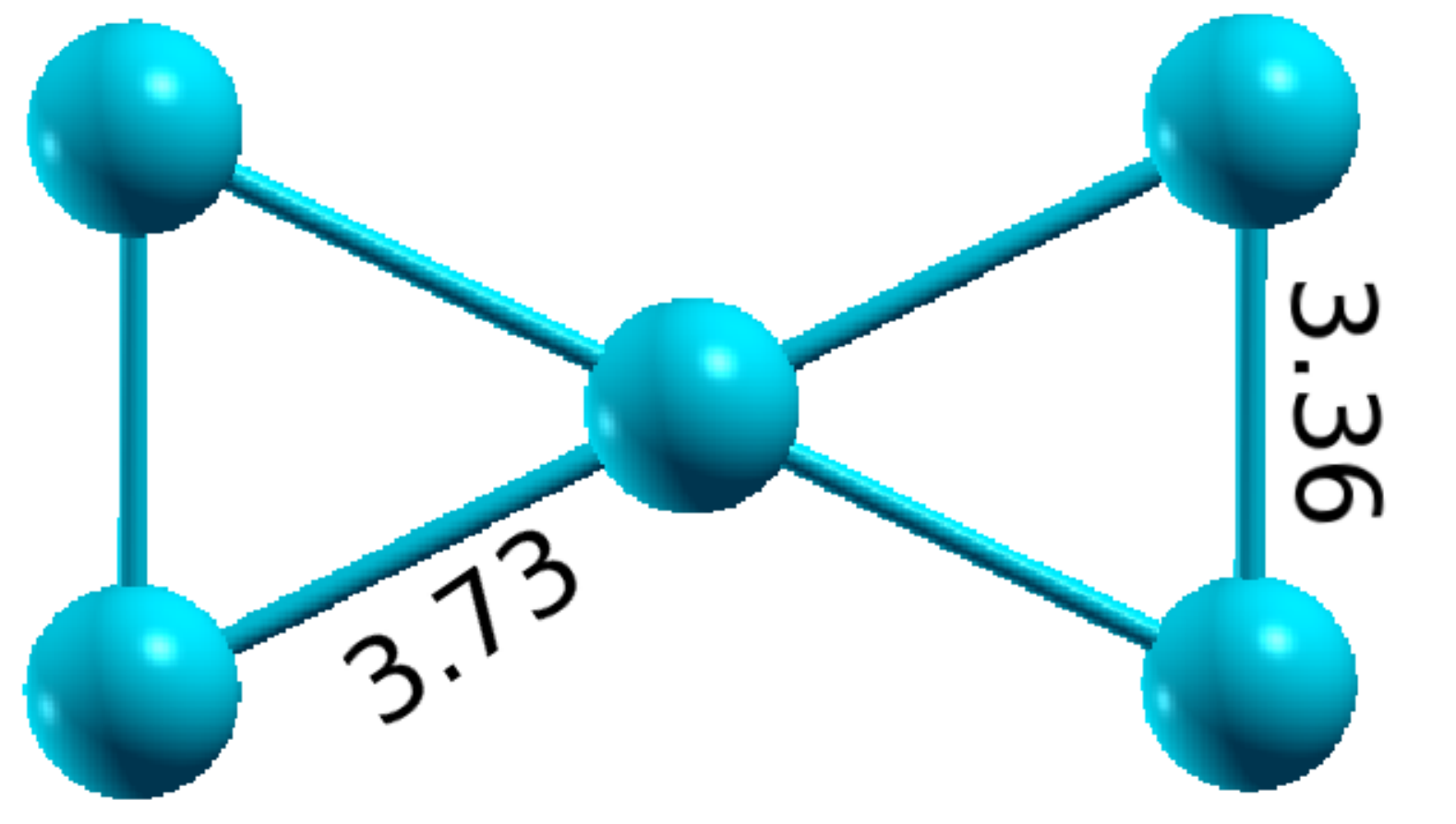}}\subfloat[\textcolor{black}{Na$_{5}^{+}$, }$\mbox{D}_{3h}$]{\includegraphics[scale=0.15]{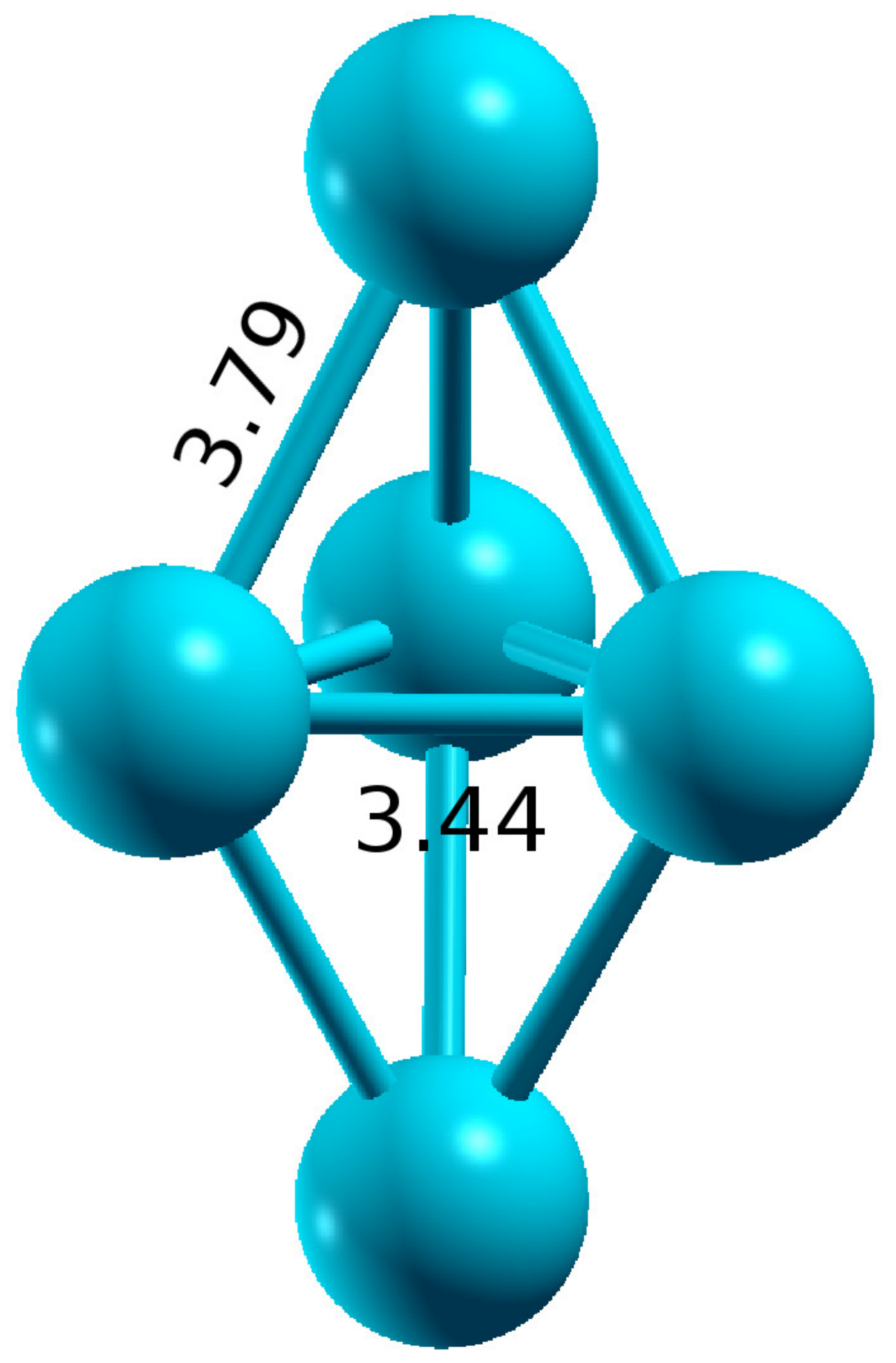}

}\subfloat[\textcolor{black}{Na$_{5}^{+}$, }$\mbox{C}_{2v}$]{\includegraphics[scale=0.2]{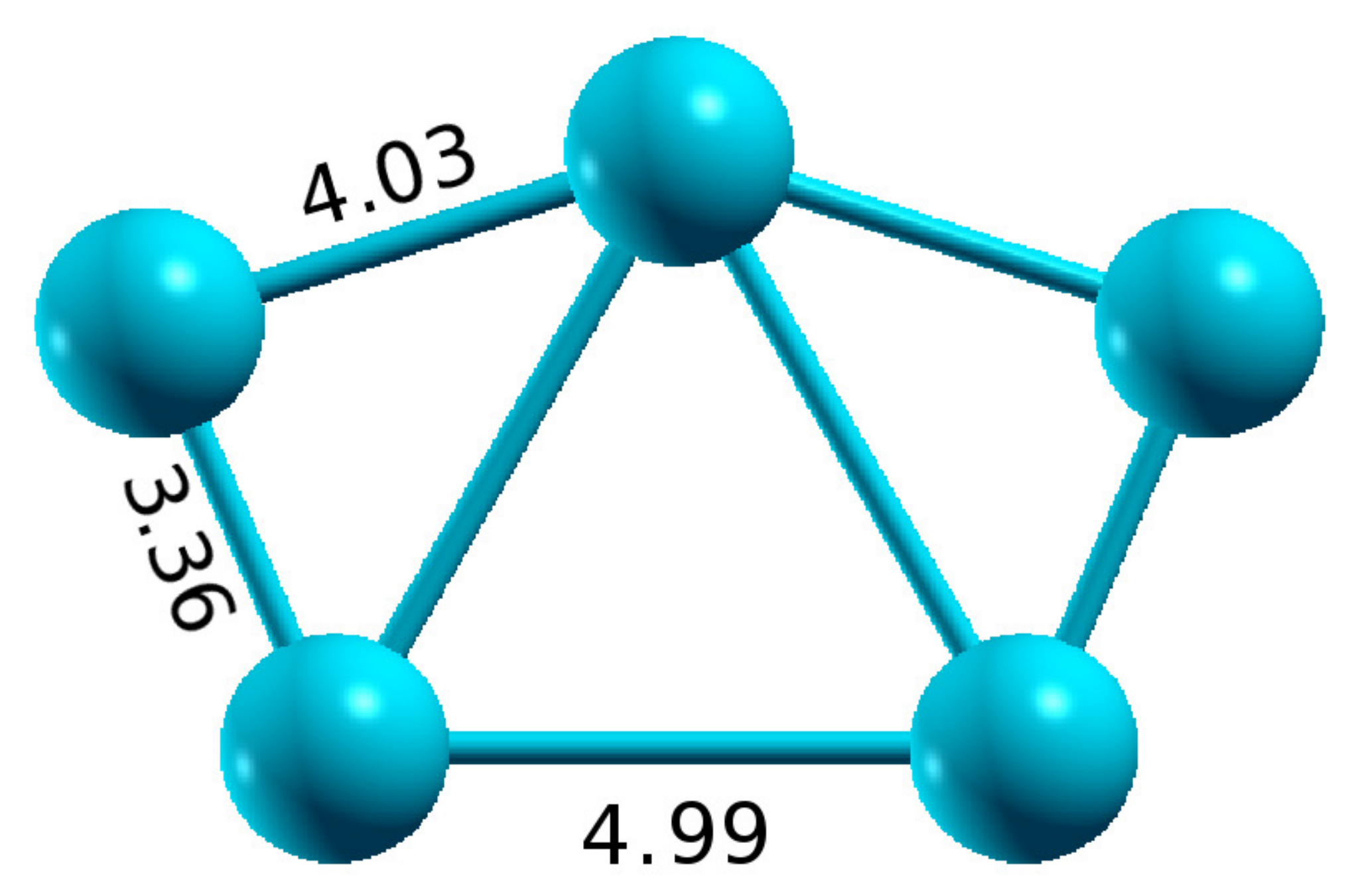}}\subfloat[\textcolor{black}{Na$_{5}$, }$\mbox{C}_{2v}-$ planar]{\includegraphics[scale=0.2]{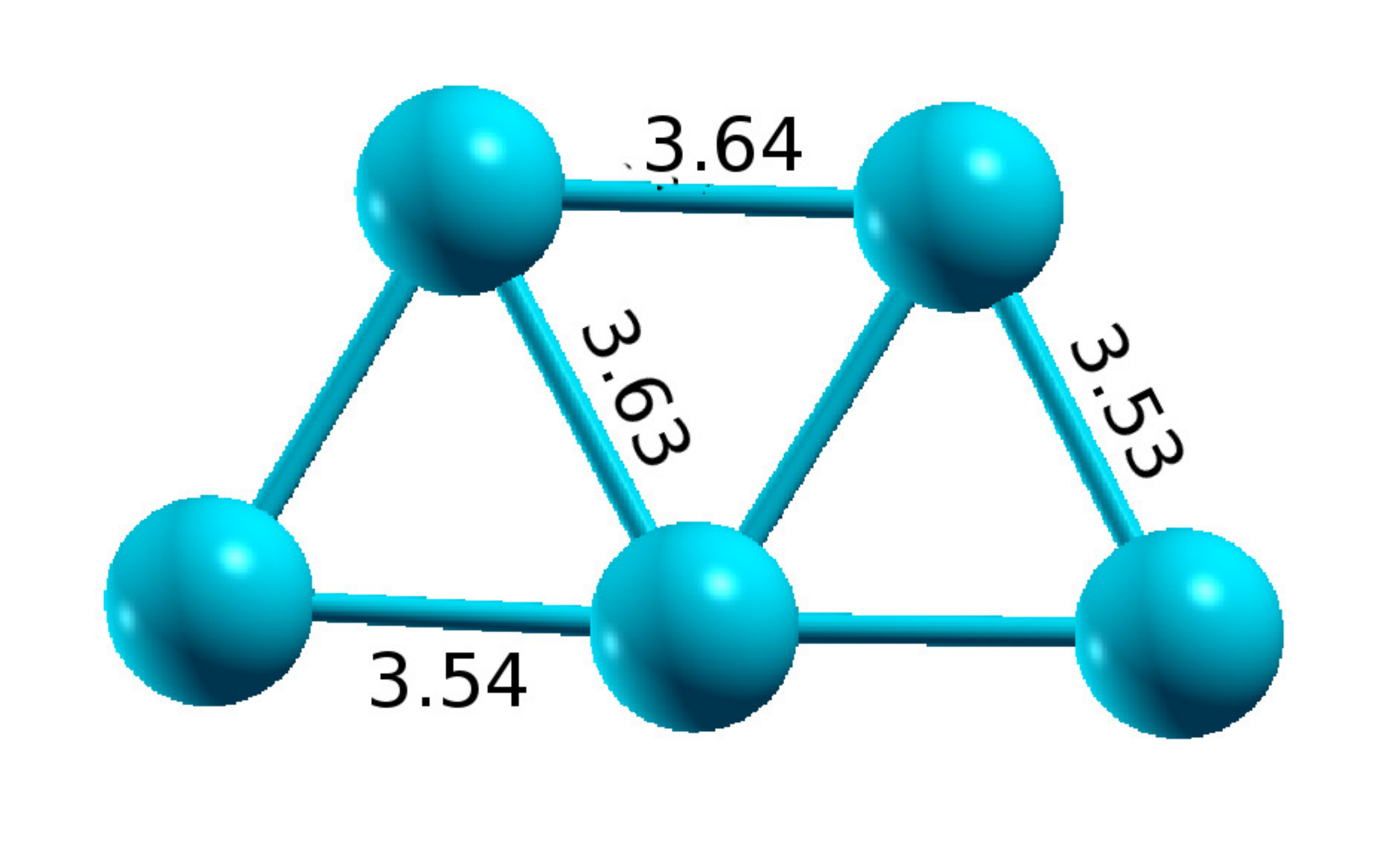}}

\subfloat[\textcolor{black}{Na$_{5}$, }$\mbox{C}_{2v}-$ bipyramidal]{\includegraphics[scale=0.24]{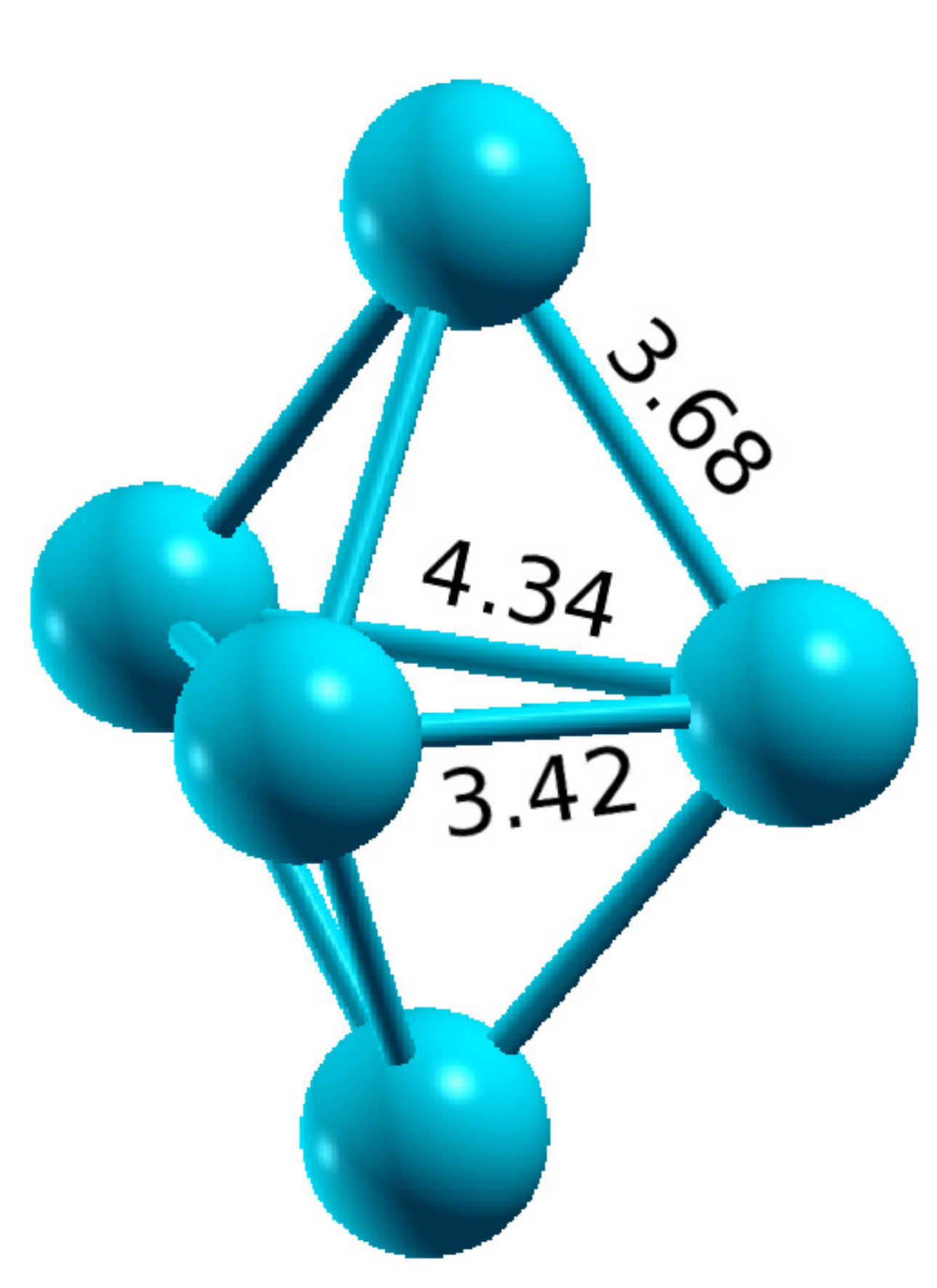}

}\subfloat[\textcolor{black}{Na$_{6}$, }$\mbox{C}_{5v}$]{\includegraphics[scale=0.24]{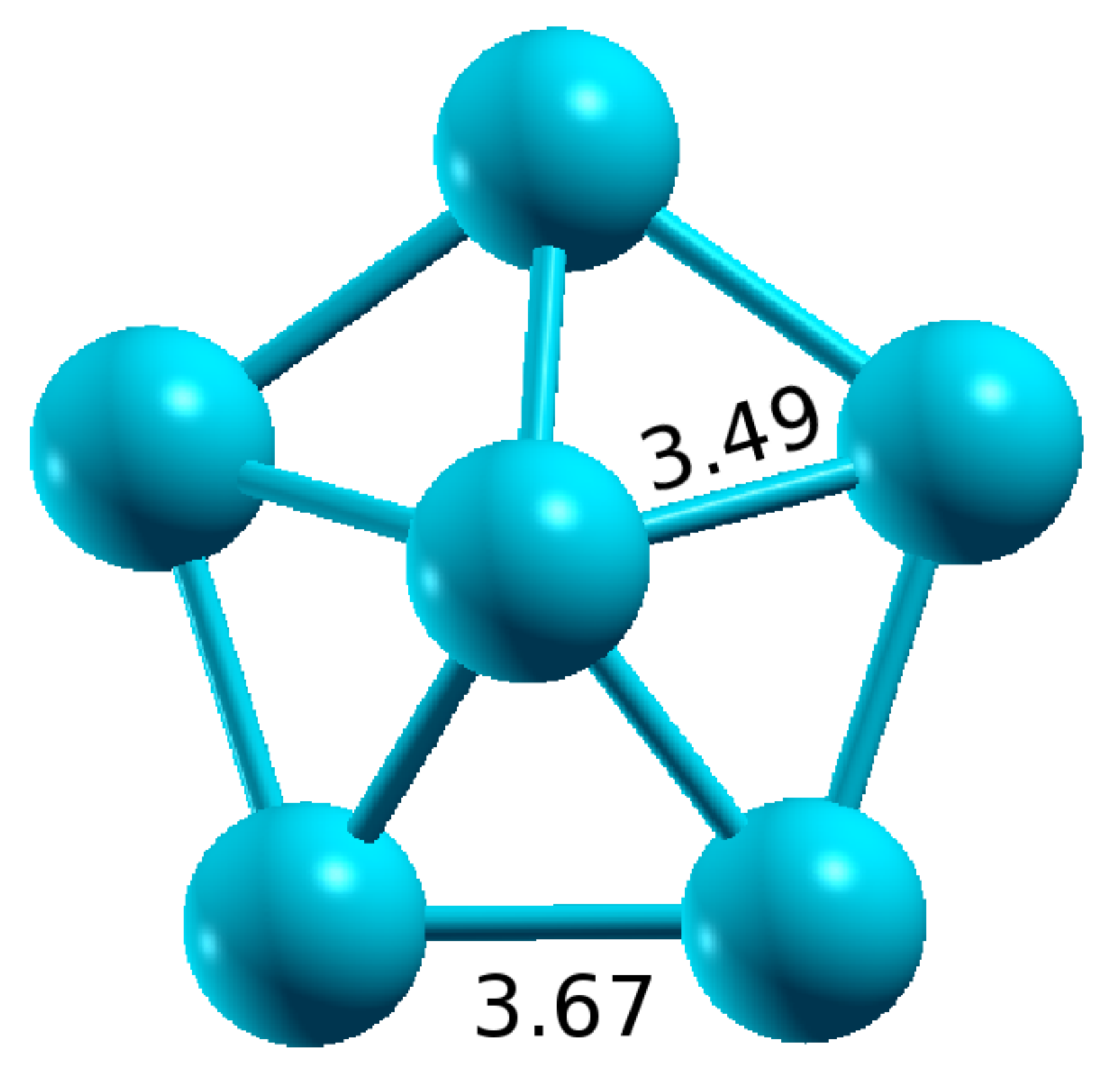}}\subfloat[\textcolor{black}{Na$_{6}$, }$\mbox{D}_{3h}$]{\includegraphics[scale=0.24]{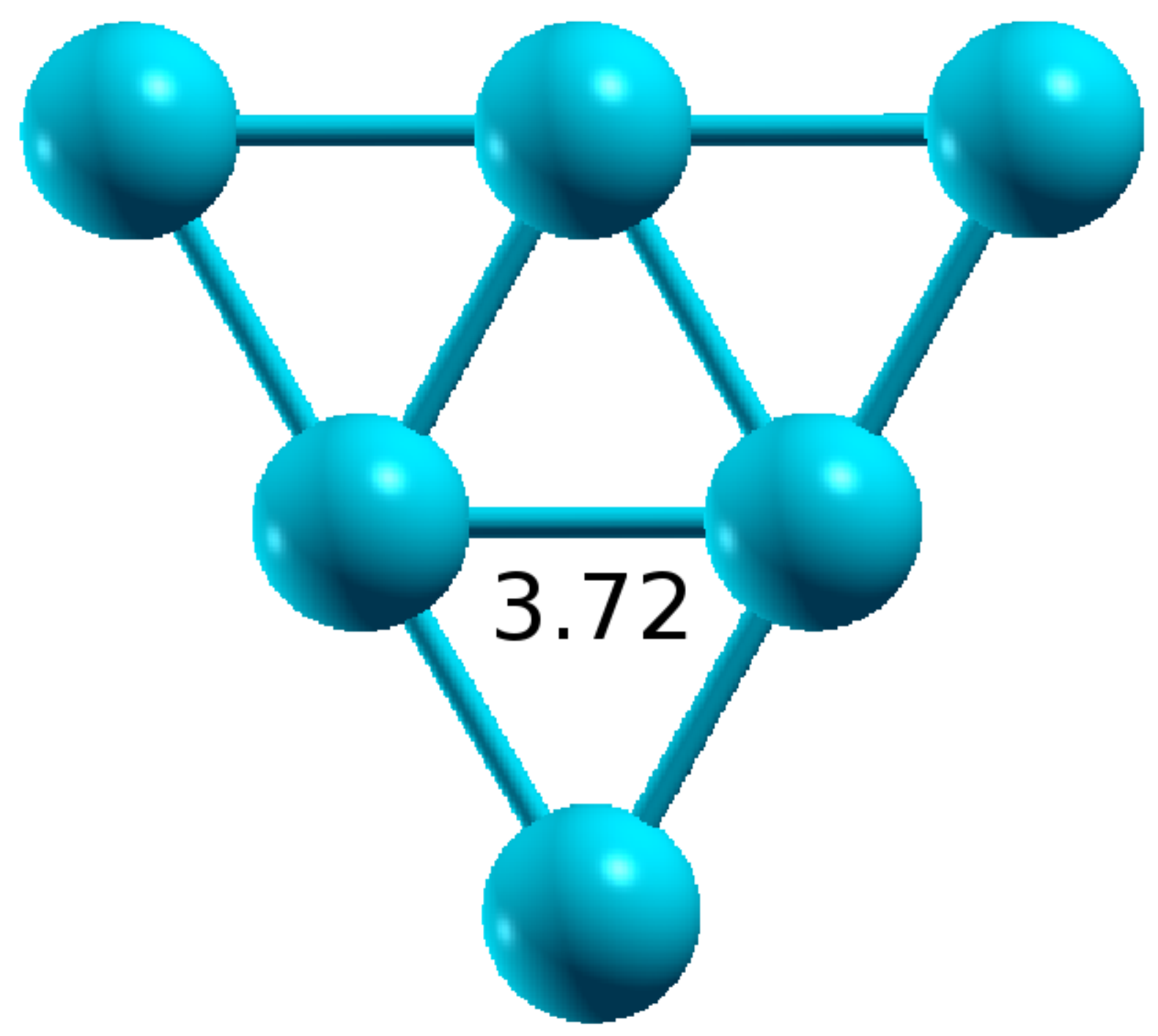}}\subfloat[\textcolor{black}{Na$_{6}$, }$\mbox{D}_{4h}$]{\includegraphics[scale=0.24]{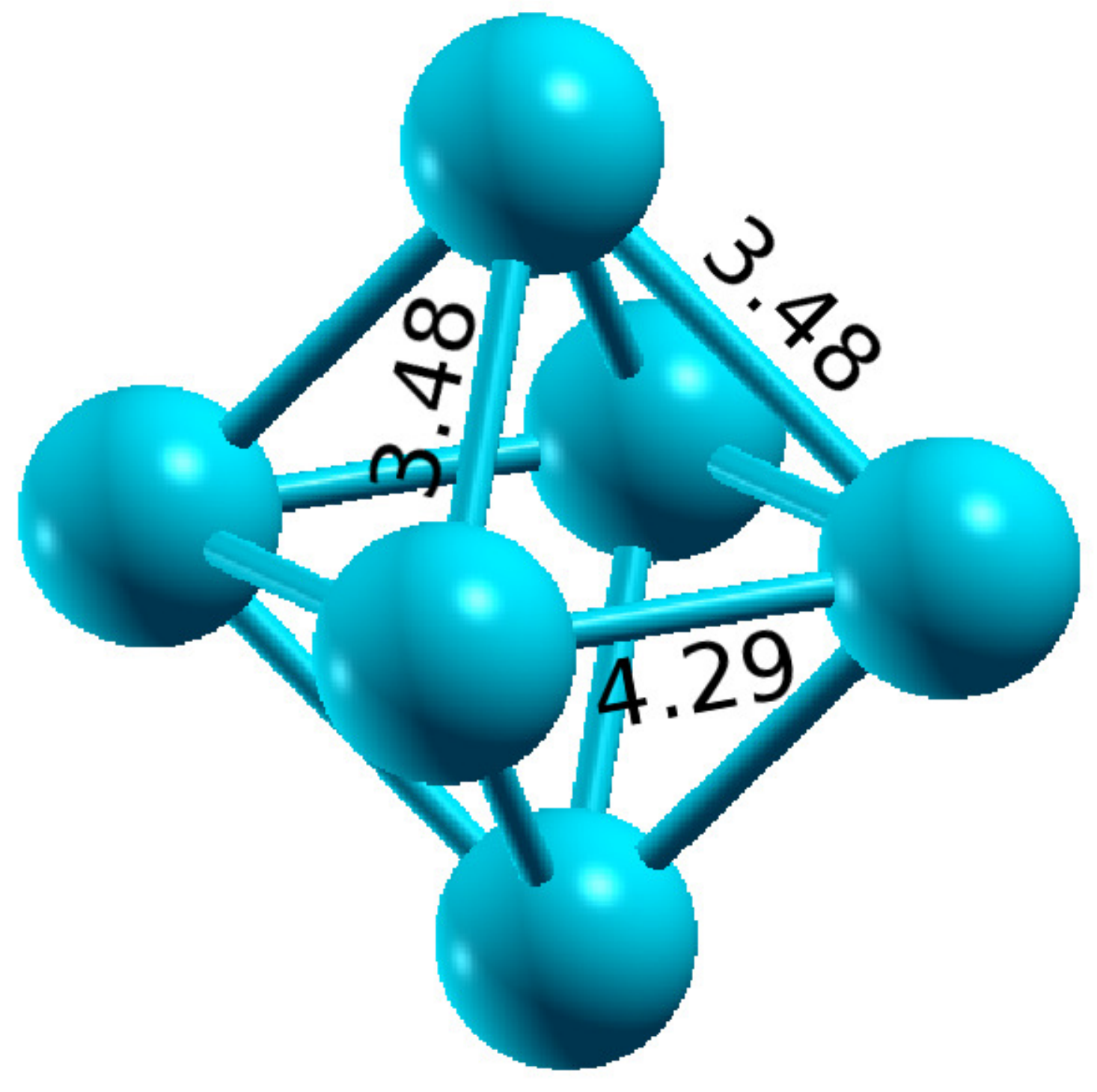}}\caption{Ground state geometries of different sodium clusters considered in
this work, with their bond lengths (\AA) indicated in the figures.}

\label{fig:Ground-state-geometries}
\end{figure}

We have performed calculations on various low-lying isomers of neutral
sodium clusters Na$_{n}$ with $n=2-6$, as well as on closed-shell
cation clusters Na$_{3}^{+}$ and Na$_{5}^{+}$. Optimized geometries
of each of those isomers are depicted in Fig. \ref{fig:Ground-state-geometries}.

Next we present and discuss the results of our calculations for different
isomers.

\subsubsection{Na\protect\textsubscript{2}}

Geometry optimization of sodium dimer was performed at the CCSD(T)
level, employing Gaussian basis set 6-311++G(3df,3pd), taken from
EMSL basis set library,\cite{emsl_bas1,emsl_bas2} and the equilibrium
bond length was obtained at 3.07 \AA\  for the singlet ground state
of symmetry $^{1}\Sigma_{g}$ (\emph{cf}. Fig. \ref{fig:Ground-state-geometries}(a)).
Pal \textit{et al}. reported an equilibrium bond length of 3.30 \AA,\cite{pal}
obtained at the HF level, which is about 0.23 \AA\  longer than
our value. Based upon \emph{ab initio} MP2 calculations Solov'yov
\textit{et al.} reported the bond length to be 3.153 \AA,\cite{solovyov}
while Ozaki and Kino,\cite{ozaki-kino-dimers-na2} based upon first
principles DFT calculations employing numerical orbitals obtained
the value 3.175 \AA. Martins \textit{et al}.\cite{martin}, reported
somewhat shorter bond length of 2.91 \AA. We note that our optimized
bond length for the sodium dimer is closer to values reported by Solov'yov
\textit{et al.},\cite{solovyov} and Ozaki and Kino,\cite{ozaki-kino-dimers-na2}
than that of Pal et al. \cite{pal}

\selectlanguage{american}%
\begin{table}[H]
\caption{\foreignlanguage{english}{Comparison of location of calculated peaks in the photoabsorption
spectrum of Na$_{2}$, with the work of other authors, and experiments.
Our calculations were performed using the frozen core FCI approach.}}

\label{tab:na2 comparison}

\begin{tabular}{|cccccccc|}
\hline 
\multirow{2}{*}{\selectlanguage{english}%
Work\selectlanguage{american}%
} & \multirow{2}{*}{Method } & \multirow{2}{*}{Bond Length (\AA)} & \multicolumn{5}{c|}{\selectlanguage{english}%
Peak Energies (Symmetry) (eV)\selectlanguage{american}%
}\tabularnewline
\cline{4-8} 
 &  &  & \selectlanguage{english}%
I($^{1}\Sigma_{u}$)\selectlanguage{american}%
 & \selectlanguage{english}%
II($^{1}\Pi_{u}$)\selectlanguage{american}%
 & \selectlanguage{english}%
\textcolor{black}{III}($^{1}\Pi_{u}$)\textcolor{black}{{} }\selectlanguage{american}%
 & \selectlanguage{english}%
IV($^{1}\Pi_{u}$)\selectlanguage{american}%
 & \selectlanguage{english}%
V($^{1}\Pi_{u}$)\selectlanguage{american}%
\tabularnewline
\hline 
\selectlanguage{english}%
This work\selectlanguage{american}%
 & \selectlanguage{english}%
FCI\selectlanguage{american}%
 & \selectlanguage{english}%
3.07\selectlanguage{american}%
 & \selectlanguage{english}%
1.87\selectlanguage{american}%
 & \selectlanguage{english}%
2.48\selectlanguage{american}%
 & \selectlanguage{english}%
3.66\selectlanguage{american}%
 & \selectlanguage{english}%
4.62\selectlanguage{american}%
 & \selectlanguage{english}%
5.46\selectlanguage{american}%
\tabularnewline
\selectlanguage{english}%
Theory (Ref.\foreignlanguage{american}{\cite{pal})}\selectlanguage{american}%
 & \selectlanguage{english}%
BSE\selectlanguage{american}%
 & \selectlanguage{english}%
3.30\selectlanguage{american}%
 & \selectlanguage{english}%
1.97\selectlanguage{american}%
 & \selectlanguage{english}%
2.56\selectlanguage{american}%
 & \selectlanguage{english}%
3.75\selectlanguage{american}%
 & \selectlanguage{english}%
\selectlanguage{american}%
 & \selectlanguage{english}%
\selectlanguage{american}%
\tabularnewline
\selectlanguage{english}%
Exp.(Ref.\cite{watson} )\selectlanguage{american}%
 & \selectlanguage{english}%
\textemdash{}\selectlanguage{american}%
 & \selectlanguage{english}%
\textemdash{}\selectlanguage{american}%
 & \selectlanguage{english}%
1.86\selectlanguage{american}%
 & \selectlanguage{english}%
2.52\selectlanguage{american}%
 & \selectlanguage{english}%
3.75\selectlanguage{american}%
 & \selectlanguage{english}%
\selectlanguage{american}%
 & \selectlanguage{english}%
\selectlanguage{american}%
\tabularnewline
\hline 
\end{tabular}
\end{table}

\selectlanguage{english}%
\begin{figure}[h]
\includegraphics[width=8cm]{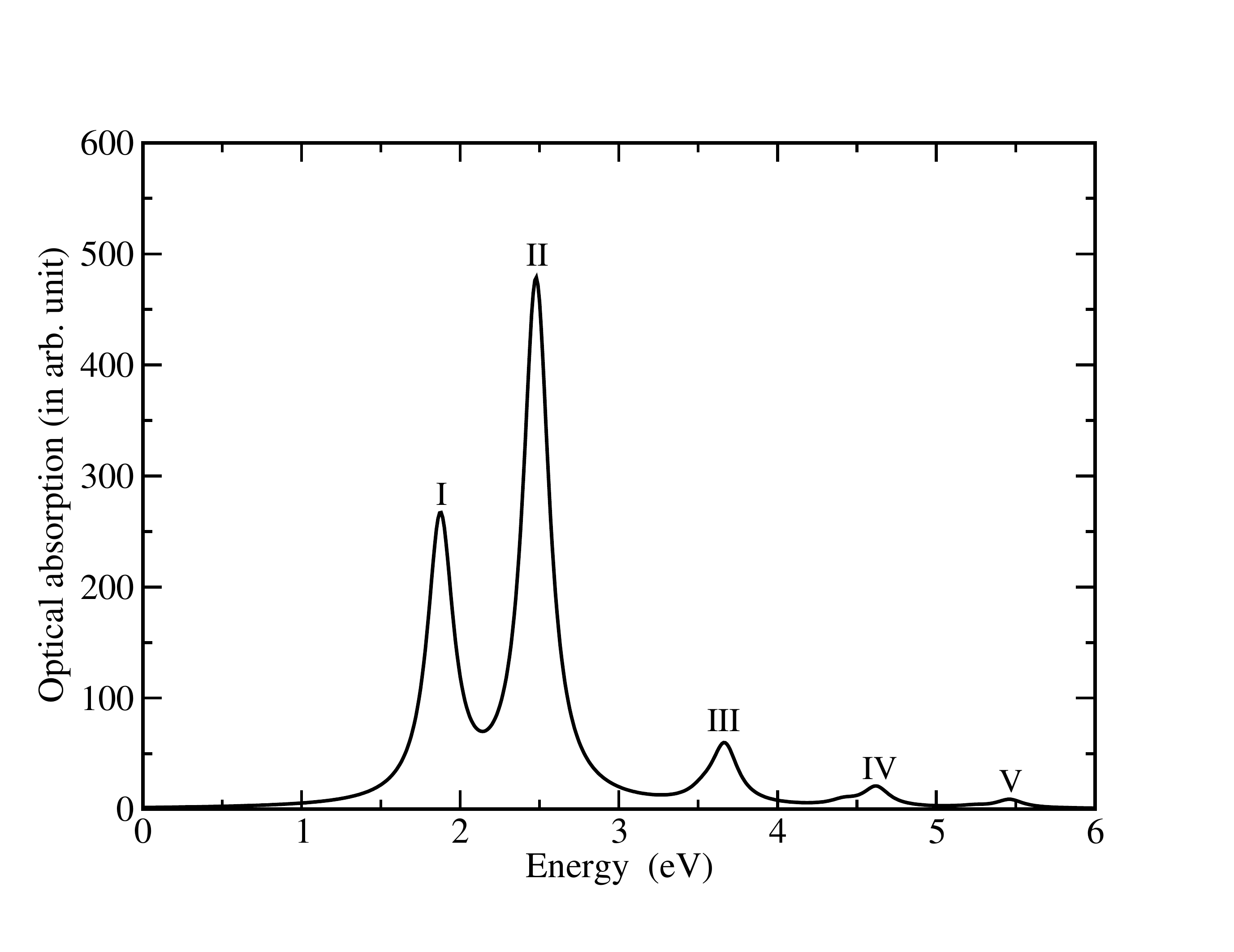}

\caption{Optical absorption spectrum of Na$_{2}$ cluster computed using the
frozen core FCI approach. For plotting the spectrum, a uniform linewidth
of 0.1 eV was used. }

\label{na2_lin}
\end{figure}

Using this bond length (3.07 \AA), we computed the optical absorption
spectrum of sodium dimer employing the FCI approach within the frozen-core
approximation, which is presented in Fig. \ref{na2_lin}. The general
qualitative features of the calculated spectrum are: (a) in the given
energy range, five peaks are observed, with most of the oscillator
strength contained in the first two peaks, (b) the first peak corresponds
to a state of $^{1}\Sigma_{u}$ symmetry, and is reached by a photon
polarized along the molecular direction (longitudinal polarization),
while the remaining peaks correspond to $^{1}\Pi_{u}$ states, accessible
by photons polarized perpendicular to the molecular direction (transverse
polarization). Next we discuss the quantitative aspects of the spectrum.

The experimental absorption spectrum of Na$_{2}$ measured by Fredrickson
and Watson,\cite{watson} and reproduced in the paper of Pal \emph{et
al}.,\cite{pal} exhibits three peaks, whose intensity profile is
in excellent agreement with the first three peaks of our calculated
spectrum (\emph{cf}. Fig. \ref{na2_lin}).The first absorption peak
in our calculated spectrum is located 1.87 eV, which is in excellent
agreement with the experimental value of 1.86 eV,\cite{watson} at
which the absorption begins. Pal \textit{et }\emph{al}.\cite{pal}
computed this peak to be located at 1.97 eV, which somewhat overestimates
the experimental value. Wave functions and other information is presented
in the Table 1 of Supporting Information, the many-particle wave function
of the excited state corresponding to this peak is dominated by $\vert H\rightarrow L\rangle$
excitation, with \foreignlanguage{american}{$\left|H\rightarrow L+3\right\rangle $}
excitation also making a significant contribution. The strongest peak
in our calculated spectrum is the second one, located at 2.48 eV,
which is also in excellent agreement with experimental value 2.52
eV reported by Fredrickson and Watson.\cite{watson} Pal \textit{et
al}.\cite{pal} computed the location of this peak to be a bit higher
at 2.56 eV. Wave function corresponding to this state exhibits strong
configuration mixing with singly-excited configurations $\left|H\rightarrow(L+2)_{1}\right\rangle $
and $\left|H\rightarrow(L+2)_{2}\right\rangle $ (orbital $L+2$ belongs
to $\pi_{u}$ irrep, and is doubly degenerate) making significant
contributions. The third peak in the experimental absorption spectrum
is much less intense as compared to the other peaks, and is located
at 3.75 eV,\cite{watson,pal} which is in good agreement with our
calculated location of peak III at 3.66 eV. The excited state corresponding
to this peak obtains its main contribution from the single excitation\textcolor{red}{{}
}$\vert H\rightarrow L+2\rangle$, $\vert H\rightarrow L+9\rangle$,
and $\vert H\rightarrow L+16\rangle$. Our calculations predict the
existence of a couple of more faint peaks located at 4.62 eV, and
5.46, which have not been reported either in the experiment,\cite{watson}
or in earlier theoretical works. The wave function of the corresponding
excited states are a strong mixture of higher-energy single excitation,
and a few doubly excited configurations. It will be interesting to
see if future experiments on sodium dimer are able to detect these
higher energy peaks, predicted in this work. 

\subsubsection{Na\protect\textsubscript{3}\protect\textsuperscript{+}}

\textcolor{black}{Next, we discuss the optical absorption spectrum
of cation Na$_{3}^{+}$, which, under the frozen-core approximation,
has only two active electrons similar to the case of sodium dimer.
The geometry of the system is well known to be an equilateral triangle
with point group $D_{3h}$, so that the electronic ground state corresponds
to the symmetry $^{1}A'_{1}$, while optically active excited states
belong to $^{1}E'$ symmetry, achievable by photons polarized in the
plane of the molecule, and $^{1}A''_{2}$ achievable through photons
polarized perpendicular to the plane of the triangle. When we optimized
the }geometry\textcolor{black}{{} of this cluster by using the 6-311++G(3df,3pd)
contracted Gaussian basis set at the full CI level, assuming frozen
cores, the optimized bond lengths of the triangle were obtained to
be 3.58 \AA,which is smaller by 0.06 \AA\  as compared to the bond
length obtained by Pal }\textit{\textcolor{black}{et al}}\textcolor{black}{.\cite{pal}
Using that bond length, in our calculated spectrum, we obtain two
dominant peaks at 2.52 eV ($^{1}E'$), and 3.15 eV ($^{1}A''_{2}$
), which are in reasonable agreement with the experimentally reported
peaks by Schmidt }\textit{\textcolor{black}{et al}}\textcolor{black}{.\cite{schmidt}
at 2.62 eV, and 3.33 eV, respectively.}\textcolor{red}{{} }\textcolor{black}{However,
Pal }\textit{\textcolor{black}{et al}}\textcolor{black}{.,\cite{pal}
using their BSE based approach, report a much better agreement with
the experiments with the computed locations of the peaks at 2.64,
and 3.36 eV, respectively.}\textcolor{red}{{} }\textcolor{black}{But,
previous theoretical work of Bona\v{c}i\'{c}-Kouteck\`{y} }\textcolor{black}{\emph{et
al}}\textcolor{black}{.,\cite{na-cation-kout-jcp-96} reports much
smaller optimized bond length 3.393 \AA\  for the ground state,
therefore, we decided to probe the influence of }geometry\textcolor{black}{{}
on optical absorption. We performed geometry optimization of the system
by going beyond the frozen-core approximation, employing the CCSD(T)
approach as implemented in the Psi4 code.\cite{psi4} Furthermore,
we also probed the influence of the basis set on the photoabsorption
spectrum of this system, and the results of this investigation are
summarized in Table \ref{tab:na3+comparison}.}\textcolor{red}{{} }\textcolor{black}{Our
best results for the absorption spectrum correspond to peak locations
2.62 eV and 3.25 eV, and were obtained from frozen-core FCI calculations
performed using 6-311++G(3df,3pd) basis set, and an optimized bond
length of 3.394 \AA\ obtained from CCSD(T) based optimization. We
note that: (a) our optimized bond length is significantly shorter
as compared to the value obtained by Pal }\textcolor{black}{\emph{et
al.}}\textcolor{black}{,\cite{pal} but is in excellent agreement
with the value obtained by Bona\v{c}i\'{c}-Kouteck\`{y} }\textcolor{black}{\emph{et
al}}\textcolor{black}{.,\cite{na-cation-kout-jcp-96} and (b) our
calculated location of the first peak is in perfect agreement with
the experiments, while the second one is underestimated by 0.08 eV.}

\begin{figure}[h]
\includegraphics[width=8cm]{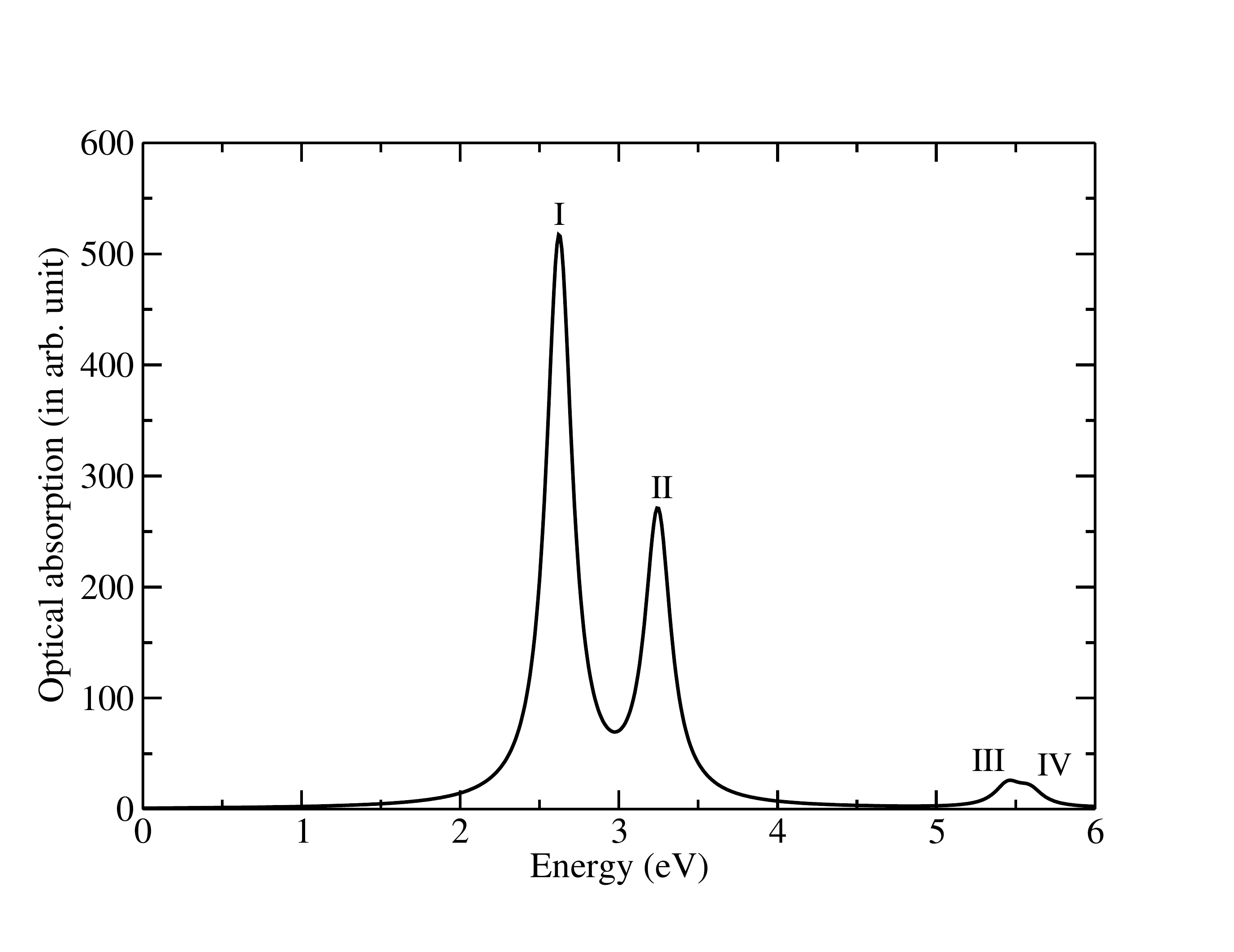}

\caption{Optical absorption spectrum of Na$_{3}$$^{+}$ cluster computed using
the frozen core full CI approach, and 6-311++G(3df,3pd) basis set.
For plotting the spectrum, a uniform linewidth of 0.1 eV was used.}

\label{na3+.eps}
\end{figure}

As far as the single-particle energy levels of the system are concerned,
the HOMO of the cluster, denoted as $H$, is non-degenerate, while
the LUMO levels, denoted $L_{1}$ and $L_{2}$, exhibit a two-fold
degeneracy. Wave functions and other information is presented in the
Table 2 of Supporting Information, many-particle wave functions of
the states giving rise to the peaks observed in the photoabsorption
spectrum are presented. From there it is obvious that the first peak,
which corresponds to a doubly degenerate state, has strong contributions
from singly excited configurations $\vert H\rightarrow L_{1}\rangle$
and $\vert H\rightarrow L_{2}\rangle$, while the wave function of
the second one is dominated by $\vert H\rightarrow L+1\rangle$ single
excitation. As compared to earlier studies, in our work, we observe
two high-energy peaks (lableld III and IV in Fig. \ref{na3+.eps})
near 5.5 eV. Peak III located at 5.45 eV belongs to an excited state
of \textcolor{black}{$^{1}E'$} symmetry, while peak IV located at
5.59 is due to a state of symmetry \textcolor{black}{$^{1}A''_{2}$.
The many-particle wave functions of these states are dominated by
single exacitations from the HOMO to higher unoccupied orbitals (}\textcolor{black}{\emph{cf}}\textcolor{black}{.
Table }2 of Supporting Information\textcolor{black}{). }These higher
energy peaks predicted by our calculations can be explored in future
experiments performed on this cluster. 

\selectlanguage{american}%
\begin{table}[H]
\caption{Comparison of location of peaks in the photoabsorption spectrum of
$Na_{3}^{+}$ calculated using various basis sets and optimized geometries,
with the theoretical works of other authors, and also with experimental
results. All our calculations were performed using the frozen core
FCI approach.}

\label{tab:na3+comparison}

\begin{tabular}{|ccccc|}
\hline 
\multirow{2}{*}{\selectlanguage{english}%
Work\selectlanguage{american}%
} & \multirow{2}{*}{Method } & \multirow{2}{*}{Bond Length (\AA)} & \multicolumn{2}{c|}{\selectlanguage{english}%
Peak Energies (Symmetry) (eV)\selectlanguage{american}%
}\tabularnewline
\cline{4-5} 
 &  &  & \selectlanguage{english}%
I(\foreignlanguage{american}{$^{1}E'$)}\selectlanguage{american}%
 & \selectlanguage{english}%
\textcolor{black}{II($^{1}A''_{2}$) }\selectlanguage{american}%
\tabularnewline
\hline 
\selectlanguage{english}%
This work$^{a}$\selectlanguage{american}%
 & \selectlanguage{english}%
FCI\selectlanguage{american}%
 & \selectlanguage{english}%
3.580\selectlanguage{american}%
 & \selectlanguage{english}%
2.52\selectlanguage{american}%
 & \selectlanguage{english}%
3.15\selectlanguage{american}%
\tabularnewline
\selectlanguage{english}%
This work$^{b}$\selectlanguage{american}%
 & FCI & \selectlanguage{english}%
 3.435\selectlanguage{american}%
 & \selectlanguage{english}%
2.60\selectlanguage{american}%
 & \selectlanguage{english}%
3.23\selectlanguage{american}%
\tabularnewline
\selectlanguage{english}%
This work$^{c}$\selectlanguage{american}%
 & \selectlanguage{english}%
FCI\selectlanguage{american}%
 & \selectlanguage{english}%
3.394\selectlanguage{american}%
 & \selectlanguage{english}%
2.60\selectlanguage{american}%
 & \selectlanguage{english}%
3.23\selectlanguage{american}%
\tabularnewline
\selectlanguage{english}%
This work$^{d}$\selectlanguage{american}%
 & \selectlanguage{english}%
FCI\selectlanguage{american}%
 & \selectlanguage{english}%
3.394\selectlanguage{american}%
 & \selectlanguage{english}%
2.62\selectlanguage{american}%
 & \selectlanguage{english}%
3.25\selectlanguage{american}%
\tabularnewline
\selectlanguage{english}%
Theory (Ref.\foreignlanguage{american}{\cite{na-cation-kout-jcp-96})}\selectlanguage{american}%
 & \selectlanguage{english}%
FCI\selectlanguage{american}%
 & \selectlanguage{english}%
3.393\selectlanguage{american}%
 & \selectlanguage{english}%
2.65\selectlanguage{american}%
 & \selectlanguage{english}%
3.37\selectlanguage{american}%
\tabularnewline
\selectlanguage{english}%
Theory (Ref.\foreignlanguage{american}{\cite{pal})}\selectlanguage{american}%
 & \selectlanguage{english}%
BSE\selectlanguage{american}%
 & \selectlanguage{english}%
3.640\selectlanguage{american}%
 & \selectlanguage{english}%
2.64\selectlanguage{american}%
 & \selectlanguage{english}%
3.36\selectlanguage{american}%
\tabularnewline
\selectlanguage{english}%
Exp.(Ref.\cite{schmidt})\selectlanguage{american}%
 & \selectlanguage{english}%
\textemdash{}\selectlanguage{american}%
 & \selectlanguage{english}%
\textemdash{}\selectlanguage{american}%
 & \selectlanguage{english}%
2.62\selectlanguage{american}%
 & \selectlanguage{english}%
3.33\selectlanguage{american}%
\tabularnewline
\hline 
\end{tabular}
\end{table}

\selectlanguage{english}%
$^{a}$Geometry optimization performed using the frozen core FCI technique,
and \foreignlanguage{american}{6-311++G(3df,3pd) basis set for all
calculations}

$^{b}$Geometry optimization performed using the CCSD(T) method using
the \foreignlanguage{american}{6-311++G(3df,3pd) basis set employed
for all calculations}

\selectlanguage{american}%
$^{c}$Geometry optimization performed using the CCSD(T) method with
CC-VTZ basis set employed for all calculations

$^{d}$Geometry optimization performed using the CCSD(T) method with
CC-VTZ basis set, and excited state calculations performed using 6-311++G(3df,3pd)
basis set.

\selectlanguage{english}%

\subsubsection{Na\protect\textsubscript{\textbf{3}}}

\begin{figure}[h]
\includegraphics[width=8cm]{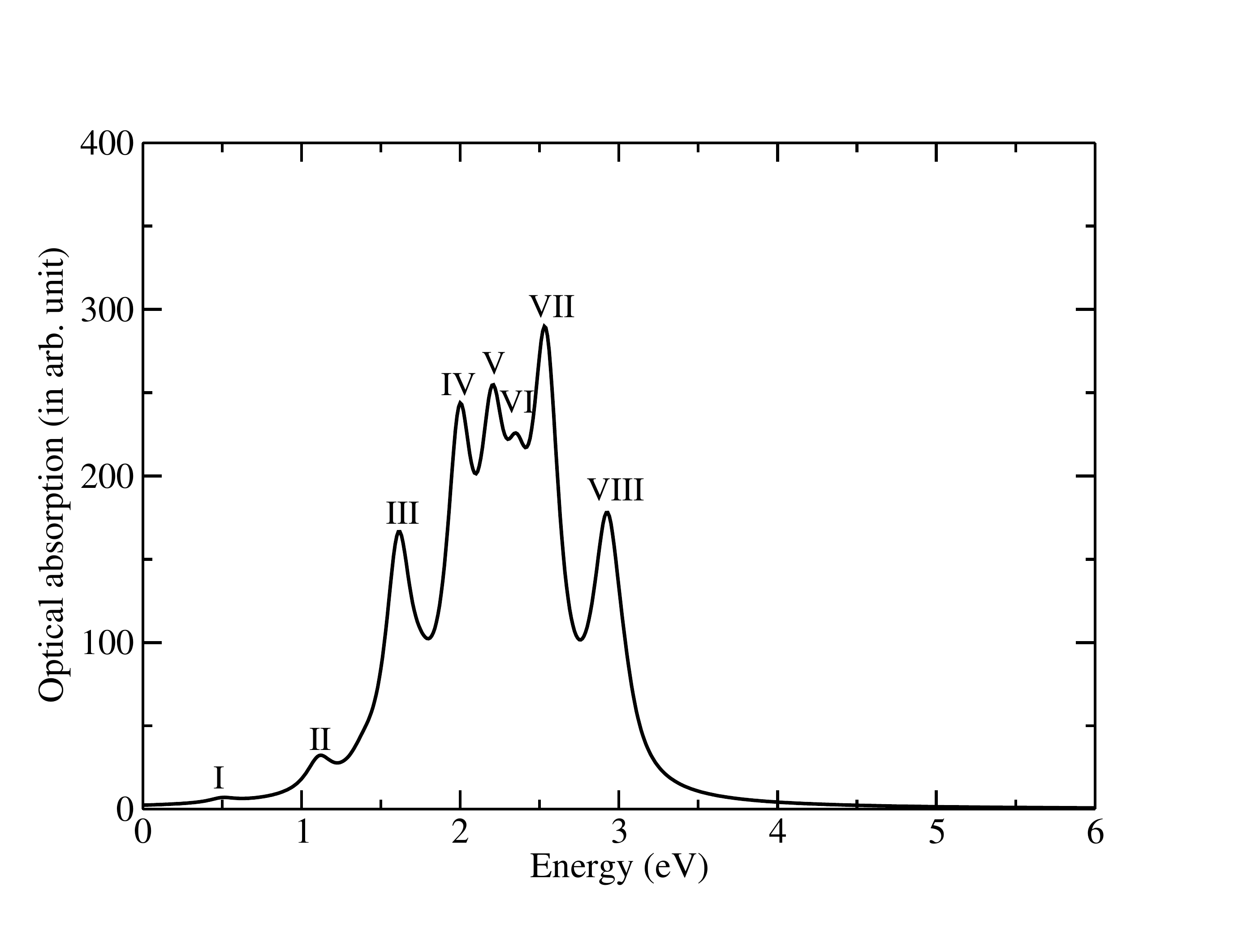}

\caption{Optical absorption spectrum of Na$_{3}$ isosceles triangular cluster
computed using the frozen core full CI approach, and 6-311++G(3df,3pd)
basis set. For plotting the spectrum, a uniform line width of 0.1
eV was used.}

\label{fig: na3_iso}
\end{figure}

\textcolor{black}{Next, we discuss optical absorption spectrum of
Na$_{3}$, which is an open-shell system, with three active electrons,
under the frozen-core approximation. The cluster possesses a symmetry
broken ground state of an isosceles triangle with point group $C_{2v}$,
as against an equilateral triangular geometry of Na$_{3}^{+}$ cluster.
In our calculations, the electronic ground state of the system was
obtained to be $^{2}B_{2}$ in agreement with previous theoretical
results.\cite{kotec} As per dipole selection rules of $C_{2v}$ point
group, with the $^{2}B_{2}$ symmetry of the ground state, linear
optical absorption can take place to the excited states of symmetries
$^{2}A_{1}$(polarized perpendicular to the plane), $^{2}A_{2}$(in-plane,
short-axis, polarized), and $^{2}B_{2}$ (in-plane, long-axis polarized).
As far as bond lengths of the cluster are concerned, several values
differing significantly from each other have been reported in the
literature. For example, Martins }\textcolor{black}{\emph{et al.,}}\textcolor{black}{\cite{martin}
based upon first-principles molecular dynamics approach, obtained
optimized bond lengths of 3.016 \AA\ (equal arms) and 3.757 \AA\ (base),
while }Calaminici \emph{et al.}\textcolor{black}{\cite{calaminici1999static}}
using first principles DFT reported lengths 3.170 \AA\ and 4.226\AA\ \textcolor{black}{for
the same bonds. Bona\v{c}i\'{c}-Kouteck\`{y} }\textcolor{black}{\emph{et
al}}\textcolor{black}{.,\cite{kotec,na3-na4-na8-kout-jcp-90} using
first-principles Hartree-Fock approach, reported two different sets
of bond lengths in two papers: (a) 3.42 }\AA\ and 4.79 \AA\textcolor{black}{,\cite{kotec}
and (b) 3.372 \AA~ and 4.509 \AA.\cite{na3-na4-na8-kout-jcp-90}}\textcolor{red}{{}
}\textcolor{black}{Therefore, we performed our own geometry optimization
for this cluster employing CCSD method with 6-311++G(3df,3pd) basis
set, and obtained values 3.249 \AA\  and 4.285 \AA, which are in
reasonably good agreement with those reported by Calaminici }\textcolor{black}{\emph{et
al.}}\textcolor{black}{\cite{calaminici1999static}}\textcolor{red}{{}
}\textcolor{black}{In order to understand the influence of geometry
on the photoabsorption spectrum, we decided to compute it for several
of these geometries, and the results of the calculations performed
using frozen-core FCI method and 6-311++G(3df,3pd) basis set, are
presented in Table \ref{tab:na3 comparison}. In the same table, we
also present }the location and symmetries of optical absorption peaks
of Na$_{3}$ measured by Wang \emph{et al.},\cite{wang} using molecular
beam photodepletion approach, while our absorption spectrum computed
using the geometry of \textcolor{black}{Calaminici }\textcolor{black}{\emph{et
al.}}\textcolor{black}{\cite{calaminici1999static} is presented in
Fig. \ref{fig: na3_iso}.}

In the measured spectrum of Na$_{3}$, Wang \emph{et al.}\cite{wang}
identified the symmetry of the ground state as the $^{2}B_{2}$, along
with seven peaks due to linear optical absorption from the ground
state, located at 1.67 eV ($^{2}A_{1}$), 1.84 eV ($^{2}A_{2}$),
2.03 eV ($^{2}A_{1}$), 2.22 eV ($^{2}B_{2}$), 2.40 eV ($^{2}B_{2}$),
2.59 eV ($^{2}A_{2}$), and 3.00 eV ($^{2}A_{2}$), where, in the
parenthesis the symmetry of the corresponding excited state is indicated.
Upon comparing this with the results of our calculations presented
in Table \textcolor{black}{\ref{tab:na3 comparison}, we conclude:
(a) peak locations predicted by our calculations are in very good
agreement with the experimental values of }Wang \emph{et al.}\cite{wang},
except for the peak of symmetry $^{2}B_{2}$ located at 1.84 eV, which
is found in none of our calculations, (b) our calculations, as well
as those of \textcolor{black}{Bona\v{c}i\'{c}-Kouteck\`{y} }\textcolor{black}{\emph{et
al}}\textcolor{black}{.,\cite{na3-na4-na8-kout-jcp-90} predict two
additional weak peaks of symmetry }$^{2}A_{1}$ located near 0.5 eV
and 1.1 eV (labeled I and II, respectively) which are not observed
in the experiment, and (c) we obtain the best agreement with the experiments
using the geometry reported by Calaminici \emph{et al.}\textcolor{black}{,\cite{calaminici1999static}
although results obtained using other geometries are also fairly close.
As far as intensity profile is concerned, our best calculation predicts
that the highest intensity peak (peak VII in Fig. \ref{fig: na3_iso})
is located at 2.53 eV and corresponds to a state of symmetry }$^{2}A_{2}$,
which is in excellent agreement with the experimental results of Wang
\emph{et al.}\cite{wang} who also assign symmetry $^{2}A_{2}$ to
their most intense peak, located at 2.59 eV. Therefore, we conclude
that our calculated optical absorption spectrum of isosceles triangle
shaped Na$_{3}$ cluster is in excellent agreement with the experimental
results.\cite{wang} Wave functions and other information is presented
in the Table 3 of Supporting Information, peak I derives its main
contribution $H\rightarrow L$ single excitation, and all the peaks
are dominated by states which are mainly composed of singly excited
configurations. 

\selectlanguage{american}%
\begin{table}[H]
\caption{\foreignlanguage{english}{Comparison of location of peaks in the photoabsorption spectrum of
Na$_{3}$ isosceles triangular cluster, calculated using various basis
sets and optimized geometries, with experimental results. All photoabsorption
calculations were performed using the frozen core FCI approach.}}

\label{tab:na3 comparison}

\begin{tabular}{|cccccccccccc|}
\hline 
\multirow{2}{*}{\selectlanguage{english}%
Work\selectlanguage{american}%
} & \multirow{2}{*}{Method } & \multirow{2}{*}{Bond Length (\AA)} & \multicolumn{9}{c|}{\selectlanguage{english}%
Peak Energies (Symmetry) (eV)\selectlanguage{american}%
}\tabularnewline
\cline{4-12} 
 &  &  & \selectlanguage{english}%
I($^{2}A_{1}$)\selectlanguage{american}%
 & \selectlanguage{english}%
II($^{2}A_{1}$)\selectlanguage{american}%
 & \selectlanguage{english}%
\textcolor{black}{III}($^{2}A_{1}$)\textcolor{black}{{} }\selectlanguage{american}%
 & \selectlanguage{english}%
($^{2}B_{2}$)\selectlanguage{american}%
 & \selectlanguage{english}%
IV($^{2}A_{1}$)\selectlanguage{american}%
 & \selectlanguage{english}%
V($^{2}B_{2}$)\selectlanguage{american}%
 & \selectlanguage{english}%
VI($^{2}B_{2}$)\selectlanguage{american}%
 & \selectlanguage{english}%
VII($^{2}A_{2}$)\selectlanguage{american}%
 & \selectlanguage{english}%
VIII($^{2}A_{2}$)\selectlanguage{american}%
\tabularnewline
\hline 
\selectlanguage{english}%
This work$^{a}$\selectlanguage{american}%
 & \selectlanguage{english}%
FCI\selectlanguage{american}%
 & \selectlanguage{english}%
3.170, 4.226 \selectlanguage{american}%
 & \selectlanguage{english}%
0.50\selectlanguage{american}%
 & \selectlanguage{english}%
1.11\selectlanguage{american}%
 & \selectlanguage{english}%
1.61\selectlanguage{american}%
 & \selectlanguage{english}%
\selectlanguage{american}%
 & \selectlanguage{english}%
1.99\selectlanguage{american}%
 & \selectlanguage{english}%
2.20\selectlanguage{american}%
 & \selectlanguage{english}%
2.35\selectlanguage{american}%
 & \selectlanguage{english}%
2.53\selectlanguage{american}%
 & \selectlanguage{english}%
2.92\selectlanguage{american}%
\tabularnewline
\selectlanguage{english}%
This work$^{b}$\selectlanguage{american}%
 & FCI & \selectlanguage{english}%
3.016, 3.757 \selectlanguage{american}%
 & \selectlanguage{english}%
0.33\selectlanguage{american}%
 & \selectlanguage{english}%
1.18\selectlanguage{american}%
 & \selectlanguage{english}%
1.57\selectlanguage{american}%
 & \selectlanguage{english}%
\selectlanguage{american}%
 & \selectlanguage{english}%
2.04\selectlanguage{american}%
 & \selectlanguage{english}%
2.25\selectlanguage{american}%
 & \selectlanguage{english}%
\selectlanguage{american}%
 & \selectlanguage{english}%
2.55\selectlanguage{american}%
 & \selectlanguage{english}%
3.00\selectlanguage{american}%
\tabularnewline
\selectlanguage{english}%
This work$^{c}$\selectlanguage{american}%
 & \selectlanguage{english}%
FCI\selectlanguage{american}%
 & \selectlanguage{english}%
3.249, 4.285 \selectlanguage{american}%
 & \selectlanguage{english}%
0.48\selectlanguage{american}%
 & \selectlanguage{english}%
1.09\selectlanguage{american}%
 & \selectlanguage{english}%
1.58\selectlanguage{american}%
 & \selectlanguage{english}%
\selectlanguage{american}%
 & \selectlanguage{english}%
1.96\selectlanguage{american}%
 & \selectlanguage{english}%
2.16\selectlanguage{american}%
 & \selectlanguage{english}%
2.33\selectlanguage{american}%
 & \selectlanguage{english}%
2.51\selectlanguage{american}%
 & \selectlanguage{english}%
2.89\selectlanguage{american}%
\tabularnewline
\selectlanguage{english}%
This work$^{d}$\selectlanguage{american}%
 & \selectlanguage{english}%
FCI\selectlanguage{american}%
 & \selectlanguage{english}%
3.372, 4.509 \selectlanguage{american}%
 & \selectlanguage{english}%
00.51\selectlanguage{american}%
 & \selectlanguage{english}%
1.05\selectlanguage{american}%
 & \selectlanguage{english}%
1.56\selectlanguage{american}%
 & \selectlanguage{english}%
\selectlanguage{american}%
 & \selectlanguage{english}%
1.93\selectlanguage{american}%
 & \selectlanguage{english}%
2.11\selectlanguage{american}%
 & \selectlanguage{english}%
2.29\selectlanguage{american}%
 & \selectlanguage{english}%
2.48\selectlanguage{american}%
 & \selectlanguage{english}%
2.84\selectlanguage{american}%
\tabularnewline
\selectlanguage{english}%
Exp.(Ref.\cite{wang})\selectlanguage{american}%
 & \selectlanguage{english}%
\textemdash{}\selectlanguage{american}%
 & \selectlanguage{english}%
\textemdash{}\selectlanguage{american}%
 & \selectlanguage{english}%
\selectlanguage{american}%
 & \selectlanguage{english}%
\selectlanguage{american}%
 & \selectlanguage{english}%
1.67\selectlanguage{american}%
 & \selectlanguage{english}%
1.84\selectlanguage{american}%
 & \selectlanguage{english}%
2.03\selectlanguage{american}%
 & \selectlanguage{english}%
2.22\selectlanguage{american}%
 & \selectlanguage{english}%
2.40\selectlanguage{american}%
 & \selectlanguage{english}%
2.59\selectlanguage{american}%
 & \selectlanguage{english}%
3.00\selectlanguage{american}%
\tabularnewline
\hline 
\end{tabular}
\end{table}

\selectlanguage{english}%
$^{a}$Geometry taken from Ref.\cite{calaminici1999static}, and excited
state calculations performed using 6-311++G(3df,3pd) basis set.

$^{b}$Geometry taken from Ref.\cite{martin}, and excited state calculations
performed using 6-311++G(3df,3pd) basis set.

$^{c}$Geometry optimization performed using the CCSD method with
6-311++G(3df,3pd) basis set, and excited state calculations performed
using 6-311++G(3df,3pd) basis set.

$^{d}$Geometry taken from Ref.\cite{koutecky}, and excited state
calculations performed using 6-311++G(3df,3pd) basis set.

\begin{figure}[h]
\includegraphics[width=8cm]{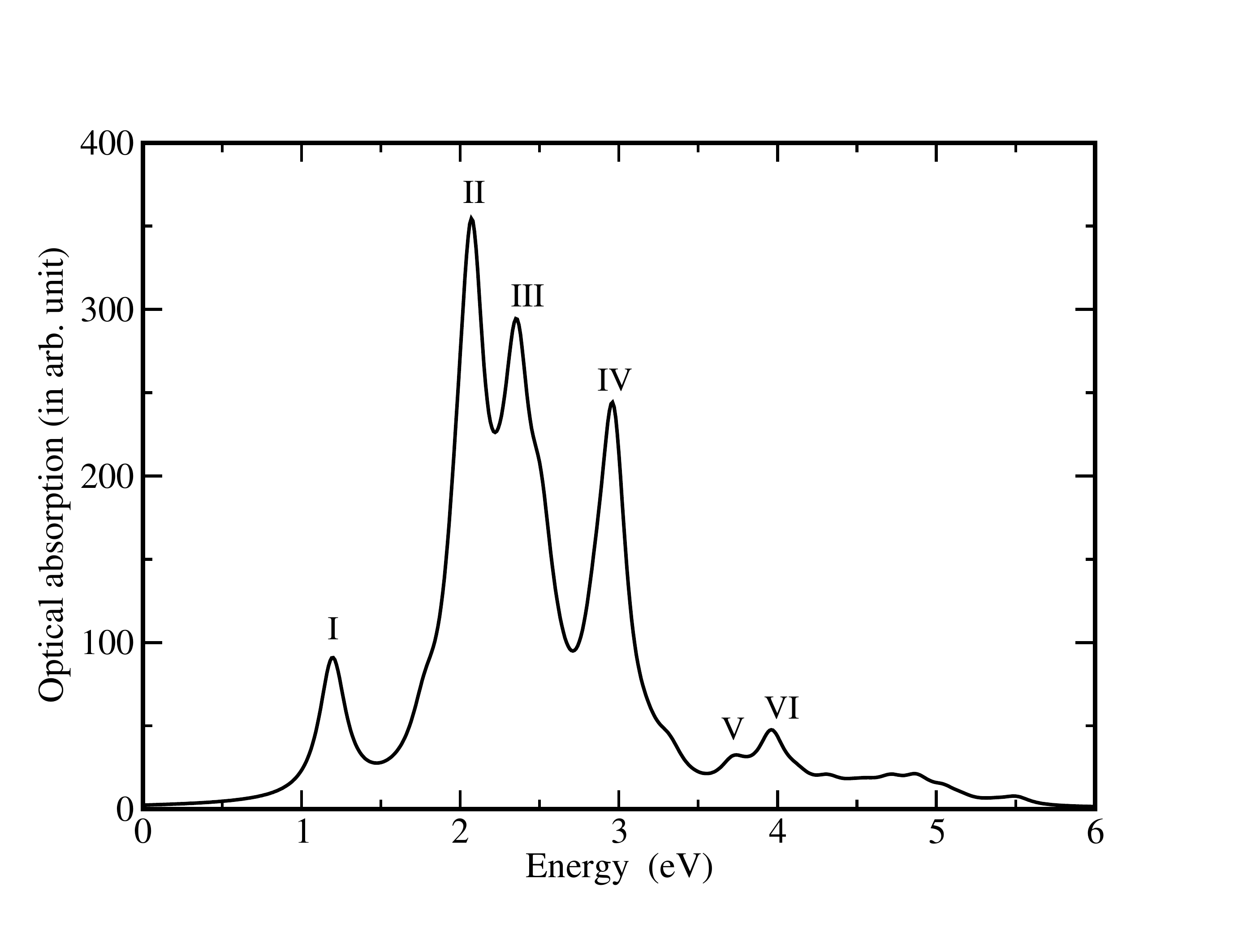}

\caption{Optical absorption spectrum of Na$_{3}$ - equilateral atomic cluster
with frozen core FCI. For plotting the spectrum, a uniform linewidth
of 0.1 eV was used. \label{fig: na3_equi}}
\end{figure}

We have also computed the optical absorption spectrum of sodium trimer,
with equilateral triangular geometry. Geometry optimization was performed
by CCSD(T) method using the 6-311++G(3df,3pd) basis set, and the optimized
bond length was obtained 3.38 \AA. Photoabsorption spectrum computed
using this optimized bond length, and the frozen-core FCI approach,
is presented in Fig. \ref{fig: na3_equi}, and exhibits six peaks
located at 1.19 eV, 2.07 eV, 2.36 eV, 2.96 eV, 3.72 eV, and 3.96 eV.
On comparing the photoabsorption spectra of isosceles and equilateral
triangles presented in Figs. \ref{fig: na3_iso} and \ref{fig: na3_equi},
we note significant qualitative and quantitative differences between
them which should allow one to differentiate between the two geometries
using absorption-based experiments. Wave functions and other information
is presented in the Table 4 of Supporting Information, we find that
the first three peaks in the absorption spectrum of the equilateral
triangular cluster correspond to photons polarized in plane of the
molecule, in complete contrast with the results for the isosceles
triangular cluster (\emph{cf}. Table 3 of Supporting Information).
As far as the wave functions of excited states contributing to the
peaks are concerned (\emph{cf}. Table 4 of Supporting Information),
first four peaks consist of states whose wave functions are dominated
by single excitations, while wave functions of those contributing
to last two peaks derive main contributions from double excitations. 

\begin{figure}[h]
\includegraphics[width=7cm]{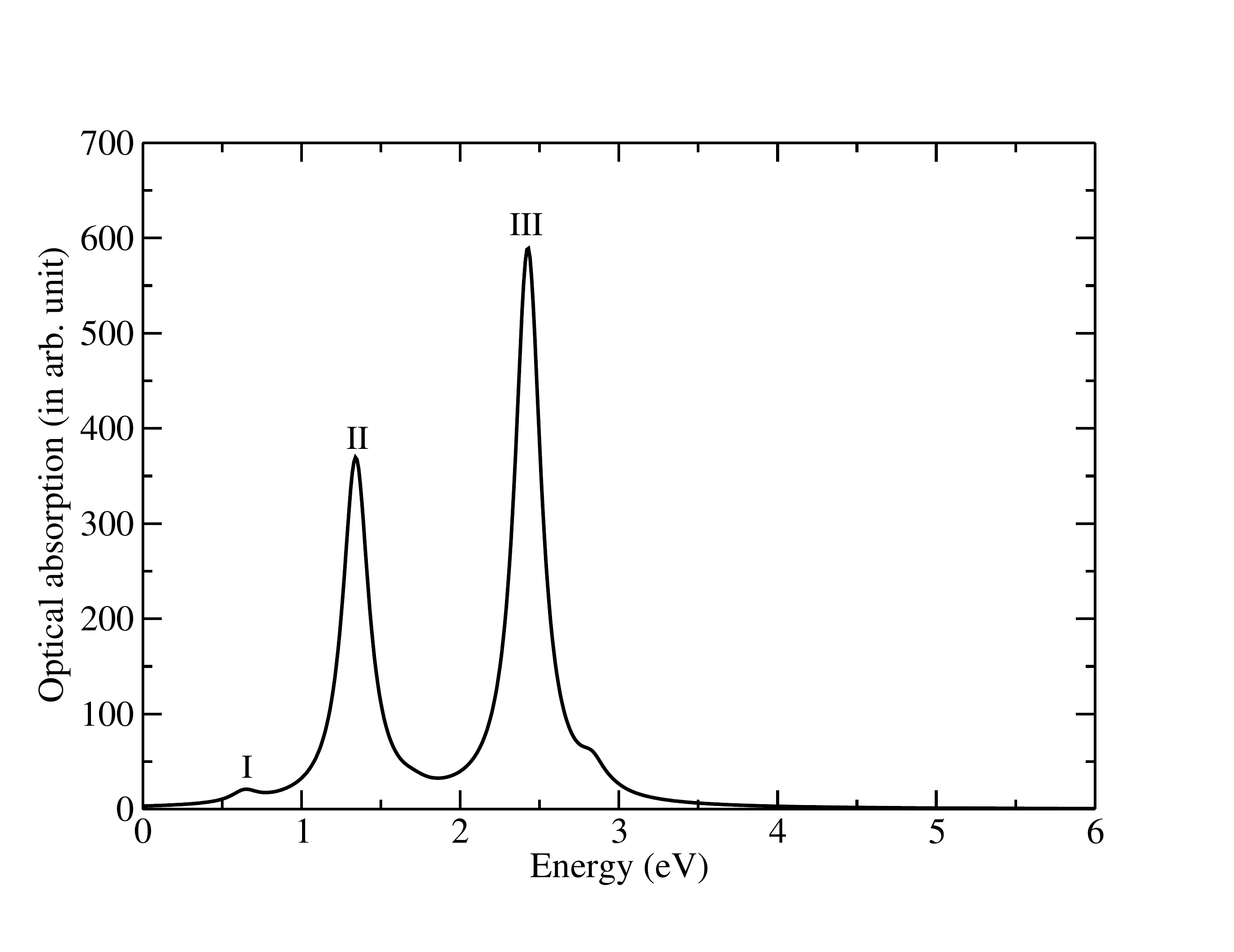}

\caption{Optical absorption spectrum of linear isomer of Na$_{3}$ cluster,
computed using the frozen core FCI approach. For plotting the spectrum,
a uniform linewidth of 0.1 eV was used.}
\label{fig: na3_lin}
\end{figure}

Apart from triangular geometries, we have also performed the optical
absorption calculations for linear Na$_{3}$ for which the ground
state has optimized bond length 3.41 Å with symmetry $^{2}\Sigma_{u}$,
of symmetry group $D_{\infty h}$. According to dipole selection rules
of group $D_{\infty h}$, optical absorption will take place to excited
states of symmetries $^{2}\Sigma_{g}$, polarized along the length
of the molecule, and doubly-degenerate $^{2}\Pi_{g}$ states, polarized
in a plane perpendicular to the molecular axis. Calculated photoabsorption
spectrum shown in Fig. \ref{fig: na3_lin} exhibits three peaks located
at 0.64 eV ($^{2}\Sigma_{g}$), 1.34 eV ($^{2}\Sigma_{g}$) and 2.43
eV ($^{2}\Pi_{g}$), of which the first peak is a rather weak one.
Wave functions and other information is presented in the Table 5 of
Supporting Information, we note that for the first two peaks, the
wave functions are dominated by singly excited configurations, while
the third peak is doubly degenerate and wave functions are dominated
by both singly and doubly excited configurations. Strong configuration
mixing exhibited by wave functions of excited states contributing
to peak III, points to the importance of electron-correlation effects
in the description of higher energy states. 

\subsubsection{Na\protect\textsubscript{\textbf{4}}}

\begin{figure}[h]
\includegraphics[width=8cm]{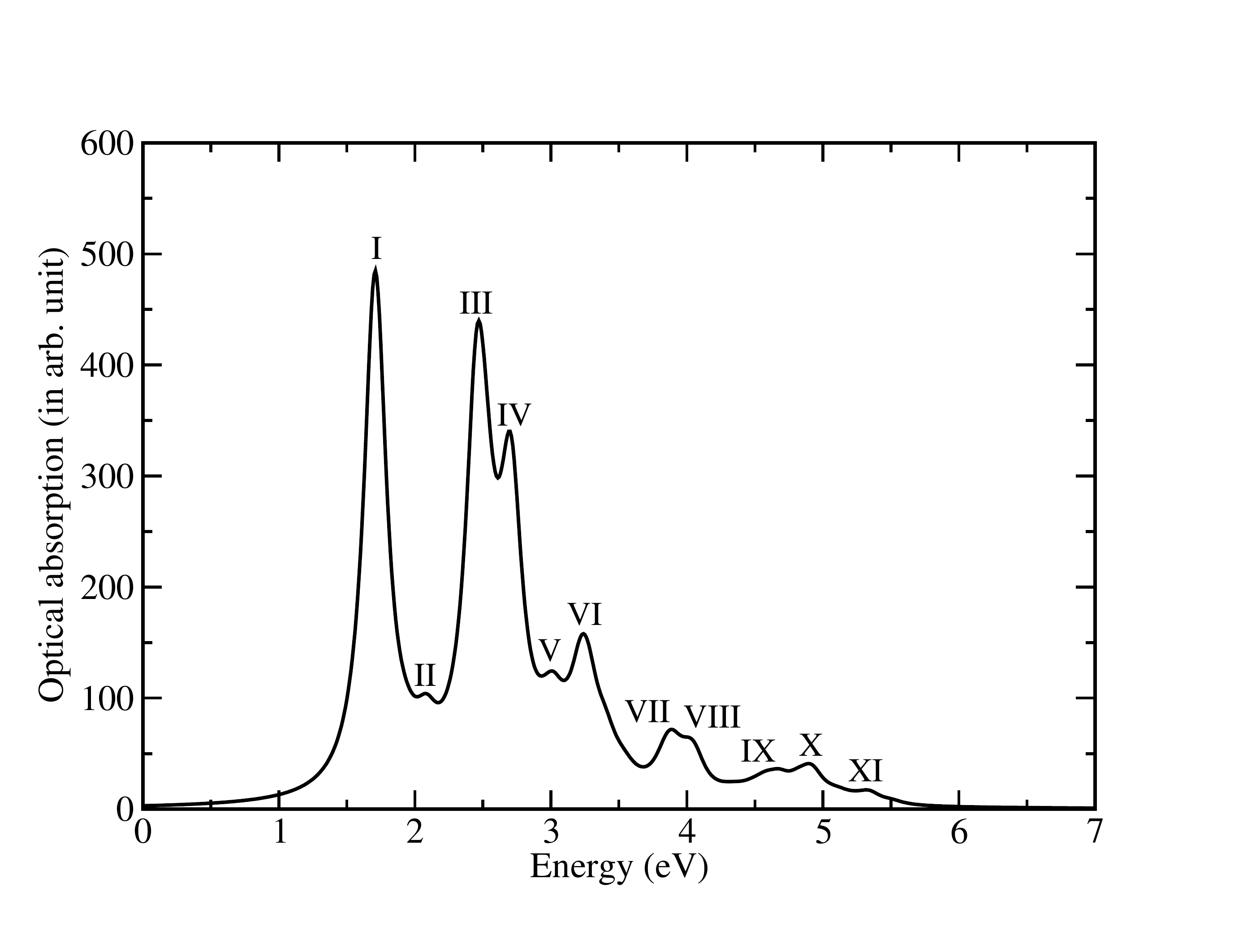}

\caption{Optical absorption spectrum of rhombus shaped Na$_{4}$ cluster, performed
using the frozen-core FCI approach. For plotting the spectrum, a uniform
line width of 0.1 eV was used.}

\label{fig:na4_rhombus_spec}
\end{figure}

Geometry optimization for this cluster was performed using 6-311++G(3df,3pd)
contracted Gaussian basis set, using the CCSD(T) approach implemented
in Psi4 computer program.\cite{psi4} The resultant rhombus-shaped
equilibrium geometry shown in Fig. \ref{fig:Ground-state-geometries}(f)
is almost identical to those reported by Solovyov \textit{et al}.,\cite{solovyov}
and Pal \textit{et al}.\cite{pal} The ground state belongs to the
$^{1}A_{g}$ irreducible representation of the $D_{2h}$ point group,
so that the excited states visible in the linear optical absorption
will correspond to the symmetries $^{1}B_{1u}$ (out of plane polarized),
$^{1}B_{2u}$ (short-diagonal polarized) and $^{1}B_{3u}$ (long-diagonal
polarized). Optical absorption calculations were performed at the
frozen-core FCI level using the same 6-311++G(3df,3pd) basis set,
and the results are presented in Fig. \ref{fig:na4_rhombus_spec}.
Wang \textit{et al}..\cite{wang,pollack} measured the optical absorption
spectrum of this cluster using the molecular beam photodepletion approach,
and found eight absorption peaks located at: 1.63 eV ($^{1}B_{2u}$),
1.80 eV ($^{1}B_{3u}$), 1.98 eV ($^{1}B_{3u}$), 2.18 eV ($^{1}B_{1u}$),
2.51 eV ($^{1}B_{2u}$), 2.63-2.78 eV ($^{1}B_{1u}$), 2.85-3.15 eV
($^{1}B_{1u}$, $^{1}B_{3u}$), 3.33 eV ($^{1}B_{2u}$), out of which
the ones at 1.80 eV and 2.51 eV are the most intense. Our calculated
spectrum (\emph{cf}. Fig. \ref{fig:na4_rhombus_spec}) shows several
peaks out of which the locations and the symmetries of the first eight
are: 1.71 eV ($^{1}B_{3u}$), 2.09 eV ($^{1}B_{1u}$), 2.47 eV ($^{1}B_{2u}$),
2.69 eV ($^{1}B_{1u}$), 3.01 eV ($^{1}B_{1u}$), 3.22/3.28 eV ($^{1}B_{2u}$/$^{1}B_{1u}$),
3.90 eV ($^{1}B_{1u}$), and 4.01 eV ($^{1}B_{1u}$), out of which
the peaks 1.71 eV and 2.47 eV are the most intense. Thus, on comparing
our theoretical results with the experimental ones,\cite{wang} we
find that the location and symmetries of the two most intense peaks
are reproduced excellently by our calculations. Furthermore, our weaker
peaks located at 2.69 eV, 3.01 eV, 3.22 eV compare very well with
experimental peaks at 2.63-2.78 eV, 2.85 eV, 3.33 eV, both in locations
and symmetries. On three weak peaks there is disagreement between
the theory and the experiments: (a) the experimental absorption starts
with a weak peak at 1.63 eV ($^{1}B_{2u}$) which is absent in our
calculations, and (b) experimental spectrum observes couple of weak
peaks corresponding to $^{1}B_{3u}$ symmetry states located at 1.98
eV and 3.15 eV, which are absent in our spectrum. Thus, out of eight
experimental peaks,\cite{wang} our theoretical calculations are able
to resolve five peaks quite well, as far as their energetic location,
point group symmetries, and intensities are concerned. Wave functions
and other information is presented in the Table 6 of Supporting Information.
From the table it is obvious that the wave functions of the excited
states behind both the two intense peaks located at 1.71 eV and 2.47
eV are dominated by single excitation. The peak at 1.71 eV derives
main contributions from the excitation $|H\rightarrow L+2\rangle$
and $|H\rightarrow L+10\rangle$, while the one at 2.47 eV is dominated
by four configurations $|H\rightarrow L+6\rangle$, $|H\rightarrow L+18\rangle$,
$|H\rightarrow L\rangle$, and $|H\rightarrow L+16\rangle$.

\begin{figure}[h]
\includegraphics[width=8cm]{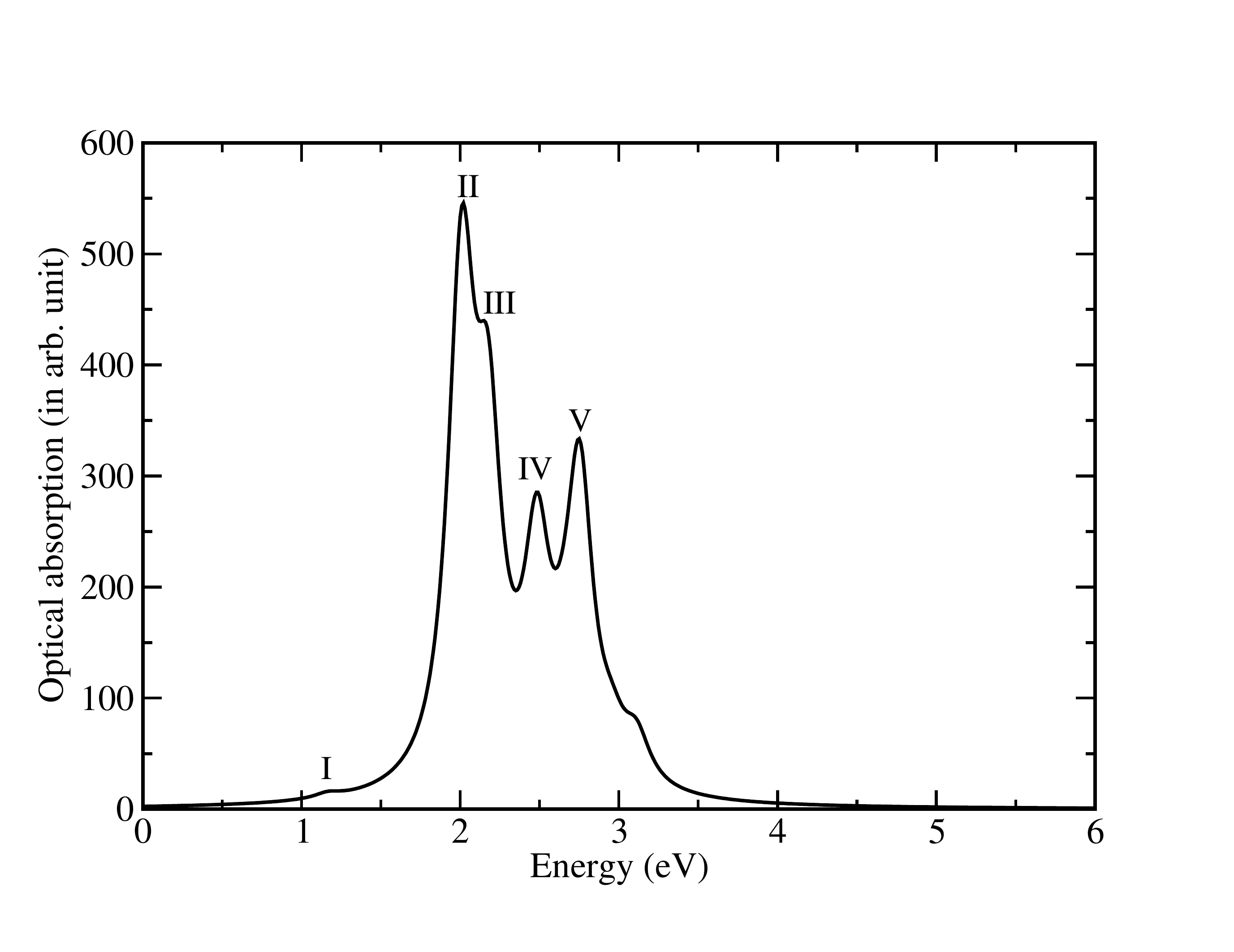}

\caption{Optical absorption spectrum of Na$_{4}$ - square-shaped atomic cluster
computed using the frozen core FCI method. For plotting the spectrum,
a uniform linewidth of 0.1 eV was used. \label{fig: na4_square_abs}}
\end{figure}

We also performed the optical absorption calculations for square-shaped
Na$_{4}$ cluster, employing the frozen-core FCI approach, and the
results are presented in Fig. \ref{fig: na4_square_abs}, along with
the excited wave functions in Table 7 of Supporting Information. For
these calculations, 6-311++G(3df,3pd) basis set was used, along with
bond length 3.48 \AA, which was obtained using all-electron CCSD(T)
geometry optimization.\cite{psi4} The ground state geometry corresponds
to symmetry $^{1}A_{1g}$ of the $D_{4h}$ point group, therefore,
linear absorption, as per dipole selection rules, will be to excited
states of symmetries $^{1}E_{u}$ (in-plane polarized), and $^{1}A_{2u}$
(perpendicularly polarized). Wave functions and other information
is presented in the Table 7 of Supporting Information, first four
absorption peaks correspond to doubly degenerate $^{1}E_{u}$ symmetry
located at 1.16 eV, 2.01 eV, 2.17 eV, and 2.48 eV, while the fifth
peak located at 2.75 eV corresponds to an excited state of symmetry
$^{1}A_{2u}$. This result is qualitatively distinct as compared to
the case of rhombus (\emph{cf.} Fig. \ref{fig:na4_rhombus_spec}),
for which, out of the first four absorption peaks, two were polarized
perpendicular to the plane of the rhombus. Another interesting aspect
of the first peak of square-shaped cluster is that the excited states
giving rise to it derive significant contributions from double excitations
(\emph{cf}. Table 7 of Supporting Information), while for the rhombus
cluster, the wave function of the first optically active state is
dominated by single excitations (\emph{cf}. Table 6 of Supporting
Information). Furthermore, compared to the rhombus shaped cluster,
optical absorption spectrum of the square-shaped cluster has relatively
lesser number of features, and also different excitation energies
for various peaks. Additionally, the first peak of the rhombus is
the most intense one of its spectrum, while the first one of the square
is the least intense one of the corresponding absorption spectrum
Therefore, peak locations, along with the intensity profiles of the
photoabsorption spectra of the two isomers can be used to distinguish
them using optical spectroscopy. 

\subsubsection{Na$_{5}{}^{+}$\protect\textsubscript{}\protect\textsuperscript{}}

For Na$_{5}^{+}$, we have considered four different isomers with
symmetries $D_{2d}$ , $D_{2h}$ , $D_{3h}$, and $C_{2v}$, as shown
in Figs. \textcolor{black}{\ref{fig:Ground-state-geometries}(h)\textemdash \ref{fig:Ground-state-geometries}(k),
respectively. Various bond lengths corresponding to the }$D_{2d}$
, $D_{2h}$ and $D_{3h}$\textcolor{black}{{} isomers indicated in those
figures were obtained by performing geometry optimization using all-electron
CCSD(T) approach implemented in the Psi4 program},\cite{psi4} while
for\textcolor{black}{{} the $C_{2v}$ isomer we used }the geometry reported
by Pal \textit{et at}.\cite{pal}. We note that Pal \emph{et al}.
considered a $C_{4v}$ isomer instead of $D_{2h}$, however, our geometry
optimization program led to the $D_{2h}$ structure even when we initialized
it with the $C_{4v}$ geometry parameters of Pal \emph{et al}. Although,
our geometry optimization predicts the $D_{2d}$ isomer to be the
lowest energy structure, but the total CCSD(T) energies (\emph{cf.}
Table \ref{tab_ener}) of all the four isomers are very close to each
other, with $D_{2d}$ and $D_{2h}$ energies being nearly degenerate,
while energies of $D_{3h}$ and $C_{2v}$ isomers being just 0.13
eV, and 0.64 eV, respectively, higher than that of the $D_{2d}$ isomer.
Given the energetic proximities of these isomers, there is a high
probability that in room temperature optical absorption experiments,
absorption spectra of several of these isomers could be observed.
Next, we discuss the calculated absorption spectra of each of these
isomers.

\begin{figure}[h]
\includegraphics[width=7cm]{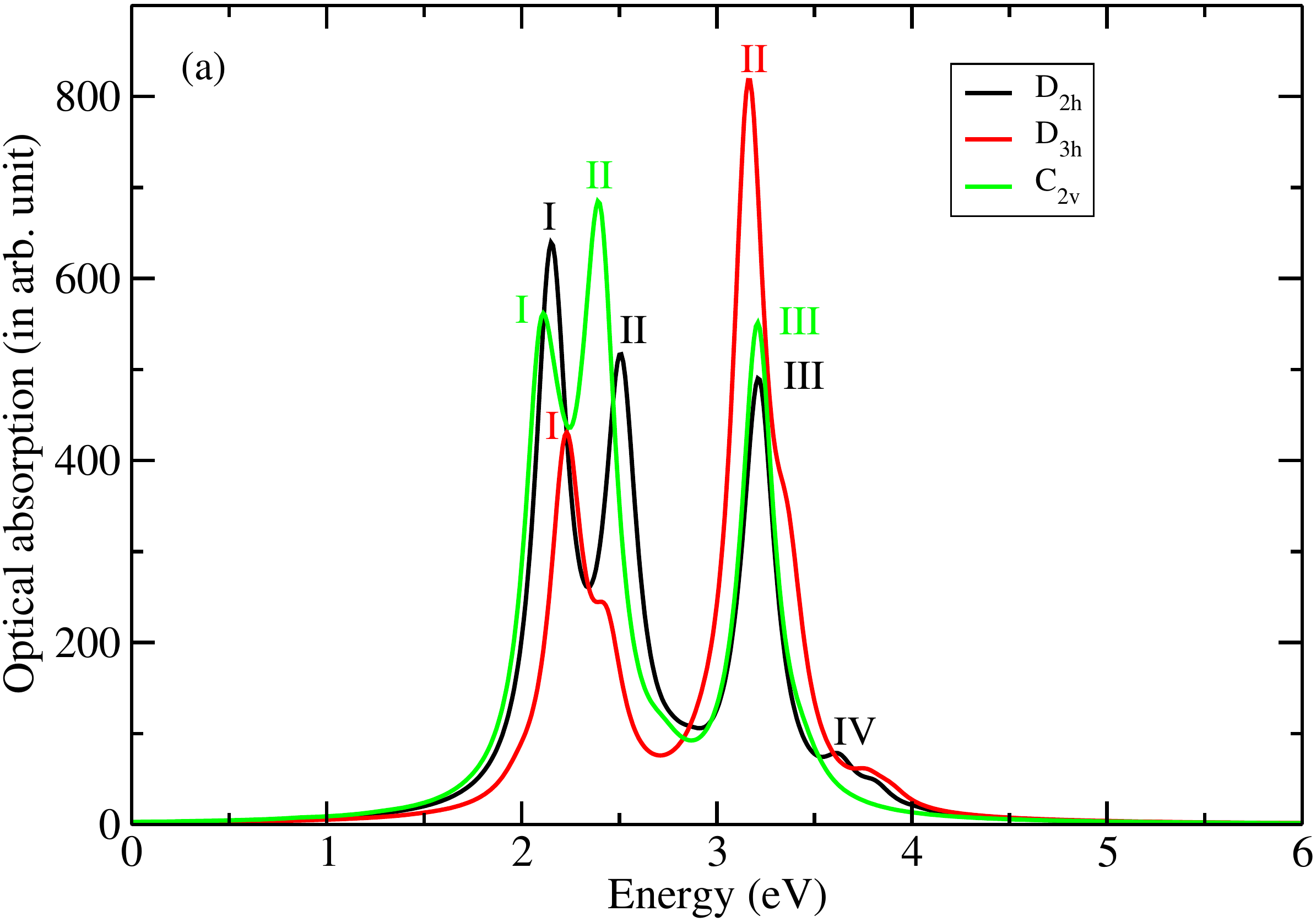}\includegraphics[width=7cm]{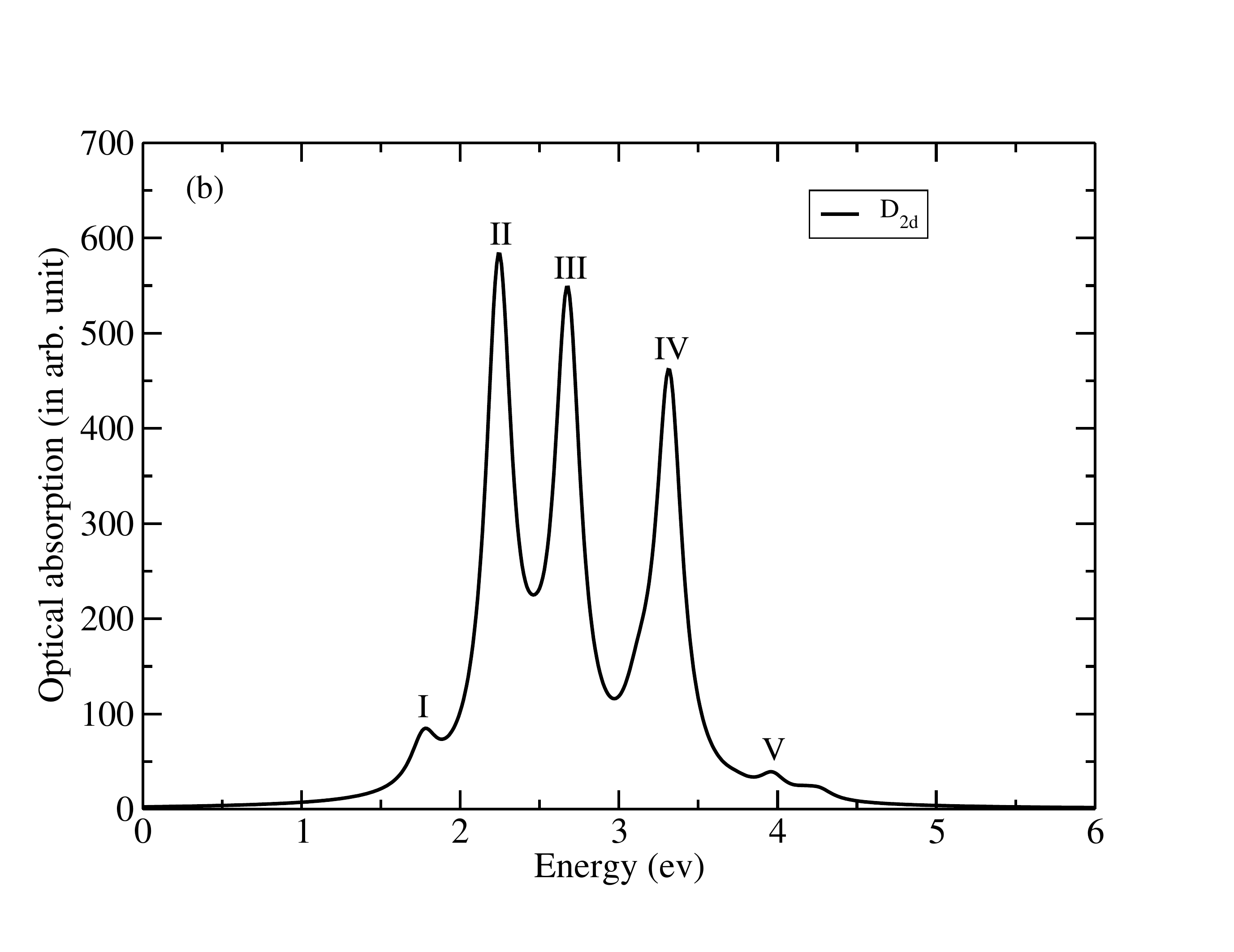}

\caption{Optical absorption spectra of various Na$_{5}$$^{+}$ isomers with:
(a) $D_{2h}$, $D_{3h}$, and $C_{2v}$ symmetries, and (b) $D_{2d}$
symmetry. These spectra were computed using the frozen core FCI approach,
employing a 6-311++G(3df,3pd) basis set. For plotting the spectra,
a uniform linewidth of 0.1 eV was used. \label{fig:na5+_opt.eps}}
\end{figure}

In order to compute the optical absorption spectra, for all isomers,
we used the 6-311++G(3df,3pd) basis set coupled with the frozen-core
FCI approach, and results of these calculations are presented in Fig.
\ref{fig:na5+_opt.eps}. Schmidt \textit{et al}. \textcolor{black}{\cite{schmidt,schmidt-na-cation-prb,schmidt-nacation-zphysd97}}
performed experimental measurements of the optical absorption spectrum
of this cluster, and found three absorption peaks located at: 2.2
eV, 2.7 eV and 3.3 eV. As far as the intensity pattern is concerned,
the measured spectrum consists of three high intensity peaks, first
one of which is the most intense, while the last one carries the least
intensity of the three.\cite{schmidt} If we compare the intensity
profiles of our theoretical spectra with the measured one, we find
perfect agreement between computed spectra of isomers of $D_{2d}$
and $D_{2h}$ symmetries, and the experiment. In Table \ref{tab:na5+ comparison}
we present the peak locations of the calculated spectra and the experimental
one, and find that the excitation energies of three major peaks in
the calculated spectrum of $D_{2d}$ isomer are in excellent agreement
with the experiments. We note that the first peak in the calculated
spectrum of $D_{2d}$ isomer located at 1.77 eV is an extremely weak
one, and has not been observed in the experiments. Furthermore, the
peak locations of the first and the third peaks in the calculated
spectrum of the $D_{2h}$ symmetry isomer are also in very good experiment
with the experiments, while that of the second peak is underestimated
by about 0.19 eV. Additionally, from Table \ref{tab:na5+ comparison}
it is obvious that it is location of the second major peak which is
sensitive to the geometry of the isomer, while locations of the first
and third peaks for all the isomers are in reasonable agreement with
the experiments. Because, for the $D_{2d}$ isomer even the location
of the second intense peak (2.67 eV) is in very good agreement with
the experimental value (2.70 eV), we believe that the experimental
result corresponds to the absorption spectrum of this isomer.

Detailed information about the excitation energies, symmetries, wave
functions, polarizations, and oscillator strengths corresponding to
various optical transitions of these isomers is presented in Tables
8\textendash 11 of the Supporting Information. We find that wave functions
of most of the optically active excited states are dominated by singly-excited
configurations. 

\textcolor{black}{Bona\v{c}i\'{c}-Kouteck\`{y} }\textcolor{black}{\emph{et
al}}\textcolor{black}{.,\cite{na-cation-kout-jcp-96}} and Pal \textit{et
at}.\cite{pal} also performed theoretical calculations of the optical
absorption spectra of various isomers of Na$_{5}^{+}$ cluster, and
in agreement with our work, they also obtained the lowest energy structure
to be the isomer with $D_{2d}$ symmetry. Various peak positions obtained
by Pal \textit{et at}.\cite{pal} are in very good quantitative agreement
with our work. As far as CI method based calculations of \textcolor{black}{Bona\v{c}i\'{c}-Kouteck\`{y}
}\textcolor{black}{\emph{et al}}\textcolor{black}{.,\cite{na-cation-kout-jcp-96}
are concerned, we not only have good quantitative agreement on the
peak locations, additionally, symmetries of the three intense peaks
match perfectly with our work. Additionally, in their calculations,\cite{na-cation-kout-jcp-96}
the first peak is a faint one located below 2 eV, in agreement with
our first weak peak of the $D_{2d}$ isomer located at 1.77 eV. }

\selectlanguage{american}%
\begin{table}[H]
\caption{\foreignlanguage{english}{Comparison of location of peaks in the photoabsorption spectrum of
Na$_{5}{}^{+}$ cluster for different isomers, calculated using 6-311++G(3df,3pd)
basis set and optimized geometries, with experimental results. All
photoabsorption calculations were performed using the frozen core
FCI approach.}}

\label{tab:na5+ comparison}

\begin{tabular}{|cccccccc|}
\hline 
\multirow{2}{*}{\selectlanguage{english}%
Work\selectlanguage{american}%
} & \multirow{2}{*}{\selectlanguage{english}%
Method \selectlanguage{american}%
} & \multirow{2}{*}{\selectlanguage{english}%
Isomers\selectlanguage{american}%
} & \multicolumn{5}{c|}{\selectlanguage{english}%
Peak Energies (Symmetry) (eV)\selectlanguage{american}%
}\tabularnewline
\cline{4-8} 
 &  &  & \selectlanguage{english}%
I\selectlanguage{american}%
 & \selectlanguage{english}%
II\selectlanguage{american}%
 & \selectlanguage{english}%
\textcolor{black}{III }\selectlanguage{american}%
 & \selectlanguage{english}%
IV\selectlanguage{american}%
 & \selectlanguage{english}%
V\selectlanguage{american}%
\tabularnewline
\hline 
\selectlanguage{english}%
This work\selectlanguage{american}%
 & \selectlanguage{english}%
FCI\selectlanguage{american}%
 & \selectlanguage{english}%
$D_{2d}$\selectlanguage{american}%
 & \selectlanguage{english}%
1.77 ($^{1}E$)\selectlanguage{american}%
 & \selectlanguage{english}%
2.24 ($^{1}B_{2}$)\selectlanguage{american}%
 & \selectlanguage{english}%
2.67 ($^{1}E$)\selectlanguage{american}%
 & \selectlanguage{english}%
3.31 ($^{1}E$)\selectlanguage{american}%
 & \selectlanguage{english}%
3.96 ($^{1}E$)\selectlanguage{american}%
\tabularnewline
\selectlanguage{english}%
This work\selectlanguage{american}%
 & \selectlanguage{english}%
FCI\selectlanguage{american}%
 & \selectlanguage{english}%
$D_{2h}$\selectlanguage{american}%
 & \selectlanguage{english}%
2.15 ($^{1}B_{2u}$)\selectlanguage{american}%
 & \selectlanguage{english}%
2.51 ($^{1}B_{3u}$)\selectlanguage{american}%
 & \selectlanguage{english}%
3.22 ($^{1}B_{1u}$)\selectlanguage{american}%
 & \selectlanguage{english}%
3.63 ($^{1}B_{3u}$)\selectlanguage{american}%
 & \selectlanguage{english}%
\selectlanguage{american}%
\tabularnewline
\selectlanguage{english}%
This work\selectlanguage{american}%
 & \selectlanguage{english}%
FCI\selectlanguage{american}%
 & \selectlanguage{english}%
$D_{3h}$\selectlanguage{american}%
 & \selectlanguage{english}%
2.22 (\textcolor{black}{$^{1}A''_{2}$})\selectlanguage{american}%
 & \selectlanguage{english}%
3.16 (\foreignlanguage{american}{$^{1}E'$})\selectlanguage{american}%
 & \selectlanguage{english}%
\selectlanguage{american}%
 & \selectlanguage{english}%
\selectlanguage{american}%
 & \selectlanguage{english}%
\selectlanguage{american}%
\tabularnewline
\selectlanguage{english}%
This work\selectlanguage{american}%
 & \selectlanguage{english}%
FCI\selectlanguage{american}%
 & \selectlanguage{english}%
$C_{2v}$\selectlanguage{american}%
 & \selectlanguage{english}%
2.10 ($^{1}B_{2}$)\selectlanguage{american}%
 & \selectlanguage{english}%
2.40 ($^{1}A_{1}$)\selectlanguage{american}%
 & \selectlanguage{english}%
3.21 ($^{1}B_{1}$)\selectlanguage{american}%
 & \selectlanguage{english}%
\selectlanguage{american}%
 & \selectlanguage{english}%
\selectlanguage{american}%
\tabularnewline
\selectlanguage{english}%
Exp.(Ref.\textcolor{black}{\cite{schmidt,schmidt-na-cation-prb,schmidt-nacation-zphysd97}})\selectlanguage{american}%
 & \selectlanguage{english}%
\textemdash{}\selectlanguage{american}%
 & \selectlanguage{english}%
\textemdash{}\selectlanguage{american}%
 & \selectlanguage{english}%
2.20\selectlanguage{american}%
 & \selectlanguage{english}%
2.70\selectlanguage{american}%
 & \selectlanguage{english}%
3.30\selectlanguage{american}%
 & \selectlanguage{english}%
\selectlanguage{american}%
 & \selectlanguage{english}%
\selectlanguage{american}%
\tabularnewline
\hline 
\end{tabular}
\end{table}
\selectlanguage{english}%

\subsubsection{Na\protect\textsubscript{\textbf{5}}}

\begin{figure}[h]
\includegraphics[width=8cm]{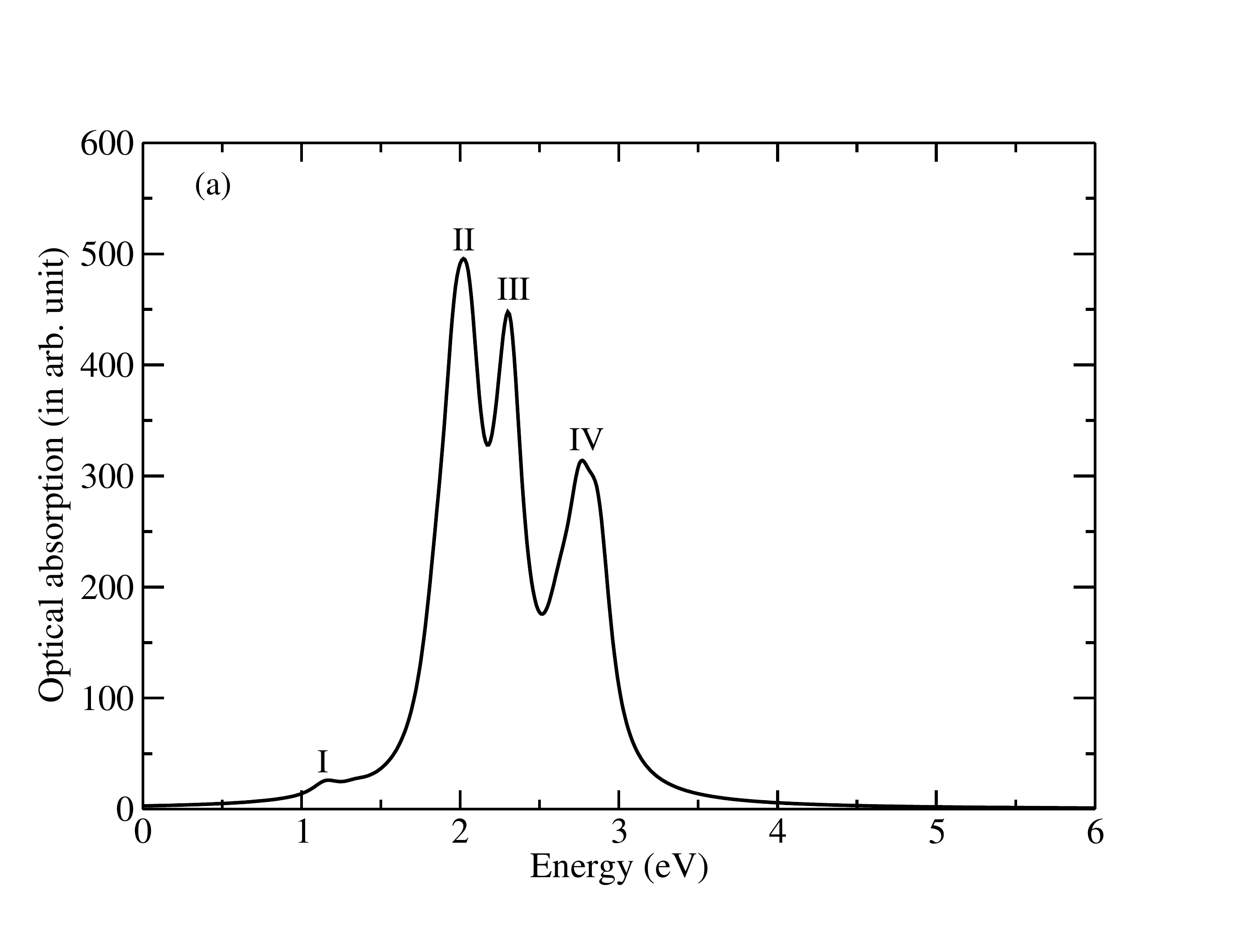}\includegraphics[width=8cm]{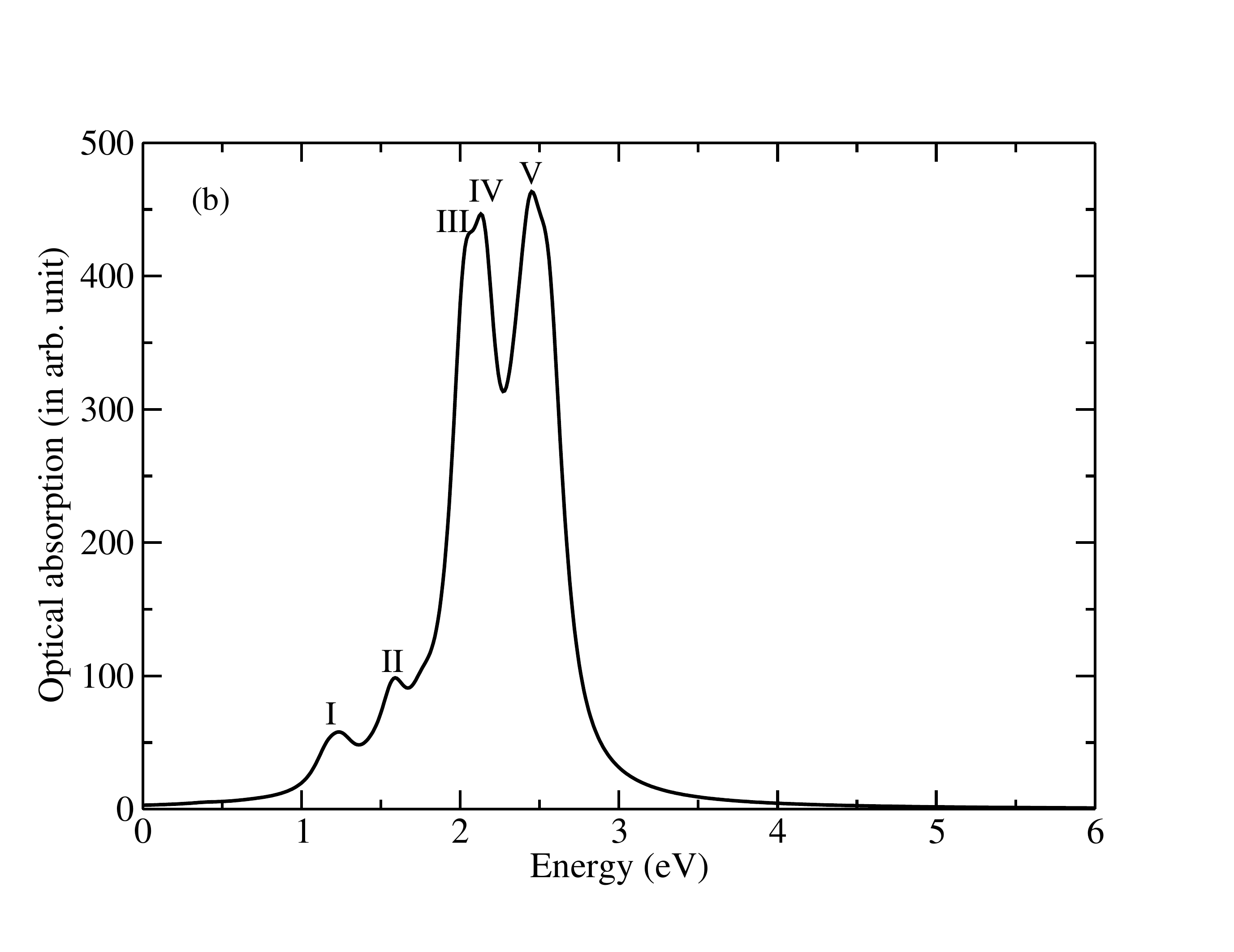}

\caption{Optical absorption spectra of Na$_{5}$ isomers with: (a) $C_{2v}$
planar structure, and (b) $C_{2v}$ trigonal bipyramidal structure.
These spectra were computed using the frozen core QCI approach, employing
a 6-311++G(3df,3pd) basis set. For plotting the spectra, a uniform
linewidth of 0.1 eV was used.\label{na5_opt.eps}}
\end{figure}

For Na$_{5}$, which is an open-shell system, we have considered two
isomers, both with\textcolor{black}{{} }$C_{2v}$ symmetry: a planar
structure shown in Fig. \ref{fig:Ground-state-geometries}(l), and
a trigonal bipyramidal structure shown in Fig. \ref{fig:Ground-state-geometries}(m).\textcolor{black}{{}
We performed geometry optimization for the planar }isomer, \textcolor{black}{by
employing all-electron CCSD approach implemented in the Psi4 code,\cite{psi4}}
and the geometrical parameters obtained are indicated in Fig. \ref{fig:Ground-state-geometries}(l).
For the bipyramid structure we used the geometry reported in paper
by \textcolor{black}{Bona\v{c}i\'{c}-Kouteck\`{y} }\textcolor{black}{\emph{et
al.}}\textcolor{black}{,\cite{bonaic_electronic_strcuture_geometry}
and we find the total energy of this structure to be only 0.11 eV
higher than that of the planar one (}\textcolor{black}{\emph{cf.}}\textcolor{black}{{}
Table \ref{tab_ener}).} Our calculated optical absorption spectra
for both the isomers is presented in Fig. \ref{na5_opt.eps}, and
it shows several absorption peaks ranging from 1.15 eV to 2.76 eV,
as summarized in Table \ref{tab:na5 comparison}. Out of these peaks,
for the planar cluster, strong absorption peaks fall in the range
1.96\textendash 2.76 eV, while for the bipyramidal isomer, they lie
in the range 2.02\textendash 2.44 eV. Experimental measurement of
photoabsorption spectrum of Na$_{5}$ cluster by Wang \emph{et al.},\cite{dahlseid}
reports only one broad and intense absorption band centered at 2.05
eV, with peaks of weaker intensities at lower energies, and questionable
data in the range 2.25\textendash 2.95 eV. Comparing these experimental
results with our calculations, for the planar isomer, peak II located
at 1.96 eV, which is also the most intense peak of the calculated
spectrum, is a candidate for the experimentally observed band center.
Similarly, for the case of bipyramidal isomer, peaks III (2.02 eV)
and IV (2.14 eV) are candidates for the observed band center. If indeed
that is the case, then the calculated lower energy low-intensity peaks
will correspond to lower energy part of the observed spectrum, which
is also weaker in intensity. However, given the uncertainties associated
with the experimental results on the photoabsorption spectrum of Na$_{5}$
cluster,\cite{dahlseid} we hope that in future more accurate measurements
will be performed on this cluster, in order to facilitate a better
comparison between experiment and theory.

As far as comparison of our computed spectra for the case of two isomers
with the theoretical calculations of \textcolor{black}{Kouteck\`{y}
and coworkers\cite{kout-na5-na6-na7} is concerned, there is good
qualitative and quantitative agreement between two sets of calculations
for the planar isomer. For example, the first major peak in the calculated
spectrum of Kouteck\`{y} and coworkers\cite{kout-na5-na6-na7} for
planar structure corresponds to two }$^{2}B_{2}$ type states located
at 1.92 and 1.97 eV, while in our spectrum it corresponds to a single
$^{2}B_{2}$ type state, located at 1.96 eV. 

Detailed information about the excited state wave functions, oscillator
strengths, and photon polarizations corresponding to various optical
transitions of the two isomers is presented in Tables 12 and 13 of
the Supporting Information. On examining the excited state wave functions
we find an interesting difference between the two isomers: for the
planar structure, the wave functions of most of the optically active
excited states are dominated by singly-excited configurations, while
for the bipyramidal structure, they are dominated by double excitations. 

\selectlanguage{american}%
\begin{table}[H]
\caption{\foreignlanguage{english}{Comparison of location of peaks in the photoabsorption spectrum of
Na$_{5}$ cluster for different isomers, calculated using 6-311++G(3df,3pd)
basis set and optimized geometries, with experimental results. All
photoabsorption calculations were performed using the frozen core
QCI approach.}}

\label{tab:na5 comparison}

\begin{tabular}{|ccccccc|}
\hline 
\multirow{2}{*}{\selectlanguage{english}%
Work\selectlanguage{american}%
} & \multirow{2}{*}{Isomers} & \multicolumn{5}{c|}{\selectlanguage{english}%
Peak Energies (Symmetry) (eV)\selectlanguage{american}%
}\tabularnewline
\cline{3-7} 
 &  & \selectlanguage{english}%
I\selectlanguage{american}%
 & \selectlanguage{english}%
II\selectlanguage{american}%
 & \selectlanguage{english}%
\textcolor{black}{III }\selectlanguage{american}%
 & \selectlanguage{english}%
IV\selectlanguage{american}%
 & \selectlanguage{english}%
V\selectlanguage{american}%
\tabularnewline
\hline 
\selectlanguage{english}%
This work\selectlanguage{american}%
 & \selectlanguage{english}%
$C_{2v}$-planar\selectlanguage{american}%
 & \selectlanguage{english}%
1.15($^{2}B_{1}$)\selectlanguage{american}%
 & \selectlanguage{english}%
1.96($^{2}B_{2}$)\selectlanguage{american}%
 & \selectlanguage{english}%
2.31($^{2}B_{1}$)\selectlanguage{american}%
 & \selectlanguage{english}%
2.76($^{2}A_{1}$)\selectlanguage{american}%
 & \selectlanguage{english}%
\selectlanguage{american}%
\tabularnewline
\selectlanguage{english}%
This work\selectlanguage{american}%
 & \selectlanguage{english}%
$C_{2v}$-pyramidal\selectlanguage{american}%
 & \selectlanguage{english}%
1.24($^{2}B_{1}$)\selectlanguage{american}%
 & \selectlanguage{english}%
1.57($^{2}A_{1}$)\selectlanguage{american}%
 & \selectlanguage{english}%
2.02($^{2}A_{1}$)\selectlanguage{american}%
 & \selectlanguage{english}%
2.14($^{2}A_{1}$) \selectlanguage{american}%
 & \selectlanguage{english}%
2.44 ($^{2}B_{1}$) \selectlanguage{american}%
\tabularnewline
\selectlanguage{english}%
Exp.(Ref.\cite{dahlseid})\selectlanguage{american}%
 & \selectlanguage{english}%
\textemdash{}\selectlanguage{american}%
 & \selectlanguage{english}%
2.05\selectlanguage{american}%
 & \selectlanguage{english}%
\selectlanguage{american}%
 & \selectlanguage{english}%
\selectlanguage{american}%
 & \selectlanguage{english}%
\selectlanguage{american}%
 & \selectlanguage{english}%
\selectlanguage{american}%
\tabularnewline
\hline 
\end{tabular}
\end{table}
\selectlanguage{english}%

\subsubsection{Na\protect\textsubscript{\textbf{6}}}

\begin{figure}[h]
\includegraphics[width=8cm]{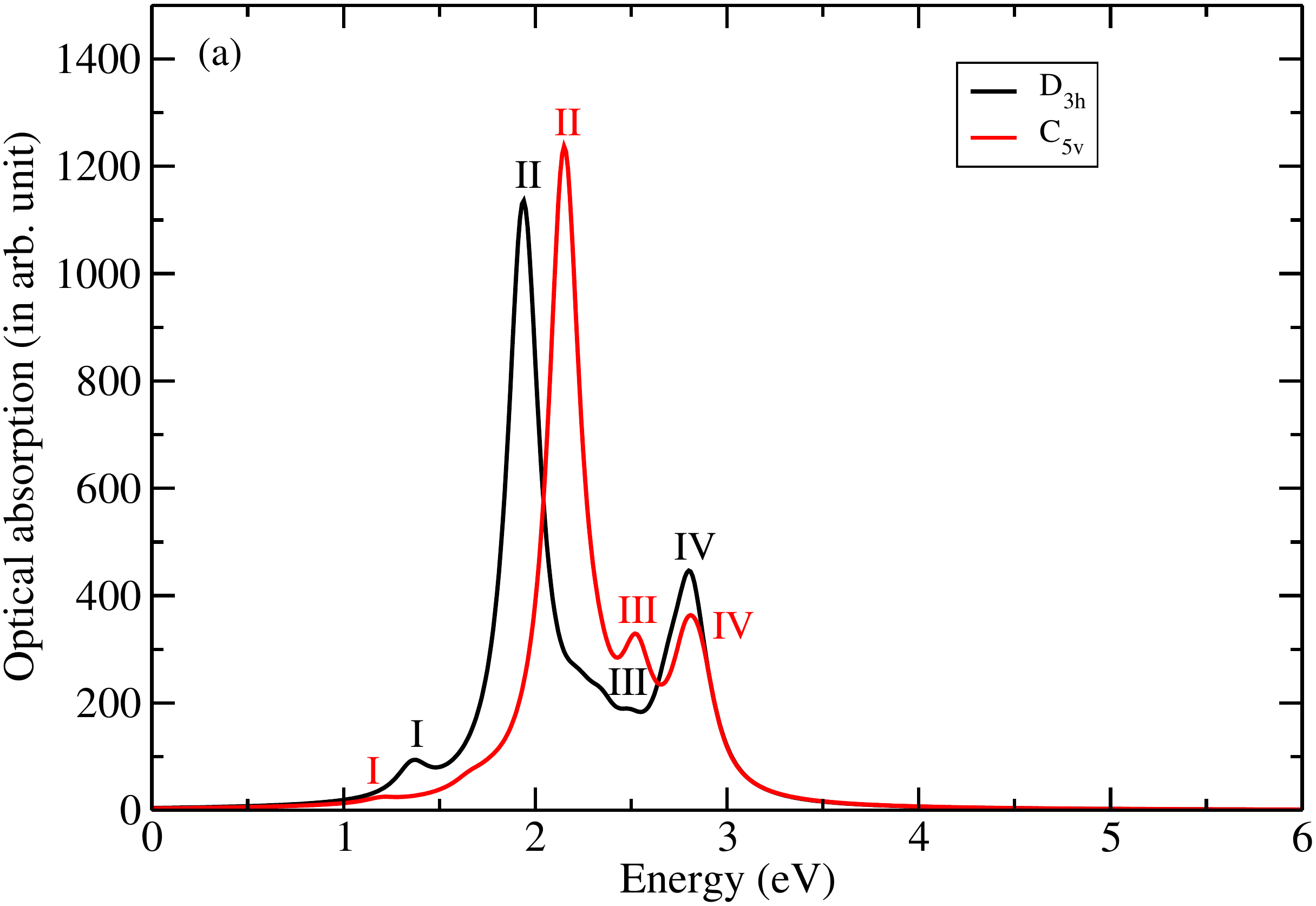}\includegraphics[width=8cm]{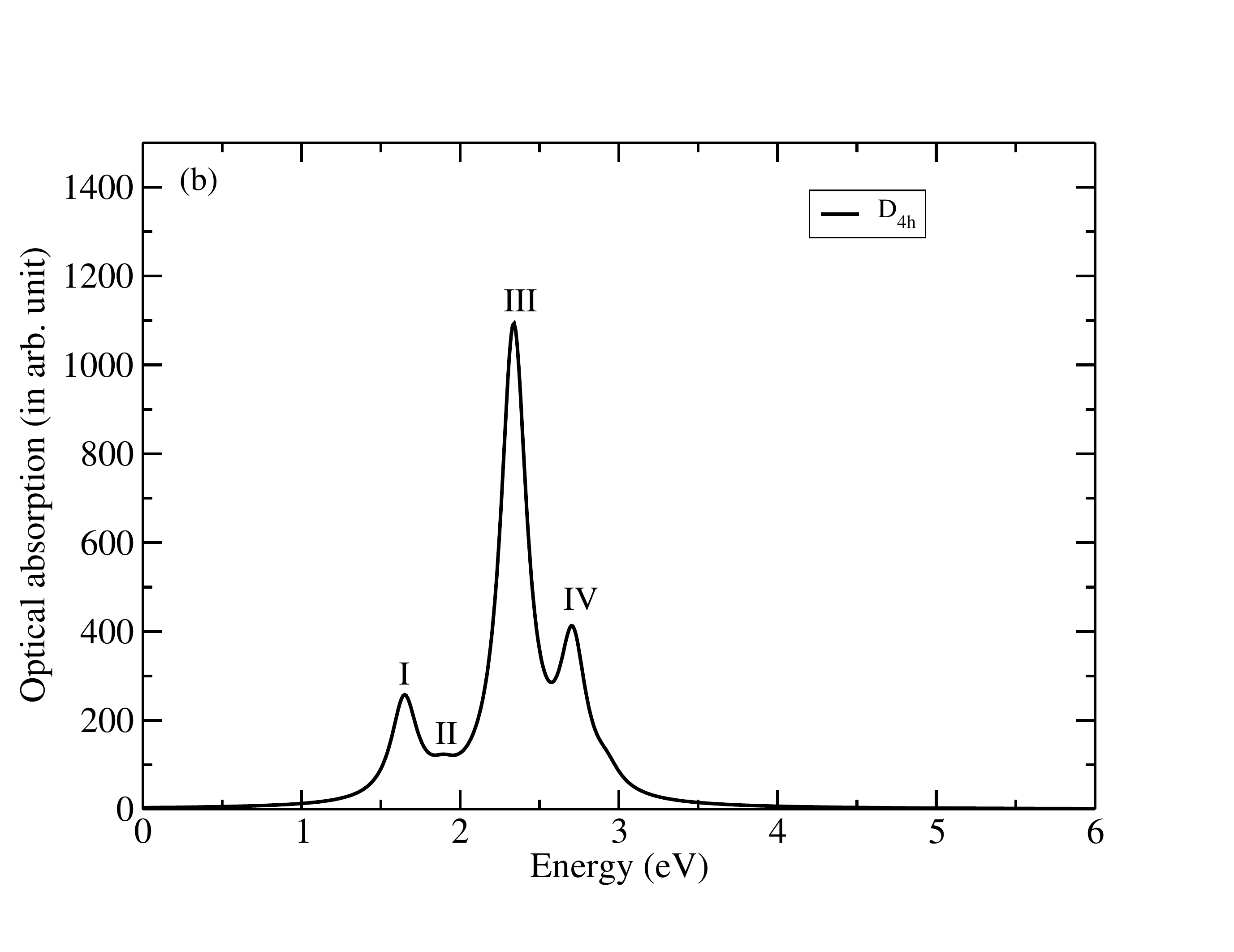}

\caption{Optical absorption spectra of various Na$_{6}$ isomers with: (a)
$D_{3h}$ and $C_{5v}$ symmetries, and (b) $D_{4h}$ symmetry. These
spectra were computed using the frozen core QCI approach, employing
a 6-311++G(3df,3pd) basis set. For plotting the spectra, a uniform
linewidth of 0.1 eV was used.\label{fig:na6_opt.eps}}
\end{figure}

Here we have considered three different types of isomers with symmetries
$C_{5v}$, $D_{4h}$, and planar ($D_{3h}$), as shown in Fig. \ref{fig:Ground-state-geometries}.
\textcolor{black}{Out of these, we performed geometry optimization
for} $C_{5v}$ and $D_{4h}$\textcolor{black}{{} structures by employing
the all-electron CCSD approach as implemented in the Psi4 code,\cite{psi4}
while for} the planar ($D_{3h}$)\textcolor{black}{{} isomer, we have
used }the geometry reported by Pal \textit{et at}.\cite{pal} For
the total energy calculations for all the three isomers, cc-PVDZ basis
set was employed.\cite{emsl_bas2} From Table \ref{tab_ener} it is
obvious that the ground state energies of all the three isomers are
quite close to each other, with the $C_{5v}$ capped pentagonal structure
being the lowest in energy. Calculations for optical absorption spectra
were performed using the frozen-core QCI approach employing 6-311++G(3df,3pd)
basis set, and the results are presented in Fig. \ref{fig:na6_opt.eps}.

Before discussing our calculated spectra, we would like to briefly
summarize the experimental results on the optical absorption of Na$_{6}$
cluster as reported by Wang \emph{et al}.\cite{dahlseid} They reported
four features in their measurement whose locations are reported in
Table \ref{tab:na6-comparison}. As far as the measured intensity
profile of the absorption spectrum is concerned, peak I was reported
to be very weak, carrying lowest intensity compared to all other peaks,
peak II was a very intense one with the highest intensity of the spectrum,
peak III was also a very weak peak but with intensity higher than
peak I, while peak IV was also intense, but with lower intensity as
compared to peak II.\cite{dahlseid} If we examine our calculated
spectra presented in Figs. \ref{fig:na6_opt.eps}(a) and \ref{fig:na6_opt.eps}(b),
we find that the computed spectra of $C_{5v}$ and $D_{3h}$ have
exactly the same intensity pattern as observed in the experiment,\cite{dahlseid}
except that the location of peak I for the $D_{3h}$ isomer is much
smaller as compared to its experimental value. On the other hand,
locations for all the peaks of the calculated spectra of $C_{5v}$
and $D_{4h}$ isomers are in very good agreement with the experiment,
except that intensity profile of the $D_{4h}$ does not match with
the experimental one, because in the calculated spectrum peak III
is the most intense one, as against peak II. Thus, considering both
the intensity profile and the peak locations, we conclude that our
calculated spectrum for the $C_{5v}$ isomer is in the best agreement
with the experiment. Thus, our conclusion disagrees with that of Wang
\emph{et al}.\cite{dahlseid} who, based upon the theoretical calculations
of \textcolor{black}{Kouteck\`{y} and coworkers,\cite{kout-na5-na6-na7}
concluded that the measured spectrum of the} Na$_{6}$ cluster\textcolor{black}{{}
}was closest to that of the $D_{3h}$ isomer. However, given the energetic
proximity of all the three structures considered, it is certainly
possible that the measured spectrum could be a mixed one, carrying
features of the absorption spectra of all the isomers.

Wave functions, along with the polarization characteristics of the
excited states contributing to the peaks in the calculated absorption
spectra of $C_{5v}$, $D_{3h}$, and $D_{4h}$ isomers are presented
in Tables 14, 15, and 16 of the Supporting Information. Upon examination
of these tables, following common features emerge: (a) all the isomers
have doubly-degenerate HOMO orbitals, (b) first three peaks in the
spectra for all the three isomers correspond to transitions polarized
``in-plane'', while the fourth peak corresponds to ``perpendicular''
polarization, and (c) wave functions of the excited states contributing
to these peaks for all the three isomers exhibit strong mixing of
singly-excited configurations with almost similar weights. According
to a criterion proposed by \textcolor{black}{Kouteck\`{y} and coworkers,\cite{plasmon-kout}
excited states whose wave functions are a mixture of a large number
of excited particle-hole configurations with almost identical weights,
signify the emergence of collective plasmonic excitations. This suggests
that the optical excitations in Na$_{6}$ cluster are plasmonic in
nature, and, therefore, we investigate this possibility further. Deshpande
}\textcolor{black}{\emph{et al.}}\textcolor{black}{{} computed the optical
absorption spectra of Na$_{8-x}$Li$_{x}$,\cite{deshpande-nali8}
and bare aluminum clusters,\cite{deshpande-alclusters} using first
principles time-dependent local-density approximation (TDLDA). For
the case of Al$_{n}$ clusters they concluded that for $n\geq6$,
presence of broad absorption bands corresponds to the emergence of
surface plasma resonances, with the increasing size.\cite{deshpande-alclusters}
For the case of Na$_{8-x}$Li$_{x}$ clusters, authors identified
the most intense peak with the plasmon resonance,\cite{deshpande-nali8}
which, for the case of Na$_{8}$ clusters, is located at $\approx$2.7
eV.\cite{deshpande-nali8} If we compare our computed absorption spectra
of various Na$_{6}$ isomers (}\textcolor{black}{\emph{cf}}\textcolor{black}{.
Fig. \ref{fig:na6_opt.eps}), with that of Na$_{8}$ cluster reported
by Deshpande et al.\cite{deshpande-nali8}, we observe the following
trends: (a) both Na$_{6}$ and Na$_{8}$ spectra are dominated by
one high intensity peak, accompanied by several peaks of weaker intensities,
(b) relative intensities of the weaker peaks for Na$_{6}$ is significantly
larger than those for Na$_{8}$. If we compare the peak locations
of the most intense peaks for these clusters, for Na$_{6}$ they are
located at 1.94 eV ($D_{3h}$), 2.15 eV ($C_{5v}$), 2.34 eV ($D_{4h}$),
while for Na$_{8}$ it is located at 2.70 eV. Thus, most intense peaks
in the absorption spectra of Na$_{6}$ isomers are at significantly
lower energies as compared to that for Na$_{8}$. However, if we allow
for the possibility that the higher energy, lower intensity, peaks
could also be due to plasmonic excitations, then we note that locations
of peak IV in the spectra of the three isomers of Na$_{6}$ at 2.81
eV ($D_{3h}$), 2.79 eV ($C_{5v}$), 2.71 eV ($D_{4h}$), are relatively
independent of their geometrical structure, and also in good agreement
with the plasmonic peak of Na$_{8}$ at 2.70 eV. Furthermore, the
excited state wave functions of these peaks presented in the Supporting
Information satisfy the configuration mixing criterion of Kouteck\`{y}
and coworkers,\cite{plasmon-kout} mentioned above. To resolve this
uncertainty regarding the location of surface plasmonic resonances
in Na clusters, it will be useful to perform such calculations in
larger clusters starting with Na$_{8}$. }

\selectlanguage{american}%
\begin{table}[H]
{\footnotesize{}\caption{\foreignlanguage{english}{Comparison of location of peaks in the photoabsorption spectrum of
Na$_{6}$ cluster for different isomers, calculated using 6-311++G(3df,3pd)
basis set and optimized geometries, with experimental results. All
photoabsorption calculations were performed using the frozen core
QCI approach. \label{tab:na6-comparison}}}
}{\footnotesize \par}

\begin{tabular}{|ccccccc|}
\hline 
\multirow{2}{*}{\selectlanguage{english}%
Work\selectlanguage{american}%
} & \multirow{2}{*}{Method } & \multirow{2}{*}{Isomers} & \multicolumn{4}{c|}{\selectlanguage{english}%
Peak Energies (Symmetry) (eV)\selectlanguage{american}%
}\tabularnewline
\cline{4-7} 
 &  &  & \selectlanguage{english}%
I\selectlanguage{american}%
 & \selectlanguage{english}%
II\selectlanguage{american}%
 & \selectlanguage{english}%
\textcolor{black}{III }\selectlanguage{american}%
 & \selectlanguage{english}%
IV\selectlanguage{american}%
\tabularnewline
\hline 
\selectlanguage{english}%
This work\selectlanguage{american}%
 & \selectlanguage{english}%
QCI\selectlanguage{american}%
 & \selectlanguage{english}%
$D_{4h}$\selectlanguage{american}%
 & \selectlanguage{english}%
1.65 ($^{1}E_{u}$)\selectlanguage{american}%
 & \selectlanguage{english}%
1.90 ($^{1}E_{u}$)\selectlanguage{american}%
 & \selectlanguage{english}%
2.34 ($^{1}E_{u}$)\selectlanguage{american}%
 & \selectlanguage{english}%
2.79($^{1}A_{2u}$)\selectlanguage{american}%
\tabularnewline
\selectlanguage{english}%
This work\selectlanguage{american}%
 & \selectlanguage{english}%
QCI\selectlanguage{american}%
 & \selectlanguage{english}%
$D_{3h}$\selectlanguage{american}%
 & \selectlanguage{english}%
1.36 (\foreignlanguage{american}{$^{1}E'$})\selectlanguage{american}%
 & \selectlanguage{english}%
1.94 (\foreignlanguage{american}{$^{1}E'$})\selectlanguage{american}%
 & \selectlanguage{english}%
2.49 (\foreignlanguage{american}{$^{1}E'$})\selectlanguage{american}%
 & \selectlanguage{english}%
2.80 (\textcolor{black}{$^{1}A''_{2}$})\selectlanguage{american}%
\tabularnewline
\selectlanguage{english}%
This work\selectlanguage{american}%
 & \selectlanguage{english}%
QCI\selectlanguage{american}%
 & \selectlanguage{english}%
$C_{5v}$\selectlanguage{american}%
 & \selectlanguage{english}%
1.19($^{1}E_{1}$)\selectlanguage{american}%
 & \selectlanguage{english}%
2.15 ($^{1}E_{1}$)\selectlanguage{american}%
 & \selectlanguage{english}%
2.52 ($^{1}E_{1}$)\selectlanguage{american}%
 & \selectlanguage{english}%
2.79 ($^{1}A_{1}$)\selectlanguage{american}%
\tabularnewline
\selectlanguage{english}%
Exp.(Ref.\cite{dahlseid})\selectlanguage{american}%
 & \selectlanguage{english}%
\textemdash{}\selectlanguage{american}%
 & \selectlanguage{english}%
\textemdash{}\selectlanguage{american}%
 & \selectlanguage{english}%
1.78\selectlanguage{american}%
 & \selectlanguage{english}%
2.08\selectlanguage{american}%
 & \selectlanguage{english}%
2.44\selectlanguage{american}%
 & \selectlanguage{english}%
2.83\selectlanguage{american}%
\tabularnewline
\hline 
\end{tabular}
\end{table}
\selectlanguage{english}%

\section{CONCLUSION }

\label{sec:conclusions}

In this work we presented\emph{ ab initio} calculations of electronic
structure and optical properties of small sodium clusters, utilizing
configuration-interaction methodology based upon Gaussian type of
basis functions. Several isomers of clusters ranging from Na$_{2}$
to Na$_{6}$ were considered, and the calculated spectra were compared
with the experimental results, where available. In most cases, we
obtained very good agreement with the experiments, thus validating
our methodology. \textcolor{black}{We also analyzed the wave functions
of the excited states corresponding to various resonances in the spectra,
and found them to be almost equal mixtures of several configurations
suggesting the possibility that those states correspond to collective
plasmonic excitations.\cite{kout-na5-na6-na7} In light of the TDLDA
calculations of optical absorption spectrum of Na$_{8}$ cluster by
Deshpande }\textcolor{black}{\emph{et al.}}\textcolor{black}{\cite{deshpande-nali8}
reporting a plasmon peak located around 2.7 eV, in future we intend
to perform similar calculations of optical absorption spectra of medium
sized Na clusters starting with Na$_{8}$ aimed at exploring the location
and nature of plasmons in these systems.}

As far as geometry optimization of various clusters is concerned,
we found that core excitations play an important role, and, therefore,
all-electron correlated calculations are essential. But, all the photoabsorption
calculations were performed using the frozen-core approximation, therefore,
good agreement between our results, and the experiments, indicates
that the models of sodium clusters assuming a single valence electron
per atom, are well founded, when it comes to low-lying excitations.
In future, we propose to perform similar studies on clusters composed
of other metal atoms such as lithium and berylium. 
\begin{acknowledgements}
We gratefully acknowledge the financial support of  Department of
Science and Technology (DST), India, under the grant \# SB/S2/CMP-066/2013.
PKP acknowledges the paid leave granted to him to pursue his Ph.D.
under the Faculty Improvement Program of University Grants Commission
(UGC), India.
\end{acknowledgements}

\section*{Author Contribution Statement}

A.S. conceptualized the problem, explained the calculation procedure
to P.K.P. and D.K.R, and wrote the paper. P.K.P. and D.K.R. performed
the calculations, and analyzed the generated data.

\bibliographystyle{/usr/share/texlive/texmf-dist/bibtex/bst/svjour/spphys}
\bibliography{na_optical}

\pagebreak{}
\end{document}


\section{Excited State Wave Functions, Energies, and Transition Moments}

\label{appb:wavefunction}

In the following tables we present the excitation energies, the many-particle
wave functions, and the transition dipole moments with respect to
the ground state, of the excited states corresponding to the peaks
in the linear absorption spectrum of Na atomic clusters, discussed
in the main text.

\newpage{}

\begin{table}
\caption{Excitation energies, $E$, and many-particle wave functions of excited
states corresponding to the peaks in the linear absorption spectrum
of Na$_{2}$ (\emph{cf}. Fig. 7, in main text), along with the oscillator
strength $f$ of the transitions. Polarization corresponding to the
transition is denoted as $\parallel$, if it is along the molecule,
and $\perp$, if it is perpendicular to it. In the wave function,
the bracketed numbers are the CI coefficients of a given electronic
configuration, and symbols $H$/$L$ denote HOMO/LUMO orbitals.}

\label{Tab:na2_wavefunction}

\begin{tabular}{|c|c|c|c|c|c|}
\hline 
Peak	 & $E$ (eV) & Symmetry & $f$ & Polarization & Wave Function\tabularnewline
\hline 
\hline 
I & 1.87 & $^{1}\Sigma_{u}$ & 0.6241 & $\parallel$ & \selectlanguage{american}%
$\left|H\rightarrow L\right\rangle $$(0.5829)$\selectlanguage{english}%
\tabularnewline
\hline 
 &  &  &  &  & \selectlanguage{american}%
$\left|H\rightarrow L+3\right\rangle $$(0.2567)$\selectlanguage{english}%
\tabularnewline
\hline 
 &  &  &  &  & \selectlanguage{american}%
$\left|H\rightarrow L;H\rightarrow L+11\right\rangle $$(0.1361)$\selectlanguage{english}%
\tabularnewline
\hline 
 &  &  &  &  & \selectlanguage{american}%
$\left|H\rightarrow L;H\rightarrow L+8\right\rangle $$(0.1292)$\selectlanguage{english}%
\tabularnewline
\hline 
II & 2.48  & $^{1}\Pi_{u}$ & 0.6411 & $\perp$ & \selectlanguage{american}%
$\left|H\rightarrow(L+2)_{1}\right\rangle $$(0.5041)$\selectlanguage{english}%
\tabularnewline
\hline 
 &  &  &  &  & \selectlanguage{american}%
$\left|H\rightarrow(L+7)_{1}\right\rangle $$(0.4251)$\selectlanguage{english}%
\tabularnewline
\hline 
 &  &  &  &  & \selectlanguage{american}%
$\left|H\rightarrow(L+2)_{1};H\rightarrow L+11\right\rangle $$(0.0848)$\selectlanguage{english}%
\tabularnewline
\hline 
 &  &  &  &  & \selectlanguage{american}%
$\left|H\rightarrow(L+7)_{2};H\rightarrow L+11\right\rangle $$(0.0601)$\selectlanguage{english}%
\tabularnewline
\hline 
 & 2.48  & $^{1}\Pi_{u}$ & 0.6411 & $\perp$ & \selectlanguage{american}%
$\left|H\rightarrow(L+2)_{2}\right\rangle $$(0.5041)$\selectlanguage{english}%
\tabularnewline
\hline 
 &  &  &  &  & \selectlanguage{american}%
$\left|H\rightarrow(L+7)_{1}\right\rangle $$(0.4251)$\selectlanguage{english}%
\tabularnewline
\hline 
 &  &  &  &  & \selectlanguage{american}%
$\left|H\rightarrow(L+2)_{2};H\rightarrow L+11\right\rangle $$(0.0848)$\selectlanguage{english}%
\tabularnewline
\hline 
 &  &  &  &  & \selectlanguage{american}%
$\left|H\rightarrow(L+7)_{1};H\rightarrow L+11\right\rangle $$(0.0601)$\selectlanguage{english}%
\tabularnewline
\hline 
III & 3.66 & $^{1}\Pi_{u}$ & 0.1221 & $\perp$ & \selectlanguage{american}%
$\left|H\rightarrow(L+7)_{1}\right\rangle $$(0.4094)$\selectlanguage{english}%
\tabularnewline
\hline 
 &  &  &  &  & \selectlanguage{american}%
$\left|H\rightarrow(L+2)_{2}\right\rangle $$(0.3656)$\selectlanguage{english}%
\tabularnewline
\hline 
 &  &  &  &  & \selectlanguage{american}%
$\left|H\rightarrow(L+7)_{2}\right\rangle $$(0.2365)$\selectlanguage{english}%
\tabularnewline
\hline 
 &  &  &  &  & \selectlanguage{american}%
$\left|H\rightarrow(L+2)_{1}\right\rangle $$(0.2130)$\selectlanguage{english}%
\tabularnewline
\hline 
 & 3.66 & $^{1}\Pi_{u}$ & 0.1221 & $\perp$ & \selectlanguage{american}%
$\left|H\rightarrow(L+7)_{2}\right\rangle $$(0.4094)$\selectlanguage{english}%
\tabularnewline
\hline 
 &  &  &  &  & \selectlanguage{american}%
$\left|H\rightarrow(L+2)_{1}\right\rangle $$(0.3656)$\selectlanguage{english}%
\tabularnewline
\hline 
 &  &  &  &  & \selectlanguage{american}%
$\left|H\rightarrow(L+7)_{1}\right\rangle $$(0.2365)$\selectlanguage{english}%
\tabularnewline
\hline 
 &  &  &  &  & \selectlanguage{american}%
$\left|H\rightarrow(L+2)_{2}\right\rangle $$(0.2130)$\selectlanguage{english}%
\tabularnewline
\hline 
IV & 4.62 & $^{1}\Pi_{u}$ & 0.0268 & $\perp$ & \selectlanguage{american}%
$\left|H\rightarrow(L+13)_{2}\right\rangle $$(0.3805)$\selectlanguage{english}%
\tabularnewline
\hline 
 &  &  &  &  & \selectlanguage{american}%
$\left|H\rightarrow(L+7)_{1};H\rightarrow L+8\right\rangle $$(0.2572)$\selectlanguage{english}%
\tabularnewline
\hline 
 &  &  &  &  & \selectlanguage{american}%
$\left|H\rightarrow(L+16)_{2}\right\rangle $$(0.2366)$\selectlanguage{english}%
\tabularnewline
\hline 
 &  &  &  &  & \selectlanguage{american}%
$\left|(L+7)_{1};H\rightarrow L+11\right\rangle $$(0.2156)$\selectlanguage{english}%
\tabularnewline
\hline 
 & 4.62 & $^{1}\Pi_{u}$ & 0.0268 & $\perp$ & \selectlanguage{american}%
$\left|H\rightarrow(L+13)_{1}\right\rangle $$(0.3805)$\selectlanguage{english}%
\tabularnewline
\hline 
 &  &  &  &  & \selectlanguage{american}%
$\left|H\rightarrow(L+7)_{2};H\rightarrow L+8\right\rangle $$(0.2572)$\selectlanguage{english}%
\tabularnewline
\hline 
 &  &  &  &  & \selectlanguage{american}%
$\left|H\rightarrow(L+16)_{1}\right\rangle $$(0.2366)$\selectlanguage{english}%
\tabularnewline
\hline 
 &  &  &  &  & \selectlanguage{american}%
$\left|(L+7)_{2};H\rightarrow L+11\right\rangle $$(0.2156)$\selectlanguage{english}%
\tabularnewline
\hline 
V & 5.46  & $^{1}\Pi_{u}$ & 0.0115 & $\perp$ & \selectlanguage{american}%
$\left|H\rightarrow(L+13)_{2}\right\rangle $$(0.3608)$\selectlanguage{english}%
\tabularnewline
\hline 
 &  &  &  &  & \selectlanguage{american}%
$\left|H\rightarrow L;H\rightarrow(L+9)_{1}\right\rangle $$(0.2150)$\selectlanguage{english}%
\tabularnewline
\hline 
 &  &  &  &  & \selectlanguage{american}%
$\left|(L+7)_{1};H\rightarrow L+8\right\rangle $$(0.2109)$\selectlanguage{english}%
\tabularnewline
\hline 
 &  &  &  &  & \selectlanguage{american}%
$\left|(L+2)_{2};H\rightarrow L+8\right\rangle $$(0.1816)$\selectlanguage{english}%
\tabularnewline
\hline 
 & 5.46  & $^{1}\Pi_{u}$ & 0.0115 & $\perp$ & \selectlanguage{american}%
$\left|H\rightarrow(L+13)_{1}\right\rangle $$(0.3608)$\selectlanguage{english}%
\tabularnewline
\hline 
 &  &  &  &  & \selectlanguage{american}%
$\left|H\rightarrow L;H\rightarrow(L+9)_{2}\right\rangle $$(0.2150)$\selectlanguage{english}%
\tabularnewline
\hline 
 &  &  &  &  & \selectlanguage{american}%
$\left|(L+7)_{2};H\rightarrow L+8\right\rangle $$(0.2109)$\selectlanguage{english}%
\tabularnewline
\hline 
 &  &  &  &  & \selectlanguage{american}%
$\left|(L+2)_{1};H\rightarrow L+8\right\rangle $$(0.1816)$\selectlanguage{english}%
\tabularnewline
\hline 
\end{tabular}
\end{table}

\begin{table}
\caption{Excitation energies ($E$), many-particle wave functions of excited
states corresponding to the peaks in the linear absorption spectrum
of Na$_{3}$$^{+}$ (\emph{cf}. Fig. 8, in main text), along with
the oscillator strengths. ``in-plane'' polarization implies photons
polarized in the plane of the triangle, while ``perpendicular''
polarization implies photons polarized perpendicular to the plane
of the triangle. For the doubly degenerate \foreignlanguage{american}{$^{1}E'$}
symmetry, wave functions of both the states are presented. Rest of
the information is same as in the caption of Table \ref{Tab:na2_wavefunction}. }

\selectlanguage{american}%
{\footnotesize{}}%
\begin{tabular}{|c|c|c|c|c|c|}
\hline 
Peak & \selectlanguage{english}%
$E$ (eV)\selectlanguage{american}%
 & Symmetry  & \selectlanguage{english}%
$f$\selectlanguage{american}%
 & Polarization & Wave Function\tabularnewline
\hline 
\hline 
\selectlanguage{english}%
I\selectlanguage{american}%
 & $2.62$ & $^{1}E'$ & $0.6254$ & \selectlanguage{english}%
in-plane\selectlanguage{american}%
 & $\left|H\rightarrow L_{1}\right\rangle $$(0.5199)$\tabularnewline
\hline 
\selectlanguage{english}%
\selectlanguage{american}%
 & \selectlanguage{english}%
\selectlanguage{american}%
 & \selectlanguage{english}%
\selectlanguage{american}%
 & \selectlanguage{english}%
\selectlanguage{american}%
 & \selectlanguage{english}%
\selectlanguage{american}%
 & $\left|H\rightarrow L_{2}\right\rangle $$(0.3951)$\tabularnewline
\hline 
\selectlanguage{english}%
\selectlanguage{american}%
 & \selectlanguage{english}%
\selectlanguage{american}%
 & \selectlanguage{english}%
\selectlanguage{american}%
 & \selectlanguage{english}%
\selectlanguage{american}%
 & \selectlanguage{english}%
\selectlanguage{american}%
 & $\left|H\rightarrow L_{1};H\rightarrow L+2\right\rangle $$(0.1031)$\tabularnewline
\hline 
\selectlanguage{english}%
\selectlanguage{american}%
 & \selectlanguage{english}%
\selectlanguage{american}%
 & \selectlanguage{english}%
\selectlanguage{american}%
 & \selectlanguage{english}%
\selectlanguage{american}%
 & \selectlanguage{english}%
\selectlanguage{american}%
 & $\left|H\rightarrow L_{2};H\rightarrow L+2\right\rangle $$(0.0784)$\tabularnewline
\hline 
\selectlanguage{english}%
\selectlanguage{american}%
 & $2.62$ & $^{1}E'$ & $0.6254$ & \selectlanguage{english}%
in-plane\selectlanguage{american}%
 & $\left|H\rightarrow L_{2}\right\rangle $$(0.5199)$\tabularnewline
\hline 
\selectlanguage{english}%
\selectlanguage{american}%
 & \selectlanguage{english}%
\selectlanguage{american}%
 & \selectlanguage{english}%
\selectlanguage{american}%
 & \selectlanguage{english}%
\selectlanguage{american}%
 & \selectlanguage{english}%
\selectlanguage{american}%
 & $\left|H\rightarrow L_{1}\right\rangle $$(0.3951)$\tabularnewline
\hline 
\selectlanguage{english}%
\selectlanguage{american}%
 & \selectlanguage{english}%
\selectlanguage{american}%
 & \selectlanguage{english}%
\selectlanguage{american}%
 & \selectlanguage{english}%
\selectlanguage{american}%
 & \selectlanguage{english}%
\selectlanguage{american}%
 & $\left|H\rightarrow L_{2};H\rightarrow L+2\right\rangle $$(0.1031)$\tabularnewline
\hline 
\selectlanguage{english}%
\selectlanguage{american}%
 & \selectlanguage{english}%
\selectlanguage{american}%
 & \selectlanguage{english}%
\selectlanguage{american}%
 & \selectlanguage{english}%
\selectlanguage{american}%
 & \selectlanguage{english}%
\selectlanguage{american}%
 & $\left|H\rightarrow L_{1};H\rightarrow L+2\right\rangle $$(0.0784)$\tabularnewline
\hline 
\selectlanguage{english}%
II\selectlanguage{american}%
 & $3.25$ & \selectlanguage{english}%
\textcolor{black}{$^{1}A''_{2}$ }\selectlanguage{american}%
 & $0.6344$ & \selectlanguage{english}%
perpendicular\selectlanguage{american}%
 & $\left|H\rightarrow L+1\right\rangle $$(0.6537)$\tabularnewline
\hline 
\selectlanguage{english}%
\selectlanguage{american}%
 & \selectlanguage{english}%
\selectlanguage{american}%
 & \selectlanguage{english}%
\selectlanguage{american}%
 & \selectlanguage{english}%
\selectlanguage{american}%
 & \selectlanguage{english}%
\selectlanguage{american}%
 & $\left|H\rightarrow L+7\right\rangle $$(0.1246)$\tabularnewline
\hline 
\selectlanguage{english}%
\selectlanguage{american}%
 & \selectlanguage{english}%
\selectlanguage{american}%
 & \selectlanguage{english}%
\selectlanguage{american}%
 & \selectlanguage{english}%
\selectlanguage{american}%
 & \selectlanguage{english}%
\selectlanguage{american}%
 & $\left|H\rightarrow L+1;H\rightarrow L+2\right\rangle $$(0.0792)$\tabularnewline
\hline 
\selectlanguage{english}%
\selectlanguage{american}%
 & \selectlanguage{english}%
\selectlanguage{american}%
 & \selectlanguage{english}%
\selectlanguage{american}%
 & \selectlanguage{english}%
\selectlanguage{american}%
 & \selectlanguage{english}%
\selectlanguage{american}%
 & $\left|H\rightarrow L+1;H\rightarrow L+17\right\rangle $$(0.0778)$\tabularnewline
\hline 
\selectlanguage{english}%
III\selectlanguage{american}%
 & $5.45$ & $^{1}E'$ & $0.0239$ & \selectlanguage{english}%
in-plane\selectlanguage{american}%
 & $\left|H\rightarrow(L+5)_{2}\right\rangle $$(0.3364)$\tabularnewline
\hline 
\selectlanguage{english}%
\selectlanguage{american}%
 & \selectlanguage{english}%
\selectlanguage{american}%
 & \selectlanguage{english}%
\selectlanguage{american}%
 & \selectlanguage{english}%
\selectlanguage{american}%
 & \selectlanguage{english}%
\selectlanguage{american}%
 & $\left|H\rightarrow(L+5)_{1}\right\rangle $$(0.2852)$\tabularnewline
\hline 
\selectlanguage{english}%
\selectlanguage{american}%
 & \selectlanguage{english}%
\selectlanguage{american}%
 & \selectlanguage{english}%
\selectlanguage{american}%
 & \selectlanguage{english}%
\selectlanguage{american}%
 & \selectlanguage{english}%
\selectlanguage{american}%
 & $\left|H\rightarrow(L+13)_{1}\right\rangle $$(0.2626)$\tabularnewline
\hline 
\selectlanguage{english}%
\selectlanguage{american}%
 & \selectlanguage{english}%
\selectlanguage{american}%
 & \selectlanguage{english}%
\selectlanguage{american}%
 & \selectlanguage{english}%
\selectlanguage{american}%
 & \selectlanguage{english}%
\selectlanguage{american}%
 & $\left|H\rightarrow L_{1};H\rightarrow L_{2}\right\rangle $$(0.2112)$\tabularnewline
\hline 
\selectlanguage{english}%
\selectlanguage{american}%
 & $5.45$ & $^{1}E'$ & $0.0239$ & \selectlanguage{english}%
in-plane\selectlanguage{american}%
 & $\left|H\rightarrow(L+5)_{1}\right\rangle $$(0.3364)$\tabularnewline
\hline 
\selectlanguage{english}%
\selectlanguage{american}%
 & \selectlanguage{english}%
\selectlanguage{american}%
 & \selectlanguage{english}%
\selectlanguage{american}%
 & \selectlanguage{english}%
\selectlanguage{american}%
 & \selectlanguage{english}%
\selectlanguage{american}%
 & $\left|H\rightarrow(L+5)_{2}\right\rangle $$(0.2852)$\tabularnewline
\hline 
\selectlanguage{english}%
\selectlanguage{american}%
 & \selectlanguage{english}%
\selectlanguage{american}%
 & \selectlanguage{english}%
\selectlanguage{american}%
 & \selectlanguage{english}%
\selectlanguage{american}%
 & \selectlanguage{english}%
\selectlanguage{american}%
 & $\left|H\rightarrow(L+13)_{2}\right\rangle $$(0.2626)$\tabularnewline
\hline 
\selectlanguage{english}%
\selectlanguage{american}%
 & \selectlanguage{english}%
\selectlanguage{american}%
 & \selectlanguage{english}%
\selectlanguage{american}%
 & \selectlanguage{english}%
\selectlanguage{american}%
 & \selectlanguage{english}%
\selectlanguage{american}%
 & $\left|H\rightarrow L_{2};H\rightarrow L_{2}\right\rangle $$(0.2112)$\tabularnewline
\hline 
\selectlanguage{english}%
IV\selectlanguage{american}%
 & $5.59$ & \selectlanguage{english}%
\textcolor{black}{$^{1}A''_{2}$ }\selectlanguage{american}%
 & $0.0349$ & \selectlanguage{english}%
perpendicular\selectlanguage{american}%
 & $\left|H\rightarrow L+7\right\rangle $$(0.4707)$\tabularnewline
\hline 
\selectlanguage{english}%
\selectlanguage{american}%
 & \selectlanguage{english}%
\selectlanguage{american}%
 & \selectlanguage{english}%
\selectlanguage{american}%
 & \selectlanguage{english}%
\selectlanguage{american}%
 & \selectlanguage{english}%
\selectlanguage{american}%
 & $\left|H\rightarrow L+16\right\rangle $$(0.3968)$\tabularnewline
\hline 
\selectlanguage{english}%
\selectlanguage{american}%
 & \selectlanguage{english}%
\selectlanguage{american}%
 & \selectlanguage{english}%
\selectlanguage{american}%
 & \selectlanguage{english}%
\selectlanguage{american}%
 & \selectlanguage{english}%
\selectlanguage{american}%
 & $\left|H\rightarrow L+1;H\rightarrow L+2\right\rangle $$(0.2122)$\tabularnewline
\hline 
\selectlanguage{english}%
\selectlanguage{american}%
 & \selectlanguage{english}%
\selectlanguage{american}%
 & \selectlanguage{english}%
\selectlanguage{american}%
 & \selectlanguage{english}%
\selectlanguage{american}%
 & \selectlanguage{english}%
\selectlanguage{american}%
 & $\left|H\rightarrow L+1\right\rangle $$(0.1175)$\tabularnewline
\hline 
\end{tabular}{\footnotesize \par}

\selectlanguage{english}%
\label{tab:na3+}
\end{table}

\begin{table}
\caption{Excitation energies, $E$, and many-particle wave functions of excited
states corresponding to the peaks in the linear absorption spectrum
of Na$_{3}$- isosceles triangular cluster (\emph{cf}. Fig. 9, in
main text), along with the oscillator strengths ($f$) of the transitions.
Possible polarizations are: (a) in plane of the triangle along the
perpendicular to the long arm (base) of the triangle (labeled: ``in-plane,
short axis''), (b) in plane of the triangle along the long arm (base)
of the triangle (labeled ``in-plane, long axis''), and (c) perpendicular
to the plane of the triangle (labeled ``perpendicular''). Rest of
the notations are same as those in the previous tables.}

\selectlanguage{american}%
{\footnotesize{}}%
\begin{tabular}{|c|c|c|c|c|c|}
\hline 
{\footnotesize{}Peak} & {\footnotesize{}E ($eV$)} & {\footnotesize{}Symmetry } & \selectlanguage{english}%
$f$\selectlanguage{american}%
 & {\footnotesize{}Polarization} & {\footnotesize{}Wave Function}\tabularnewline
\hline 
\hline 
\selectlanguage{english}%
I\selectlanguage{american}%
 & $0.4997$ & \selectlanguage{english}%
$^{2}A_{1}$\selectlanguage{american}%
 & $0.0065$ & \selectlanguage{english}%
perpendicular\selectlanguage{american}%
 & $\left|H\rightarrow L\right\rangle $$(0.7815)$\tabularnewline
\hline 
\selectlanguage{english}%
\selectlanguage{american}%
 & \selectlanguage{english}%
\selectlanguage{american}%
 & \selectlanguage{english}%
\selectlanguage{american}%
 & \selectlanguage{english}%
\selectlanguage{american}%
 & \selectlanguage{english}%
\selectlanguage{american}%
 & $\left|H\rightarrow L+5\right\rangle $$(0.3322)$\tabularnewline
\hline 
\selectlanguage{english}%
\selectlanguage{american}%
 & \selectlanguage{english}%
\selectlanguage{american}%
 & \selectlanguage{english}%
\selectlanguage{american}%
 & \selectlanguage{english}%
\selectlanguage{american}%
 & \selectlanguage{english}%
\selectlanguage{american}%
 & $\left|H\rightarrow L+4\right\rangle $$(0.2014)$\tabularnewline
\hline 
\selectlanguage{english}%
\selectlanguage{american}%
 & \selectlanguage{english}%
\selectlanguage{american}%
 & \selectlanguage{english}%
\selectlanguage{american}%
 & \selectlanguage{english}%
\selectlanguage{american}%
 & \selectlanguage{english}%
\selectlanguage{american}%
 & $\left|H\rightarrow L+20\right\rangle $$(0.1985)$\tabularnewline
\hline 
\selectlanguage{english}%
II\selectlanguage{american}%
 & $1.1108$ & \selectlanguage{english}%
$^{2}A_{1}$\selectlanguage{american}%
 & $0.0472$ & \selectlanguage{english}%
perpendicular\selectlanguage{american}%
 & $\left|H-1\rightarrow H\right\rangle $$(0.6154)$\tabularnewline
\hline 
\selectlanguage{english}%
\selectlanguage{american}%
 & \selectlanguage{english}%
\selectlanguage{american}%
 & \selectlanguage{english}%
\selectlanguage{american}%
 & \selectlanguage{english}%
\selectlanguage{american}%
 & \selectlanguage{english}%
\selectlanguage{american}%
 & $\left|H\rightarrow L+14\right\rangle $$(0.4608)$\tabularnewline
\hline 
\selectlanguage{english}%
\selectlanguage{american}%
 & \selectlanguage{english}%
\selectlanguage{american}%
 & \selectlanguage{english}%
\selectlanguage{american}%
 & \selectlanguage{english}%
\selectlanguage{american}%
 & \selectlanguage{english}%
\selectlanguage{american}%
 & $\left|H\rightarrow L+1\right\rangle $$(0.2831)$\tabularnewline
\hline 
\selectlanguage{english}%
\selectlanguage{american}%
 & \selectlanguage{english}%
\selectlanguage{american}%
 & \selectlanguage{english}%
\selectlanguage{american}%
 & \selectlanguage{english}%
\selectlanguage{american}%
 & \selectlanguage{english}%
\selectlanguage{american}%
 & $\left|H\rightarrow L+4\right\rangle $$(0.2079)$\tabularnewline
\hline 
\selectlanguage{english}%
III\selectlanguage{american}%
 & $1.6101$ & \selectlanguage{english}%
$^{2}A_{1}$\selectlanguage{american}%
 & $0.3336$ & \selectlanguage{english}%
perpendicular\selectlanguage{american}%
 & $\left|H\rightarrow L+1\right\rangle $$(0.5273)$\tabularnewline
\hline 
\selectlanguage{english}%
\selectlanguage{american}%
 & \selectlanguage{english}%
\selectlanguage{american}%
 & \selectlanguage{english}%
\selectlanguage{american}%
 & \selectlanguage{english}%
\selectlanguage{american}%
 & \selectlanguage{english}%
\selectlanguage{american}%
 & $\left|H-1\rightarrow H\right\rangle $$(0.4143)$\tabularnewline
\hline 
\selectlanguage{english}%
\selectlanguage{american}%
 & \selectlanguage{english}%
\selectlanguage{american}%
 & \selectlanguage{english}%
\selectlanguage{american}%
 & \selectlanguage{english}%
\selectlanguage{american}%
 & \selectlanguage{english}%
\selectlanguage{american}%
 & $\left|H\rightarrow L+10\right\rangle $$(0.3208)$\tabularnewline
\hline 
\selectlanguage{english}%
\selectlanguage{american}%
 & \selectlanguage{english}%
\selectlanguage{american}%
 & \selectlanguage{english}%
\selectlanguage{american}%
 & \selectlanguage{english}%
\selectlanguage{american}%
 & \selectlanguage{english}%
\selectlanguage{american}%
 & $\left|H\rightarrow L+16\right\rangle $$(0.3053)$\tabularnewline
\hline 
\selectlanguage{english}%
IV\selectlanguage{american}%
 & $1.9944$ & \selectlanguage{english}%
$^{2}A_{1}$\selectlanguage{american}%
 & $0.4487$ & \selectlanguage{english}%
perpendicular\selectlanguage{american}%
 & $\left|H\rightarrow L+16\right\rangle $$(0.4769)$\tabularnewline
\hline 
\selectlanguage{english}%
\selectlanguage{american}%
 & \selectlanguage{english}%
\selectlanguage{american}%
 & \selectlanguage{english}%
\selectlanguage{american}%
 & \selectlanguage{english}%
\selectlanguage{american}%
 & \selectlanguage{english}%
\selectlanguage{american}%
 & $\left|H-1\rightarrow H\right\rangle $$(0.4698)$\tabularnewline
\hline 
\selectlanguage{english}%
\selectlanguage{american}%
 & \selectlanguage{english}%
\selectlanguage{american}%
 & \selectlanguage{english}%
\selectlanguage{american}%
 & \selectlanguage{english}%
\selectlanguage{american}%
 & \selectlanguage{english}%
\selectlanguage{american}%
 & $\left|H\rightarrow L+4\right\rangle $$(0.3876)$\tabularnewline
\hline 
\selectlanguage{english}%
\selectlanguage{american}%
 & \selectlanguage{english}%
\selectlanguage{american}%
 & \selectlanguage{english}%
\selectlanguage{american}%
 & \selectlanguage{english}%
\selectlanguage{american}%
 & \selectlanguage{english}%
\selectlanguage{american}%
 & $\left|H\rightarrow L+5\right\rangle $$(0.2437)$\tabularnewline
\hline 
\selectlanguage{english}%
V\selectlanguage{american}%
 & $2.2019$ & \selectlanguage{english}%
$^{2}B_{2}$\selectlanguage{american}%
 & $0.3970$ & \selectlanguage{english}%
in-plane, long axis\selectlanguage{american}%
 & $\left|H-1\rightarrow L\right\rangle $$(0.4238)$\tabularnewline
\hline 
\selectlanguage{english}%
\selectlanguage{american}%
 & \selectlanguage{english}%
\selectlanguage{american}%
 & \selectlanguage{english}%
\selectlanguage{american}%
 & \selectlanguage{english}%
\selectlanguage{american}%
 & \selectlanguage{english}%
\selectlanguage{american}%
 & $\left|H\rightarrow L+3\right\rangle $$(0.3707)$\tabularnewline
\hline 
\selectlanguage{english}%
\selectlanguage{american}%
 & \selectlanguage{english}%
\selectlanguage{american}%
 & \selectlanguage{english}%
\selectlanguage{american}%
 & \selectlanguage{english}%
\selectlanguage{american}%
 & \selectlanguage{english}%
\selectlanguage{american}%
 & $\left|H\rightarrow L+15\right\rangle $$(0.3521)$\tabularnewline
\hline 
\selectlanguage{english}%
\selectlanguage{american}%
 & \selectlanguage{english}%
\selectlanguage{american}%
 & \selectlanguage{english}%
\selectlanguage{american}%
 & \selectlanguage{english}%
\selectlanguage{american}%
 & \selectlanguage{english}%
\selectlanguage{american}%
 & $\left|H-1\rightarrow H;H\rightarrow L\right\rangle $$(0.2580)$\tabularnewline
\hline 
\selectlanguage{english}%
VI\selectlanguage{american}%
 & $2.3539$ & \selectlanguage{english}%
$^{2}B_{2}$\selectlanguage{american}%
 & $0.2252$ & \selectlanguage{english}%
in-plane, long axis\selectlanguage{american}%
 & $\left|H\rightarrow L+6\right\rangle $$(0.4592)$\tabularnewline
\hline 
\selectlanguage{english}%
\selectlanguage{american}%
 & \selectlanguage{english}%
\selectlanguage{american}%
 & \selectlanguage{english}%
\selectlanguage{american}%
 & \selectlanguage{english}%
\selectlanguage{american}%
 & \selectlanguage{english}%
\selectlanguage{american}%
 & $\left|H\rightarrow L+19\right\rangle $$(0.4428)$\tabularnewline
\hline 
\selectlanguage{english}%
\selectlanguage{american}%
 & \selectlanguage{english}%
\selectlanguage{american}%
 & \selectlanguage{english}%
\selectlanguage{american}%
 & \selectlanguage{english}%
\selectlanguage{american}%
 & \selectlanguage{english}%
\selectlanguage{american}%
 & $\left|H\rightarrow L+3\right\rangle $$(0.4043)$\tabularnewline
\hline 
\selectlanguage{english}%
\selectlanguage{american}%
 & \selectlanguage{english}%
\selectlanguage{american}%
 & \selectlanguage{english}%
\selectlanguage{american}%
 & \selectlanguage{english}%
\selectlanguage{american}%
 & \selectlanguage{english}%
\selectlanguage{american}%
 & $\left|H-1\rightarrow L\right\rangle $$(0.2602)$\tabularnewline
\hline 
\selectlanguage{english}%
VII\selectlanguage{american}%
 & $2.5314$ & \selectlanguage{english}%
$^{2}A_{2}$\selectlanguage{american}%
 & $0.4157$ & \selectlanguage{english}%
in-plane, short axis\selectlanguage{american}%
 & $\left|H\rightarrow L+7\right\rangle $$(0.5306)$\tabularnewline
\hline 
\selectlanguage{english}%
\selectlanguage{american}%
 & \selectlanguage{english}%
\selectlanguage{american}%
 & \selectlanguage{english}%
\selectlanguage{american}%
 & \selectlanguage{english}%
\selectlanguage{american}%
 & \selectlanguage{english}%
\selectlanguage{american}%
 & $\left|H-1\rightarrow L+2\right\rangle $$(0.3960)$\tabularnewline
\hline 
\selectlanguage{english}%
\selectlanguage{american}%
 & \selectlanguage{english}%
\selectlanguage{american}%
 & \selectlanguage{english}%
\selectlanguage{american}%
 & \selectlanguage{english}%
\selectlanguage{american}%
 & \selectlanguage{english}%
\selectlanguage{american}%
 & $\left|H-1\rightarrow L+11\right\rangle $$(0.3172)$\tabularnewline
\hline 
\selectlanguage{english}%
\selectlanguage{american}%
 & \selectlanguage{english}%
\selectlanguage{american}%
 & \selectlanguage{english}%
\selectlanguage{american}%
 & \selectlanguage{english}%
\selectlanguage{american}%
 & \selectlanguage{english}%
\selectlanguage{american}%
 & $\left|H\rightarrow L+17\right\rangle $$(0.3119)$\tabularnewline
\hline 
\selectlanguage{english}%
VIII\selectlanguage{american}%
 & $2.9243$ & \selectlanguage{english}%
$^{2}A_{2}$\selectlanguage{american}%
 & $0.2998$ & \selectlanguage{english}%
in-plane, short axis\selectlanguage{american}%
 & $\left|H\rightarrow L+2;H-1\rightarrow H\right\rangle $$(0.4590)$\tabularnewline
\hline 
\selectlanguage{english}%
\selectlanguage{american}%
 & \selectlanguage{english}%
\selectlanguage{american}%
 & \selectlanguage{english}%
\selectlanguage{american}%
 & \selectlanguage{english}%
\selectlanguage{american}%
 & \selectlanguage{english}%
\selectlanguage{american}%
 & $\left|H-1\rightarrow L+11\right\rangle $$(0.3984)$\tabularnewline
\hline 
\selectlanguage{english}%
\selectlanguage{american}%
 & \selectlanguage{english}%
\selectlanguage{american}%
 & \selectlanguage{english}%
\selectlanguage{american}%
 & \selectlanguage{english}%
\selectlanguage{american}%
 & \selectlanguage{english}%
\selectlanguage{american}%
 & $\left|H-1\rightarrow L+2\right\rangle $$(0.3328)$\tabularnewline
\hline 
\selectlanguage{english}%
\selectlanguage{american}%
 & \selectlanguage{english}%
\selectlanguage{american}%
 & \selectlanguage{english}%
\selectlanguage{american}%
 & \selectlanguage{english}%
\selectlanguage{american}%
 & \selectlanguage{english}%
\selectlanguage{american}%
 & $\left|H-1\rightarrow H;H\rightarrow L+11\right\rangle $$(0.2538)$\tabularnewline
\hline 
\end{tabular}{\footnotesize \par}

\selectlanguage{english}%
\label{tab:na3-iso}
\end{table}

\begin{table}
\caption{Excitation energies, $E$, and many-particle wave functions of excited
states corresponding to the peaks in the linear absorption spectrum
of Na$_{3}$ equilateral triangle-shaped cluster (\emph{cf}. Fig.
10, in main text), along with the oscillator strengths $f$ of the
transitions. Possible polarizations are: (a) in plane of the triangle
(labeled ``in-plane''), and (b) perpendicular to the plane of the
triangle (labeled ``perpendicular''). Rest of the notations are
same as those in the previous tables.}

\begin{tabular}{|c|c|c|c|c|}
\hline 
Peak & $E$ (eV) & $f$ & Polarization & Wave Function\tabularnewline
\hline 
\hline 
I & 1.1925 & 0.1755 & in-plane & $\vert H\rightarrow L+1\rangle$(0.5108)\tabularnewline
\hline 
 &  &  &  & $\vert H\rightarrow L+14\rangle$(0.3805)\tabularnewline
\hline 
 &  &  &  & $\vert H\rightarrow L+16\rangle$(0.3394)\tabularnewline
\hline 
 &  &  &  & $\vert H-1\rightarrow H\rangle$(0.3102)\tabularnewline
\hline 
 & 1.2130 & 0.0250 & in-plane & $\vert H-1\rightarrow L\rangle$(0.5475)\tabularnewline
\hline 
 &  &  &  & $\vert H\rightarrow L+15\rangle$(0.5307)\tabularnewline
\hline 
 &  &  &  & $\vert H\rightarrow L+1\rangle$(0.3235)\tabularnewline
\hline 
 &  &  &  & $\vert H\rightarrow L+16\rangle$(0.2604)\tabularnewline
\hline 
II & 2.0724 & 0.3470 & in-plane & $\vert H-1\rightarrow L\rangle$(0.5838)\tabularnewline
\hline 
 &  &  &  & $\vert H\rightarrow L+15\rangle$(0.3604)\tabularnewline
\hline 
 &  &  &  & $\vert H\rightarrow L+1\rangle$(0.3591)\tabularnewline
\hline 
 &  &  &  & $\vert H-1\rightarrow L+3\rangle$(0.2239)\tabularnewline
\hline 
 & 2.0743 & 0.3254 & in-plane & $\vert H\rightarrow L+7\rangle$(0.4402)\tabularnewline
\hline 
 &  &  &  & $\vert H-1\rightarrow H\rangle$(0.3985)\tabularnewline
\hline 
 &  &  &  & $\vert H\rightarrow L;H-1\rightarrow L\rangle$(0.3696)\tabularnewline
\hline 
 &  &  &  & $\vert H\rightarrow L+16\rangle$(0.3401)\tabularnewline
\hline 
III & 2.3622 & 0.4090 & in-plane & $\vert H-1\rightarrow L\rangle$(0.7926)\tabularnewline
\hline 
 &  &  &  & $\vert H-1\rightarrow L+3\rangle$(0.2389)\tabularnewline
\hline 
 &  &  &  & $\vert H-1\rightarrow L+19\rangle$(0.1587)\tabularnewline
\hline 
 &  &  &  & $\vert H-1\rightarrow L;H-1\rightarrow L+22\rangle$(0.1571)\tabularnewline
\hline 
IV & 2.9634 & 0.4938 & perpendicular & $\vert H-1\rightarrow L+2\rangle$(0.5975)\tabularnewline
\hline 
 &  &  &  & $\vert H-1\rightarrow L+11\rangle$(0.4493)\tabularnewline
\hline 
 &  &  &  & $\vert H-1\rightarrow H\rangle$(0.2695)\tabularnewline
\hline 
 &  &  &  & $\vert H\rightarrow L+18\rangle$(0.2600)\tabularnewline
\hline 
V & 3.7223 & 0.0307 & in-plane & $\vert H\rightarrow L+3\rangle$(0.4445)\tabularnewline
\hline 
 &  &  &  & $\vert H\rightarrow L+4;H-1\rightarrow L\rangle$(0.3352)\tabularnewline
\hline 
 &  &  &  & $\vert H-1\rightarrow L+19\rangle$(0.2945)\tabularnewline
\hline 
 &  &  &  & $\vert H-1\rightarrow L;H\rightarrow L+20\rangle$(0.2847)\tabularnewline
\hline 
VI & 3.9622 & 0.0634 & perpendicular & $\vert H\rightarrow L+7\rangle$(0.5908)\tabularnewline
\hline 
 &  &  &  & $\vert H\rightarrow L+1;H-1\rightarrow L+11\rangle$(0.2889)\tabularnewline
\hline 
 &  &  &  & $\vert H\rightarrow L+1;H-1\rightarrow L+2\rangle$(0.2815)\tabularnewline
\hline 
 &  &  &  & $\vert H-1\rightarrow L+11\rangle$(0.2295)\tabularnewline
\hline 
\end{tabular}\label{tab:na3-equi}
\end{table}

\begin{table}
\caption{Excitation energies, $E$, and many-particle wave functions of excited
states corresponding to the peaks in the photoabsorption spectrum
of the linear isomer of Na$_{3}$ (\emph{cf}. Fig. 11, in main text),
along with the oscillator strength ($f$) of the transitions. Polarization
corresponding to the transition is denoted as $\parallel$, if it
is along the length of the molecule, and $\perp$, if it is perpendicular
to it. Rest of the notations are same as those in the previous tables.\label{tab: na3-lin}}

\begin{tabular}{|c|c|c|c|c|c|}
\hline 
Peak & $E$ (eV) & Symmetry & $f$ & Polarization & Wave Function\tabularnewline
\hline 
\hline 
I & 0.6414 & $^{2}\Sigma_{g}$ & 0.0287 & $\parallel$ & $\vert H-1\rightarrow H\rangle$(0.8136)\tabularnewline
\hline 
 &  &  &  &  & $\vert H\rightarrow L\rangle$(0.2675)\tabularnewline
\hline 
 &  &  &  &  & $\vert H\rightarrow L+4\rangle$(0.2305)\tabularnewline
\hline 
 &  &  &  &  & $\vert H\rightarrow L+5\rangle$(0.1569)\tabularnewline
\hline 
II & 1.3414 & $^{2}\Sigma_{g}$ & 0.8917 & $\parallel$ & $\vert H\rightarrow L\rangle$(0.6016)\tabularnewline
\hline 
 &  &  &  &  & $\vert H\rightarrow L+4\rangle$(0.4124)\tabularnewline
\hline 
 &  &  &  &  & $\vert H-1\rightarrow H\rangle$(0.3893)\tabularnewline
\hline 
 &  &  &  &  & $\vert H-1\rightarrow H;H-1\rightarrow L\rangle$(0.1857)\tabularnewline
\hline 
III & 2.4268 & $^{2}\Pi_{g}$ & 0.6994 & $\perp$ & $\vert H-1\rightarrow(L+2)_{1}\rangle$(0.4762)\tabularnewline
\hline 
 &  &  &  &  & $\vert H\rightarrow(L+3)_{1}\rangle$(0.3944)\tabularnewline
\hline 
 &  &  &  &  & $\vert H\rightarrow(L+11)_{1}\rangle$(0.3595)\tabularnewline
\hline 
 &  &  &  &  & $\vert H\rightarrow(L+2)_{1};H-1\rightarrow H\rangle$(0.3094)\tabularnewline
\hline 
 & 2.4268 & $^{2}\Pi_{g}$ & 0.6994 & $\perp$ & $\vert H-1\rightarrow(L+2)_{2}\rangle$(0.4762)\tabularnewline
\hline 
 &  &  &  &  & $\vert H\rightarrow(L+3)_{2}\rangle$(0.3944)\tabularnewline
\hline 
 &  &  &  &  & $\vert H\rightarrow(L+11)_{2}\rangle$(0.3595)\tabularnewline
\hline 
 &  &  &  &  & $\vert H\rightarrow(L+2)_{2};H-1\rightarrow H\rangle$(0.3094)\tabularnewline
\hline 
\end{tabular}\label{tab:na3-lin}
\end{table}

\begin{table}
\caption{Excitation energies ($E$), and many-particle wave functions of excited
states corresponding to the peaks in the linear absorption spectrum
of rhombus-shape Na$_{4}$ cluster (\emph{cf}. Fig. 12, in main text),
along with the oscillator strengths $f$ of the transitions. The cluster
is assumed to lie in the $xy$ plane, with $x$ and $y$ axes along
the short, and long the diagonals, respectively, while the $z$ axis
is assumed perpendicular to the plane of the molecule. Rest of the
notations are same as in the previous tables.}

\begin{tabular}{|c|c|c|c|c|c|}
\hline 
Peak & State & $E$(eV) & $f$ & Polarization & Wave Function\tabularnewline
\hline 
\hline 
I & $B_{3u}$ & 1.712 & 1.1477 & $y$ & $\vert H\rightarrow L+2\rangle$(0.7232)\tabularnewline
\hline 
 &  &  &  &  & $\vert H\rightarrow L+10\rangle$(0.3832)\tabularnewline
\hline 
 &  &  &  &  & $\vert H\rightarrow L+20\rangle$(0.1768)\tabularnewline
\hline 
 &  &  &  &  & $\vert H\rightarrow L+8\rangle$(0.1720)\tabularnewline
\hline 
 &  &  &  &  & $\vert H\rightarrow L+2;H\rightarrow L+24\rangle$(0.1330)\tabularnewline
\hline 
II & $B_{1u}$ & 2.090 & 0.0857 & $z$ & $\vert H\rightarrow L+22\rangle$(0.4537)\tabularnewline
\hline 
 &  &  &  &  & $\vert H\rightarrow L+9\rangle$(0.3546)\tabularnewline
\hline 
 &  &  &  &  & $\vert H\rightarrow L+8;H\rightarrow L+1\rangle$(0.2771)\tabularnewline
\hline 
III & $B_{2u}$ & 2.470 & 0.7719 & $x$ & $\vert H\rightarrow L+6\rangle$(0.5839)\tabularnewline
\hline 
 &  &  &  &  & $\vert H\rightarrow L+18\rangle$(0.3464)\tabularnewline
\hline 
 &  &  &  &  & $\vert H\rightarrow L\rangle$(0.3052)\tabularnewline
\hline 
 &  &  &  &  & $\vert H\rightarrow L+16\rangle$(0.3018)\tabularnewline
\hline 
 &  &  &  &  & $\vert H\rightarrow L+2;H\rightarrow L+3\rangle$(0.1852)\tabularnewline
\hline 
IV & $B_{1u}$ & 2.699 & 0.6171 & $z$ & $\vert H\rightarrow L+9\rangle$(0.5504)\tabularnewline
\hline 
 &  &  &  &  & $\vert H-1\rightarrow L+1\rangle$(0.3739)\tabularnewline
\hline 
 &  &  &  &  & $\vert H\rightarrow L+22\rangle$(0.3032)\tabularnewline
\hline 
 &  &  &  &  & $\vert H-1\rightarrow L+31\rangle$(0.2119)\tabularnewline
\hline 
 &  &  &  &  & $\vert H\rightarrow L+2;H\rightarrow L+1\rangle$(0.1626)\tabularnewline
\hline 
V & $B_{1u}$ & 3.010 & 0.1093 & $z$ & $\vert H-1\rightarrow L+1\rangle$(0.4178)\tabularnewline
\hline 
 &  &  &  &  & $\vert H\rightarrow L+8;H\rightarrow L+1\rangle$(0.2934)\tabularnewline
\hline 
 &  &  &  &  & $\vert H\rightarrow L+8;H\rightarrow L+12\rangle$(0.2740)\tabularnewline
\hline 
VI & $B_{2u}$  & 3.220 & 0.1641 & $x$ & $\vert H\rightarrow L+6\rangle$(0.5619)\tabularnewline
\hline 
 &  &  &  &  & $\vert H\rightarrow L+18\rangle$(0.4104)\tabularnewline
\hline 
 &  &  &  &  & $\vert H\rightarrow L+16\rangle$(0.3258)\tabularnewline
\hline 
VI & $B_{1u}$ & 3.280  & 0.1371 & $z$ & $\vert H\rightarrow L+9\rangle$(0.3237)\tabularnewline
\hline 
 &  &  &  &  & $\vert H-1\rightarrow L+1\rangle$(0.3092)\tabularnewline
\hline 
 &  &  &  &  & $\vert H\rightarrow L+19\rangle$(0.2290)\tabularnewline
\hline 
 &  &  &  &  & $\vert H\rightarrow L+5;H\rightarrow L+12\rangle$(0.2049)\tabularnewline
\hline 
 &  &  &  &  & $\vert H\rightarrow L+5;H\rightarrow L+1\rangle$(0.1911)\tabularnewline
\hline 
VII & $B_{1u}$ & 3.900 & 0.0641 & $z$ & $\vert H-1\rightarrow L+23\rangle$(0.3563)\tabularnewline
\hline 
 &  &  &  &  & $\vert H-1\rightarrow L+5\rangle$(0.2860)\tabularnewline
\hline 
 &  &  &  &  & $\vert H\rightarrow L+24;H\rightarrow L+22\rangle$(0.2348)\tabularnewline
\hline 
 &  &  &  &  & $\vert H\rightarrow L+8;H\rightarrow L+23\rangle$(0.1733)\tabularnewline
\hline 
 &  &  &  &  & $\vert H\rightarrow L+7\rangle$(0.1712)\tabularnewline
\hline 
VIII & $B_{1u}$ & 4.010 & 0.0532 & $z$ & $\vert H-1\rightarrow L+12\rangle$(0.4309)\tabularnewline
\hline 
 &  &  &  &  & $\vert H-1\rightarrow L+1\rangle$(0.3721)\tabularnewline
\hline 
 &  &  &  &  & $\vert H-1\rightarrow L+5\rangle$(0.2459)\tabularnewline
\hline 
IX & $B_{1u}$ & 4.670 & 0.0332 & $z$ & $\vert H\rightarrow L+8;H\rightarrow L+1\rangle$(0.3437)\tabularnewline
\hline 
 &  &  &  &  & $\vert H-1\rightarrow L+12\rangle$(0.3051)\tabularnewline
\hline 
 &  &  &  &  & $\vert H\rightarrow L+35\rangle$(0.2758)\tabularnewline
\hline 
X & $B_{1u}$ & 4.912 & 0.0452 & $z$ & $\vert H\rightarrow L+35\rangle$(0.3861)\tabularnewline
\hline 
 &  &  &  &  & $\vert H\rightarrow L+22;H\rightarrow L+25\rangle$(0.1866)\tabularnewline
\hline 
 &  &  &  &  & $\vert H\rightarrow L+22;H\rightarrow L+11\rangle$(0.1425)\tabularnewline
\hline 
 &  &  &  &  & $\vert H\rightarrow L+10;H\rightarrow L+12\rangle$(0.1409)\tabularnewline
\hline 
 &  &  &  &  & $\vert H-1\rightarrow L+25;H\rightarrow L+12\rangle$(0.1349)\tabularnewline
\hline 
XI & $B_{1u}$ & 5.360 & 0.0221 & $z$ & $\vert H\rightarrow L+33;H\rightarrow L+5\rangle$(0.2490)\tabularnewline
\hline 
 &  &  &  &  & $\vert H\rightarrow L+33;H\rightarrow L+1\rangle$(0.2339)\tabularnewline
\hline 
 &  &  &  &  & $\vert H-1\rightarrow L+29\rangle$(0.2222)\tabularnewline
\hline 
\end{tabular}\label{tab:na4-rhombus}
\end{table}

\begin{table}
\caption{Excitation energies, $E$, and many-particle wave functions of excited
states corresponding to the peaks in the linear absorption spectrum
of Na$_{4}$, square-shaped cluster (\emph{cf}. Fig. 13, in main text),
along with the oscillator strengths of the transitions. Rest of the
information is same as in previous tables. }

\selectlanguage{american}%
{\footnotesize{}}%
\begin{tabular}{|c|c|c|c|c|c|}
\hline 
Peak & \selectlanguage{english}%
$E$ (eV)\selectlanguage{american}%
 & Symmetry  & \selectlanguage{english}%
$f$\selectlanguage{american}%
 & Polarization & Wave Function\tabularnewline
\hline 
\hline 
\selectlanguage{english}%
I\selectlanguage{american}%
 & \selectlanguage{english}%
1.164 \selectlanguage{american}%
 & \selectlanguage{english}%
$^{1}E_{u}$\selectlanguage{american}%
 & \selectlanguage{english}%
0.0107 \selectlanguage{american}%
 & \selectlanguage{english}%
in-plane\selectlanguage{american}%
 & $\left|H\rightarrow L;H\rightarrow L+18\right\rangle $$(0.4144)$\tabularnewline
\hline 
\selectlanguage{english}%
\selectlanguage{american}%
 & \selectlanguage{english}%
\selectlanguage{american}%
 & \selectlanguage{english}%
\selectlanguage{american}%
 & \selectlanguage{english}%
\selectlanguage{american}%
 & \selectlanguage{english}%
\selectlanguage{american}%
 & $\left|H\rightarrow L;H\rightarrow L+4\right\rangle $$(0.3638)$\tabularnewline
\hline 
\selectlanguage{english}%
\selectlanguage{american}%
 & \selectlanguage{english}%
\selectlanguage{american}%
 & \selectlanguage{english}%
\selectlanguage{american}%
 & \selectlanguage{english}%
\selectlanguage{american}%
 & \selectlanguage{english}%
\selectlanguage{american}%
 & $\left|H-1\rightarrow L\right\rangle $$(0.3327)$\tabularnewline
\hline 
\selectlanguage{english}%
\selectlanguage{american}%
 & \selectlanguage{english}%
\selectlanguage{american}%
 & \selectlanguage{english}%
\selectlanguage{american}%
 & \selectlanguage{english}%
\selectlanguage{american}%
 & \selectlanguage{english}%
\selectlanguage{american}%
 & $\left|H\rightarrow L+16\right\rangle $$(0.2510)$\tabularnewline
\hline 
\selectlanguage{english}%
\selectlanguage{american}%
 & \selectlanguage{english}%
1.166\selectlanguage{american}%
 & \selectlanguage{english}%
$^{1}E_{u}$\selectlanguage{american}%
 & \selectlanguage{english}%
0.0110 \selectlanguage{american}%
 & \selectlanguage{english}%
in-plane\selectlanguage{american}%
 & $\left|H\rightarrow L+1\right\rangle $$(0.4755)$\tabularnewline
\hline 
\selectlanguage{english}%
\selectlanguage{american}%
 & \selectlanguage{english}%
\selectlanguage{american}%
 & \selectlanguage{english}%
\selectlanguage{american}%
 & \selectlanguage{english}%
\selectlanguage{american}%
 & \selectlanguage{english}%
\selectlanguage{american}%
 & $\left|H\rightarrow L+9\right\rangle $$(0.3948)$\tabularnewline
\hline 
\selectlanguage{english}%
\selectlanguage{american}%
 & \selectlanguage{english}%
\selectlanguage{american}%
 & \selectlanguage{english}%
\selectlanguage{american}%
 & \selectlanguage{english}%
\selectlanguage{american}%
 & \selectlanguage{english}%
\selectlanguage{american}%
 & $\left|H-1\rightarrow L;H\rightarrow L\right\rangle $$(0.3165)$\tabularnewline
\hline 
\selectlanguage{english}%
\selectlanguage{american}%
 & \selectlanguage{english}%
\selectlanguage{american}%
 & \selectlanguage{english}%
\selectlanguage{american}%
 & \selectlanguage{english}%
\selectlanguage{american}%
 & \selectlanguage{english}%
\selectlanguage{american}%
 & $\left|H\rightarrow L+18\right\rangle $$(0.2633)$\tabularnewline
\hline 
\selectlanguage{english}%
II\selectlanguage{american}%
 & \selectlanguage{english}%
2.005\selectlanguage{american}%
 & \selectlanguage{english}%
$^{1}E_{u}$\selectlanguage{american}%
 & \selectlanguage{english}%
1.0985 \selectlanguage{american}%
 & \selectlanguage{english}%
in-plane\selectlanguage{american}%
 & $\left|H\rightarrow L+4\right\rangle $$(0.4848)$\tabularnewline
\hline 
\selectlanguage{english}%
\selectlanguage{american}%
 & \selectlanguage{english}%
\selectlanguage{american}%
 & \selectlanguage{english}%
\selectlanguage{american}%
 & \selectlanguage{english}%
\selectlanguage{american}%
 & \selectlanguage{english}%
\selectlanguage{american}%
 & $\left|H\rightarrow L+13\right\rangle $$(0.3791)$\tabularnewline
\hline 
\selectlanguage{english}%
\selectlanguage{american}%
 & \selectlanguage{english}%
\selectlanguage{american}%
 & \selectlanguage{english}%
\selectlanguage{american}%
 & \selectlanguage{english}%
\selectlanguage{american}%
 & \selectlanguage{english}%
\selectlanguage{american}%
 & $\left|H\rightarrow L+4\right\rangle $$(0.3777)$\tabularnewline
\hline 
\selectlanguage{english}%
\selectlanguage{american}%
 & \selectlanguage{english}%
\selectlanguage{american}%
 & \selectlanguage{english}%
\selectlanguage{american}%
 & \selectlanguage{english}%
\selectlanguage{american}%
 & \selectlanguage{english}%
\selectlanguage{american}%
 & $\left|H\rightarrow L+18\right\rangle $$(0.2474)$\tabularnewline
\hline 
\selectlanguage{english}%
\selectlanguage{american}%
 & \selectlanguage{english}%
2.015 \selectlanguage{american}%
 & \selectlanguage{english}%
$^{1}E_{u}$\selectlanguage{american}%
 & \selectlanguage{english}%
1.0709 \selectlanguage{american}%
 & \selectlanguage{english}%
in-plane\selectlanguage{american}%
 & $\left|H\rightarrow L;H\rightarrow L+1\right\rangle $$(0.5937)$\tabularnewline
\hline 
\selectlanguage{english}%
\selectlanguage{american}%
 & \selectlanguage{english}%
\selectlanguage{american}%
 & \selectlanguage{english}%
\selectlanguage{american}%
 & \selectlanguage{english}%
\selectlanguage{american}%
 & \selectlanguage{english}%
\selectlanguage{american}%
 & $\left|H-1\rightarrow L\right\rangle $$(0.2880)$\tabularnewline
\hline 
\selectlanguage{english}%
\selectlanguage{american}%
 & \selectlanguage{english}%
\selectlanguage{american}%
 & \selectlanguage{english}%
\selectlanguage{american}%
 & \selectlanguage{english}%
\selectlanguage{american}%
 & \selectlanguage{english}%
\selectlanguage{american}%
 & $\left|H\rightarrow L;H\rightarrow L+9\right\rangle $$(0.2826)$\tabularnewline
\hline 
\selectlanguage{english}%
\selectlanguage{american}%
 & \selectlanguage{english}%
\selectlanguage{american}%
 & \selectlanguage{english}%
\selectlanguage{american}%
 & \selectlanguage{english}%
\selectlanguage{american}%
 & \selectlanguage{english}%
\selectlanguage{american}%
 & $\left|H\rightarrow L+1;H\rightarrow L+22\right\rangle $$(0.2122)$\tabularnewline
\hline 
\selectlanguage{english}%
III\selectlanguage{american}%
 & \selectlanguage{english}%
2.175\selectlanguage{american}%
 & \selectlanguage{english}%
$^{1}E_{u}$\selectlanguage{american}%
 & \selectlanguage{english}%
0.6775\selectlanguage{american}%
 & \selectlanguage{english}%
in-plane\selectlanguage{american}%
 & $\left|H\rightarrow L+5\right\rangle $$(0.4769)$\tabularnewline
\hline 
\selectlanguage{english}%
\selectlanguage{american}%
 & \selectlanguage{english}%
\selectlanguage{american}%
 & \selectlanguage{english}%
\selectlanguage{american}%
 & \selectlanguage{english}%
\selectlanguage{american}%
 & \selectlanguage{english}%
\selectlanguage{american}%
 & $\left|H\rightarrow L;H\rightarrow L+4\right\rangle $$(0.4122)$\tabularnewline
\hline 
\selectlanguage{english}%
\selectlanguage{american}%
 & \selectlanguage{english}%
\selectlanguage{american}%
 & \selectlanguage{english}%
\selectlanguage{american}%
 & \selectlanguage{english}%
\selectlanguage{american}%
 & \selectlanguage{english}%
\selectlanguage{american}%
 & $\left|H\rightarrow L+16\right\rangle $$(0.3209)$\tabularnewline
\hline 
\selectlanguage{english}%
\selectlanguage{american}%
 & \selectlanguage{english}%
\selectlanguage{american}%
 & \selectlanguage{english}%
\selectlanguage{american}%
 & \selectlanguage{english}%
\selectlanguage{american}%
 & \selectlanguage{english}%
\selectlanguage{american}%
 & $\left|H\rightarrow L;H\rightarrow L+13\right\rangle $$(0.2071)$\tabularnewline
\hline 
\selectlanguage{english}%
\selectlanguage{american}%
 & \selectlanguage{english}%
2.177\selectlanguage{american}%
 & \selectlanguage{english}%
$^{1}E_{u}$\selectlanguage{american}%
 & \selectlanguage{english}%
0.6833 \selectlanguage{american}%
 & \selectlanguage{english}%
in-plane\selectlanguage{american}%
 & $\left|H\rightarrow L;H\rightarrow(L+5\right\rangle $$(0.4283)$\tabularnewline
\hline 
\selectlanguage{english}%
\selectlanguage{american}%
 & \selectlanguage{english}%
\selectlanguage{american}%
 & \selectlanguage{english}%
\selectlanguage{american}%
 & \selectlanguage{english}%
\selectlanguage{american}%
 & \selectlanguage{english}%
\selectlanguage{american}%
 & $\left|H\rightarrow L+1\right\rangle $$(0.3461)$\tabularnewline
\hline 
\selectlanguage{english}%
\selectlanguage{american}%
 & \selectlanguage{english}%
\selectlanguage{american}%
 & \selectlanguage{english}%
\selectlanguage{american}%
 & \selectlanguage{english}%
\selectlanguage{american}%
 & \selectlanguage{english}%
\selectlanguage{american}%
 & $\left|H\rightarrow L+4\right\rangle $$(0.3302)$\tabularnewline
\hline 
\selectlanguage{english}%
\selectlanguage{american}%
 & \selectlanguage{english}%
\selectlanguage{american}%
 & \selectlanguage{english}%
\selectlanguage{american}%
 & \selectlanguage{english}%
\selectlanguage{american}%
 & \selectlanguage{english}%
\selectlanguage{american}%
 & $\left|H\rightarrow L+9\right\rangle $$(0.2805)$\tabularnewline
\hline 
\selectlanguage{english}%
IV\selectlanguage{american}%
 & \selectlanguage{english}%
2.482\selectlanguage{american}%
 & \selectlanguage{english}%
$^{1}E_{u}$\selectlanguage{american}%
 & \selectlanguage{english}%
0.4637 \selectlanguage{american}%
 & \selectlanguage{english}%
in-plane\selectlanguage{american}%
 & $\left|H-1\rightarrow L\right\rangle $$(0.6015)$\tabularnewline
\hline 
\selectlanguage{english}%
\selectlanguage{american}%
 & \selectlanguage{english}%
\selectlanguage{american}%
 & \selectlanguage{english}%
\selectlanguage{american}%
 & \selectlanguage{english}%
\selectlanguage{american}%
 & \selectlanguage{english}%
\selectlanguage{american}%
 & $\left|H\rightarrow L+5\right\rangle $$(0.3705)$\tabularnewline
\hline 
\selectlanguage{english}%
\selectlanguage{american}%
 & \selectlanguage{english}%
\selectlanguage{american}%
 & \selectlanguage{english}%
\selectlanguage{american}%
 & \selectlanguage{english}%
\selectlanguage{american}%
 & \selectlanguage{english}%
\selectlanguage{american}%
 & $\left|H\rightarrow L;H\rightarrow L+1\right\rangle $$(0.1639)$\tabularnewline
\hline 
\selectlanguage{english}%
\selectlanguage{american}%
 & \selectlanguage{english}%
\selectlanguage{american}%
 & \selectlanguage{english}%
\selectlanguage{american}%
 & \selectlanguage{english}%
\selectlanguage{american}%
 & \selectlanguage{english}%
\selectlanguage{american}%
 & $\left|H\rightarrow L+4\right\rangle $$(0.1523)$\tabularnewline
\hline 
\selectlanguage{english}%
\selectlanguage{american}%
 & \selectlanguage{english}%
2.485\selectlanguage{american}%
 & \selectlanguage{english}%
$^{1}E_{u}$\selectlanguage{american}%
 & \selectlanguage{english}%
0.4375 \selectlanguage{american}%
 & \selectlanguage{english}%
in-plane\selectlanguage{american}%
 & $\left|H-1\rightarrow L;H\rightarrow L\right\rangle $$(0.4970)$\tabularnewline
\hline 
\selectlanguage{english}%
\selectlanguage{american}%
 & \selectlanguage{english}%
\selectlanguage{american}%
 & \selectlanguage{english}%
\selectlanguage{american}%
 & \selectlanguage{english}%
\selectlanguage{american}%
 & \selectlanguage{english}%
\selectlanguage{american}%
 & $\left|H\rightarrow L;H\rightarrow L+5\right\rangle $$(0.3376)$\tabularnewline
\hline 
\selectlanguage{english}%
\selectlanguage{american}%
 & \selectlanguage{english}%
\selectlanguage{american}%
 & \selectlanguage{english}%
\selectlanguage{american}%
 & \selectlanguage{english}%
\selectlanguage{american}%
 & \selectlanguage{english}%
\selectlanguage{american}%
 & $\left|H\rightarrow L;H-1\rightarrow L+22\right\rangle $$(0.2057)$\tabularnewline
\hline 
\selectlanguage{english}%
\selectlanguage{american}%
 & \selectlanguage{english}%
\selectlanguage{american}%
 & \selectlanguage{english}%
\selectlanguage{american}%
 & \selectlanguage{english}%
\selectlanguage{american}%
 & \selectlanguage{english}%
\selectlanguage{american}%
 & $\left|H\rightarrow L+5\right\rangle $$(0.1761)$\tabularnewline
\hline 
\selectlanguage{english}%
V\selectlanguage{american}%
 & \selectlanguage{english}%
2.752\selectlanguage{american}%
 & \selectlanguage{english}%
$A_{2u}$\selectlanguage{american}%
 & \selectlanguage{english}%
0.6634 \selectlanguage{american}%
 & \selectlanguage{english}%
perpendicular\selectlanguage{american}%
 & $\left|H\rightarrow L+8\right\rangle $$(0.4383)$\tabularnewline
\hline 
\selectlanguage{english}%
\selectlanguage{american}%
 & \selectlanguage{english}%
\selectlanguage{american}%
 & \selectlanguage{english}%
\selectlanguage{american}%
 & \selectlanguage{english}%
\selectlanguage{american}%
 & \selectlanguage{english}%
\selectlanguage{american}%
 & $\left|H\rightarrow L;H\rightarrow L+7\right\rangle $$(0.3968)$\tabularnewline
\hline 
\selectlanguage{english}%
\selectlanguage{american}%
 & \selectlanguage{english}%
\selectlanguage{american}%
 & \selectlanguage{english}%
\selectlanguage{american}%
 & \selectlanguage{english}%
\selectlanguage{american}%
 & \selectlanguage{english}%
\selectlanguage{american}%
 & $\left|H-1\rightarrow L+2\right\rangle $$(0.2655)$\tabularnewline
\hline 
\selectlanguage{english}%
\selectlanguage{american}%
 & \selectlanguage{english}%
\selectlanguage{american}%
 & \selectlanguage{english}%
\selectlanguage{american}%
 & \selectlanguage{english}%
\selectlanguage{american}%
 & \selectlanguage{english}%
\selectlanguage{american}%
 & $\left|H\rightarrow L+23\right\rangle $$(0.2365)$\tabularnewline
\hline 
\end{tabular}{\footnotesize{}\label{tab-Na4-square}}{\footnotesize \par}\selectlanguage{english}%
\end{table}

\begin{table}
\caption{Excitation energies, $E$, and many-particle wave functions of excited
states corresponding to the peaks in the linear absorption spectrum
of Na$_{5}$$^{+}$ $D_{2d}$ structure (\emph{cf}. Fig. 14(b), in
main text), along with the oscillator strength $f$ of the transitions.
In our calculations, the two triangles of the structure are assumed
to lie in the $yz$ and $xz$ planes, while $z$-axis is along the
common perpendicular bisectors of the two triangles. Thus, $xy$-plane
is perpendicular to the planes of both the triangles. Rest of the
information is same as in previous tables. }

\begin{tabular}{|c|c|c|c|c|c|}
\hline 
Peak	 & $E$ (eV) & Symmetry & $f$ & Polarization & Wave Function\tabularnewline
\hline 
\hline 
I & 1.770952 & $^{1}E$ & 0.0636 & $xy$ & \selectlanguage{american}%
$\left|H\rightarrow L_{1}\right\rangle $$(0.6165)$\selectlanguage{english}%
\tabularnewline
\hline 
 &  &  &  &  & \selectlanguage{american}%
$\left|H\rightarrow L_{2}\right\rangle $$(0.5691)$\selectlanguage{english}%
\tabularnewline
\hline 
 &  &  &  &  & \selectlanguage{american}%
$\left|H\rightarrow L_{1};H-1\rightarrow L+1\right\rangle $$(0.1363)$\selectlanguage{english}%
\tabularnewline
\hline 
 &  &  &  &  & \selectlanguage{american}%
$\left|H\rightarrow L_{2};H-1\rightarrow L+1\right\rangle $$(0.1258)$\selectlanguage{english}%
\tabularnewline
\hline 
 & 1.770952 & $^{1}E$ & 0.0636 & $xy$ & \selectlanguage{american}%
$\left|H\rightarrow L_{2}\right\rangle $$(0.6165)$\selectlanguage{english}%
\tabularnewline
\hline 
 &  &  &  &  & \selectlanguage{american}%
$\left|H\rightarrow L_{1}\right\rangle $$(0.5691)$\selectlanguage{english}%
\tabularnewline
\hline 
 &  &  &  &  & \selectlanguage{american}%
$\left|H\rightarrow L_{2};H-1\rightarrow L+1\right\rangle $$(0.1363)$\selectlanguage{english}%
\tabularnewline
\hline 
 &  &  &  &  & \selectlanguage{american}%
$\left|H\rightarrow L_{1};H-1\rightarrow L+1\right\rangle $$(0.1258)$\selectlanguage{english}%
\tabularnewline
\hline 
II & 2.243351 & $^{1}B_{2}$ & 1.3390 & $z$ & \selectlanguage{american}%
$\left|H\rightarrow(L+1)\right\rangle $$(0.8954)$\selectlanguage{english}%
\tabularnewline
\hline 
 &  &  &  &  & \selectlanguage{american}%
$\left|H-1\rightarrow(L+3)\right\rangle $$(0.1582)$\selectlanguage{english}%
\tabularnewline
\hline 
 &  &  &  &  & \selectlanguage{american}%
$\left|H\rightarrow(L+1);H\rightarrow(L+3)\right\rangle $$(0.1158)$\selectlanguage{english}%
\tabularnewline
\hline 
 &  &  &  &  & \selectlanguage{american}%
$\left|H\rightarrow(L+8)\right\rangle $$(0.0905)$\selectlanguage{english}%
\tabularnewline
\hline 
III & 2.6766435 & $^{1}E$ & 0.6144 & $xy$ & \selectlanguage{american}%
$\left|H\rightarrow(L+2){}_{2}\right\rangle $$(0.4478)$\selectlanguage{english}%
\tabularnewline
\hline 
 &  &  &  &  & \selectlanguage{american}%
$\left|H-1\rightarrow L{}_{1}\right\rangle $$(0.4014)$\selectlanguage{english}%
\tabularnewline
\hline 
 &  &  &  &  & \selectlanguage{american}%
$\left|H-1\rightarrow L{}_{2}\right\rangle $$(0.3116)$\selectlanguage{english}%
\tabularnewline
\hline 
 &  &  &  &  & \selectlanguage{american}%
$\left|H-1\rightarrow L{}_{1}\right\rangle $$(0.2792)$\selectlanguage{english}%
\tabularnewline
\hline 
 & 2.6766435 & $^{1}E$ & 0.6144 & $xy$ & \selectlanguage{american}%
$\left|H\rightarrow(L+2){}_{1}\right\rangle $$(0.4478)$\selectlanguage{english}%
\tabularnewline
\hline 
 &  &  &  &  & \selectlanguage{american}%
$\left|H-1\rightarrow L{}_{2}\right\rangle $$(0.4014)$\selectlanguage{english}%
\tabularnewline
\hline 
 &  &  &  &  & \selectlanguage{american}%
$\left|H-1\rightarrow L{}_{1}\right\rangle $$(0.3116)$\selectlanguage{english}%
\tabularnewline
\hline 
 &  &  &  &  & \selectlanguage{american}%
$\left|H-1\rightarrow L{}_{2}\right\rangle $$(0.2792)$\selectlanguage{english}%
\tabularnewline
\hline 
IV & 3.316498 & $^{1}E$ & 0.5247 & $xy$ & \selectlanguage{american}%
$\left|H-1\rightarrow(L+2){}_{2}\right\rangle $$(0.3950)$\selectlanguage{english}%
\tabularnewline
\hline 
 &  &  &  &  & \selectlanguage{american}%
$\left|H-1\rightarrow(L+2){}_{1}\right\rangle $$(0.3233)$\selectlanguage{english}%
\tabularnewline
\hline 
 &  &  &  &  & \selectlanguage{american}%
$\left|H\rightarrow(L+2)_{2}\right\rangle $$(0.3050)$\selectlanguage{english}%
\tabularnewline
\hline 
 &  &  &  &  & \selectlanguage{american}%
$\left|H\rightarrow(L+2)_{1}\right\rangle $$(0.2495)$\selectlanguage{english}%
\tabularnewline
\hline 
 & 3.316498 & $^{1}E$ & 0.5247 & $xy$ & \selectlanguage{american}%
$\left|H-1\rightarrow(L+2){}_{1}\right\rangle $$(0.3950)$\selectlanguage{english}%
\tabularnewline
\hline 
 &  &  &  &  & \selectlanguage{american}%
$\left|H-1\rightarrow(L+2){}_{2}\right\rangle $$(0.3233)$\selectlanguage{english}%
\tabularnewline
\hline 
 &  &  &  &  & \selectlanguage{american}%
$\left|H\rightarrow(L+2)_{1}\right\rangle $$(0.3050)$\selectlanguage{english}%
\tabularnewline
\hline 
 &  &  &  &  & \selectlanguage{american}%
$\left|H\rightarrow(L+2)_{2}\right\rangle $$(0.2495)$\selectlanguage{english}%
\tabularnewline
\hline 
V & 3.966769 & $^{1}E$ & 0.0250 & $xy$ & \selectlanguage{american}%
$\left|H-1\rightarrow(L+2){}_{1}\right\rangle $$(0.3039)$\selectlanguage{english}%
\tabularnewline
\hline 
 &  &  &  &  & \selectlanguage{american}%
$\left|H-1\rightarrow(L+2){}_{2}\right\rangle $$(0.3025)$\selectlanguage{english}%
\tabularnewline
\hline 
 &  &  &  &  & \selectlanguage{american}%
$\left|H\rightarrow L_{1};H\rightarrow L+3\right\rangle $$(0.2164)$\selectlanguage{english}%
\tabularnewline
\hline 
 &  &  &  &  & \selectlanguage{american}%
$\left|H\rightarrow L_{2};H\rightarrow L+3\right\rangle $$(0.2154)$\selectlanguage{english}%
\tabularnewline
\hline 
 & 3.966769 & $^{1}E$ & 0.0250 & $xy$ & \selectlanguage{american}%
$\left|H-1\rightarrow(L+2){}_{2}\right\rangle $$(0.3039)$\selectlanguage{english}%
\tabularnewline
\hline 
 &  &  &  &  & \selectlanguage{american}%
$\left|H-1\rightarrow(L+2){}_{1}\right\rangle $$(0.3025)$\selectlanguage{english}%
\tabularnewline
\hline 
 &  &  &  &  & \selectlanguage{american}%
$\left|H\rightarrow L_{2};H\rightarrow L+3\right\rangle $$(0.2164)$\selectlanguage{english}%
\tabularnewline
\hline 
 &  &  &  &  & \selectlanguage{american}%
$\left|H\rightarrow L_{1};H\rightarrow L+3\right\rangle $$(0.2154)$\selectlanguage{english}%
\tabularnewline
\hline 
\end{tabular}\label{tab-na5+-d2d}
\end{table}

\begin{table}
\caption{Excitation energies, $E$, and many-particle wave functions of excited
states corresponding to the peaks in the linear absorption spectrum
of Na$_{5}{}^{+}$\textbf{\textit{ $D_{2h}$}} structure (\emph{cf}.
Fig. 14(a), in main text), along with the oscillator strength $f$
of the transitions. Rest of the information is same as in previous
tables. }

\begin{tabular}{|c|c|c|c|c|c|}
\hline 
\selectlanguage{american}%
Peak\selectlanguage{english}%
 & $E$ (eV) & \selectlanguage{american}%
Symmetry \selectlanguage{english}%
 & $f$ & \selectlanguage{american}%
Polarization\selectlanguage{english}%
 & \selectlanguage{american}%
Wave Function\selectlanguage{english}%
\tabularnewline
\hline 
\hline 
I & 2.149 & $^{1}B_{2u}$ & 1.3471 & in-plane & \selectlanguage{american}%
$\left|H\rightarrow L+1\right\rangle $$(0.8855)$\selectlanguage{english}%
\tabularnewline
\hline 
 &  &  &  &  & \selectlanguage{american}%
$\left|H-1\rightarrow H;H\rightarrow L+5\right\rangle $$(0.1484)$\selectlanguage{english}%
\tabularnewline
\hline 
 &  &  &  &  & \selectlanguage{american}%
$\left|H\rightarrow L+1;H\rightarrow L+5\right\rangle $$(0.1148)$\selectlanguage{english}%
\tabularnewline
\hline 
 &  &  &  &  & \selectlanguage{american}%
$\left|H\rightarrow L+13\right\rangle $$(0.0863)$\selectlanguage{english}%
\tabularnewline
\hline 
II & 2.508 & $^{1}B_{3u}$ & 1.1284  & in-plane & \selectlanguage{american}%
$\left|H-1\rightarrow H;H\rightarrow L\right\rangle $$(0.6883)$\selectlanguage{english}%
\tabularnewline
\hline 
 &  &  &  &  & \selectlanguage{american}%
$\left|H\rightarrow L+2\right\rangle $$(0.4827)$\selectlanguage{english}%
\tabularnewline
\hline 
 &  &  &  &  & \selectlanguage{american}%
$\left|H\rightarrow L;H\rightarrow L+1\right\rangle $$(0.1960)$\selectlanguage{english}%
\tabularnewline
\hline 
 &  &  &  &  & \selectlanguage{american}%
$\left|H-1\rightarrow L;H-1\rightarrow L+1\right\rangle $$(0.1770)$\selectlanguage{english}%
\tabularnewline
\hline 
III & 3.215 & $^{1}B_{1u}$ & 1.1321  & perpendicular & \selectlanguage{american}%
$\left|H-1\rightarrow H;H\rightarrow L+3\right\rangle $$(0.6405)$\selectlanguage{english}%
\tabularnewline
\hline 
 &  &  &  &  & \selectlanguage{american}%
$\left|H\rightarrow L+6\right\rangle $$(0.5829)$\selectlanguage{english}%
\tabularnewline
\hline 
 &  &  &  &  & \selectlanguage{american}%
$\left|H-1\rightarrow L+1;H-1\rightarrow L+3\right\rangle $$(0.1497)$\selectlanguage{english}%
\tabularnewline
\hline 
 &  &  &  &  & \selectlanguage{american}%
$\left|H\rightarrow L+20\right\rangle $$(0.1264)$\selectlanguage{english}%
\tabularnewline
\hline 
IV & 3.626 & $^{1}B_{3u}$ & 0.0826  & in-plane & \selectlanguage{american}%
$\left|H-1\rightarrow H;H\rightarrow L+7\right\rangle $$(0.3928)$\selectlanguage{english}%
\tabularnewline
\hline 
 &  &  &  &  & \selectlanguage{american}%
$\left|H\rightarrow L+31\right\rangle $$(0.3100)$\selectlanguage{english}%
\tabularnewline
\hline 
 &  &  &  &  & \selectlanguage{american}%
$\left|H-1\rightarrow L+1;H\rightarrow L+2\right\rangle $$(0.2732)$\selectlanguage{english}%
\tabularnewline
\hline 
 &  &  &  &  & \selectlanguage{american}%
$\left|H\rightarrow L;H\rightarrow L+1\right\rangle $$(0.2664)$\selectlanguage{english}%
\tabularnewline
\hline 
\end{tabular}\label{tab-na5+-d2h}
\end{table}

\begin{table}
\caption{Excitation energies, $E$, and many-particle wave functions of excited
states corresponding to the peaks in the linear absorption spectrum
of Na$_{5}$$^{+}$ $C_{2v}$ structure (\emph{cf}. Fig. 14(a), in
main text), along with the oscillator strength $f$ of the transitions.
Rest of the information is same as in previous tables. }

\begin{tabular}{|c|c|c|c|c|c|}
\hline 
Peak	 & $E$ (eV) & Symmetry & $f$ & Polarization & Wave Function\tabularnewline
\hline 
\hline 
I & 2.096397 & $^{1}B_{2}$ & 1.0294 & in-plane & \selectlanguage{american}%
$\left|H\rightarrow L+1\right\rangle $$(0.8189)$\selectlanguage{english}%
\tabularnewline
\hline 
 &  &  &  &  & \selectlanguage{american}%
$\left|H\rightarrow L+1;H\rightarrow L+2\right\rangle $$(0.1855)$\selectlanguage{english}%
\tabularnewline
\hline 
 &  &  &  &  & \selectlanguage{american}%
$\left|H\rightarrow L;H\rightarrow L+2\right\rangle $$(0.1769)$\selectlanguage{english}%
\tabularnewline
\hline 
 &  &  &  &  & \selectlanguage{american}%
$\left|H\rightarrow L;H-1\rightarrow L\right\rangle $$(0.1612)$\selectlanguage{english}%
\tabularnewline
\hline 
II & 2.400917 & $^{1}A_{1}$ & 1.3692 & in-plane & \selectlanguage{american}%
$\left|H\rightarrow L+2\right\rangle $$(0.6908)$\selectlanguage{english}%
\tabularnewline
\hline 
 &  &  &  &  & \selectlanguage{american}%
$\left|H-1\rightarrow L\right\rangle $$(0.5297)$\selectlanguage{english}%
\tabularnewline
\hline 
 &  &  &  &  & \selectlanguage{american}%
$\left|H-1\rightarrow L;H-1\rightarrow L;H\rightarrow L+2\right\rangle $$(0.1551)$\selectlanguage{english}%
\tabularnewline
\hline 
 &  &  &  &  & \selectlanguage{american}%
$\left|H-1\rightarrow L;H\rightarrow L+2;H\rightarrow L+2\right\rangle $$(0.1309)$\selectlanguage{english}%
\tabularnewline
\hline 
III & 3.208789 & $^{1}B_{1}$ & 1.2917 & perpendicular & \selectlanguage{american}%
$\left|H\rightarrow L+5\right\rangle $$(0.6425)$\selectlanguage{english}%
\tabularnewline
\hline 
 &  &  &  &  & \selectlanguage{american}%
$\left|H-1\rightarrow L+3\right\rangle $$(0.6185)$\selectlanguage{english}%
\tabularnewline
\hline 
 &  &  &  &  & \selectlanguage{american}%
$\left|H\rightarrow L+20\right\rangle $$(0.1186)$\selectlanguage{english}%
\tabularnewline
\hline 
 &  &  &  &  & \selectlanguage{american}%
$\left|H-1\rightarrow L+7\right\rangle $$(0.1026)$\selectlanguage{english}%
\tabularnewline
\hline 
\end{tabular}\label{tab-na5+-c2v}
\end{table}

\begin{table}
\caption{Excitation energies, $E$, and many-particle wave functions of excited
states corresponding to the peaks in the linear absorption spectrum
Na of Na$_{5}{}^{+}$\textit{ D$_{3h}$} structure (\emph{cf}. Fig.
14(a), in main text), along with the oscillator strength $f$ of the
transitions In the wave function, the bracketed numbers are the CI
coefficients of a given electronic configuration. Given the trigonal
bipyramidal structure of the cluster (\emph{cf}. Fig. 6(j)), and \textit{$^{1}A_{1}^{'}$}
symmetry of the ground state, optically active excited states are
of doubly degenerate $^{1}E^{'}$ symmetry with polarization along
the common base of the pyramids, and of $^{1}A_{2}^{''}$ symmetry,
with polarization perpendicular to the base. We refer to their polarizations
as ``in-plane'', and ``perpendicular'', respectively. Rest of
the information is same as in previous tables. }

\begin{tabular}{|c|c|c|c|c|c|}
\hline 
\selectlanguage{american}%
Peak\selectlanguage{english}%
 & $E$ (eV) & \selectlanguage{american}%
Symmetry \selectlanguage{english}%
 & $f$ & \selectlanguage{american}%
Polarization\selectlanguage{english}%
 & \selectlanguage{american}%
Wave Function\selectlanguage{english}%
\tabularnewline
\hline 
\hline 
I & 2.227 & \textcolor{black}{$^{1}A_{2}^{''}$} & 0.9568 & perpendicular & \selectlanguage{american}%
$\left|H\rightarrow L+1\right\rangle $$(0.7991)$\selectlanguage{english}%
\tabularnewline
\hline 
 &  &  &  &  & \selectlanguage{american}%
$\left|H\rightarrow L+6\right\rangle $$(0.3394)$\selectlanguage{english}%
\tabularnewline
\hline 
 &  &  &  &  & \selectlanguage{american}%
$\left|H\rightarrow L+8\right\rangle $$(0.1986)$\selectlanguage{english}%
\tabularnewline
\hline 
 &  &  &  &  & \selectlanguage{american}%
$\left|H\rightarrow L_{2};H\rightarrow(L+3)_{1}\right\rangle $$(0.1332)$\selectlanguage{english}%
\tabularnewline
\hline 
II & 3.163 & \selectlanguage{american}%
$^{1}E'$\selectlanguage{english}%
 & 1.0761 & in-plane & \selectlanguage{american}%
$\left|H\rightarrow(L+3)_{1}\right\rangle $$(0.7281)$\selectlanguage{english}%
\tabularnewline
\hline 
 &  &  &  &  & \selectlanguage{american}%
$\left|H-1\rightarrow H;H\rightarrow L{}_{2}\right\rangle $$(0.3847)$\selectlanguage{english}%
\tabularnewline
\hline 
 &  &  &  &  & \selectlanguage{american}%
$\left|H\rightarrow L_{2};H\rightarrow L+1\right\rangle $$(0.2311)$\selectlanguage{english}%
\tabularnewline
\hline 
 &  &  &  &  & \selectlanguage{american}%
$\left|H\rightarrow L_{2};H\rightarrow L+2\right\rangle $$(0.2020)$\selectlanguage{english}%
\tabularnewline
\hline 
 & 3.163 & \selectlanguage{american}%
$^{1}E'$\selectlanguage{english}%
 & 1.0761 & in-plane & \selectlanguage{american}%
$\left|H\rightarrow(L+3)_{2}\right\rangle $$(0.7281)$\selectlanguage{english}%
\tabularnewline
\hline 
 &  &  &  &  & \selectlanguage{american}%
$\left|H-1\rightarrow H;H\rightarrow L{}_{1}\right\rangle $$(0.3847)$\selectlanguage{english}%
\tabularnewline
\hline 
 &  &  &  &  & \selectlanguage{american}%
$\left|H\rightarrow L_{1};H\rightarrow L+1\right\rangle $$(0.2311)$\selectlanguage{english}%
\tabularnewline
\hline 
 &  &  &  &  & \selectlanguage{american}%
$\left|H\rightarrow L_{1};H\rightarrow L+2\right\rangle $$(0.2019)$\selectlanguage{english}%
\tabularnewline
\hline 
\end{tabular}\label{tab-na5+-d3h}
\end{table}

\begin{table}
\caption{Excitation energies, $E$, and many-particle wave functions of excited
states corresponding to the peaks in the linear absorption spectrum
of Na$_{5}$- planar ($C_{2v}$) structure (\emph{cf}. Fig. 15(a),
in main text), along with the oscillator strengths $f$ of the transitions.
Given the \textit{$^{2}A_{1}^{'}$} symmetry of the ground state,
optically active excited states are of $^{2}B_{1}$ and $^{2}A_{1}$
type with polarizations in plane of the cluster, and $^{2}B_{2}$
type, with polarization perpendicular to it. We refer to these polarizations
as ``in-plane'', and ``perpendicular'', respectively. Rest of
the information is same as in previous tables. }

\begin{tabular}{|c|c|c|c|c|c|}
\hline 
Peak	 & $E$ (eV) & Symmetry & $f$ & Polarization & Wave Function\tabularnewline
\hline 
\hline 
I & 1.15 & $^{2}B_{1}$ & 0.0210  & in-plane & \selectlanguage{american}%
$\left|H-2\rightarrow H)\right\rangle $$(0.4052)$\selectlanguage{english}%
\tabularnewline
\hline 
 &  &  &  &  & \selectlanguage{american}%
$\left|H\rightarrow L\right\rangle $$(0.2818)$\selectlanguage{english}%
\tabularnewline
\hline 
 &  &  &  &  & \selectlanguage{american}%
$\left|H-1\rightarrow L+2\right\rangle $$(0.2694)$\selectlanguage{english}%
\tabularnewline
\hline 
 &  &  &  &  & \selectlanguage{american}%
$\left|H-1\rightarrow L+10\right\rangle $$(0.2577)$\selectlanguage{english}%
\tabularnewline
\hline 
II & 1.96 & $^{2}A_{1}$ & 0.5896 & in-plane & \selectlanguage{american}%
$\left|H-1\rightarrow L\right\rangle $$(0.5565)$\selectlanguage{english}%
\tabularnewline
\hline 
 &  &  &  &  & \selectlanguage{american}%
$\left|H-1\rightarrow L+21\right\rangle $$0.2734)$\selectlanguage{english}%
\tabularnewline
\hline 
 &  &  &  &  & \selectlanguage{american}%
$\left|H-1\rightarrow L+8\right\rangle $$(0.2634)$\selectlanguage{english}%
\tabularnewline
\hline 
 &  &  &  &  & \selectlanguage{american}%
$\left|H-1\rightarrow L+4\right\rangle $$(0.1669)$\selectlanguage{english}%
\tabularnewline
\hline 
III & 2.31 & $^{2}B_{1}$ & 0.7687  & in-plane & \selectlanguage{american}%
$\left|H\rightarrow L+4\right\rangle $$(0.3359)$\selectlanguage{english}%
\tabularnewline
\hline 
 &  &  &  &  & \selectlanguage{american}%
$\left|H-1\rightarrow L+2\right\rangle $$(0.3075)$\selectlanguage{english}%
\tabularnewline
\hline 
 &  &  &  &  & \selectlanguage{american}%
$\left|H-2\rightarrow H)\right\rangle $$(0.2867)$\selectlanguage{english}%
\tabularnewline
\hline 
 &  &  &  &  & \selectlanguage{american}%
$\left|H\rightarrow L+14\right\rangle $$(0.2278)$\selectlanguage{english}%
\tabularnewline
\hline 
IV & 2.76 & $^{2}B_{2}$ & 0.2170  & perpendicular & \selectlanguage{american}%
$\left|H\rightarrow L+7\right\rangle $$(0.3108)$\selectlanguage{english}%
\tabularnewline
\hline 
 &  &  &  &  & \selectlanguage{american}%
$\left|H-1\rightarrow L+6\right\rangle $$(0.2691)$\selectlanguage{english}%
\tabularnewline
\hline 
 &  &  &  &  & \selectlanguage{american}%
$\left|H-1\rightarrow L+1;H\rightarrow L+2\right\rangle $$(0.2060)$\selectlanguage{english}%
\tabularnewline
\hline 
 &  &  &  &  & \selectlanguage{american}%
$\left|H-1\rightarrow L+1;H\rightarrow L+10\right\rangle $$(0.2016)$\selectlanguage{english}%
\tabularnewline
\hline 
\end{tabular}\label{tab:na5-planar}
\end{table}

\begin{table}
\caption{Excitation energies, $E$, and many-particle wave functions of excited
states corresponding to the peaks in the linear absorption spectrum
of Na$_{5}$ bipyramid ($C_{2v}$) structure (\emph{cf}. Fig. 15(b),
in main text), along with the oscillator strengths $f$ of the transitions.
Given the trigonal bipyramidal structure of the cluster (\emph{cf}.
Fig. 6(m)), and \textit{$^{2}B_{1}$} symmetry of the ground state,
optically active excited states are of $^{2}A_{2}$ and $^{2}B_{1}$
symmetry with polarization along the common base of the pyramids,
and of $^{2}A_{1}$ symmetry, with polarization perpendicular to the
base. We refer to their polarizations as ``in-plane'', and ``perpendicular'',
respectively. Rest of the information is same as in previous tables. }

\begin{tabular}{|c|c|c|c|c|c|}
\hline 
Peak & $E$ (eV) & Symmetry & $f$ & Polarization & Wave Function\tabularnewline
\hline 
\hline 
I & 1.24 & $^{2}B_{1}$ & 0.0515 & in-plane & $\vert H-1\rightarrow H;H-1\text{\textrightarrow}L\rangle$(0.4700)\tabularnewline
\hline 
 &  &  &  &  & $\vert H\text{\textrightarrow}L+16\rangle$(0.3747)\tabularnewline
\hline 
 &  &  &  &  & $\vert H-1\rightarrow H;H-2\text{\textrightarrow}H-1\rangle$(0.3036)\tabularnewline
\hline 
 &  &  &  &  & $\vert H\text{\textrightarrow}L+2\rangle$(0.2931)\tabularnewline
\hline 
II & 1.57 & $^{2}A_{1}$ & 0.2306 & perpendicular & $\vert H-1\rightarrow H;H\text{\textrightarrow}L+16\rangle$(0.5187)\tabularnewline
\hline 
 &  &  &  &  & $\vert H-1\rightarrow H;H\text{\textrightarrow}L+1\rangle$(0.4365)\tabularnewline
\hline 
 &  &  &  &  & $\vert H-1\rightarrow H;H\text{\textrightarrow}L+2\rangle$(0.3958)\tabularnewline
\hline 
 &  &  &  &  & $\vert H\text{\textrightarrow}L+6\rangle$(0.4517)\tabularnewline
\hline 
 &  &  &  &  & $\vert H\text{\textrightarrow}L+19\rangle$(0.4123)\tabularnewline
\hline 
 &  &  &  &  & $\vert H-1\rightarrow H;H-2\text{\textrightarrow}L+1\rangle$(0.4070)\tabularnewline
\hline 
III & 2.02 & $^{2}A_{1}$ & 0.6558 & perpendicular & $\vert H-1\rightarrow H;H\text{\textrightarrow}L+2\rangle$(0.4562)\tabularnewline
\hline 
 &  &  &  &  & $\vert H-1\rightarrow H;H\text{\textrightarrow}L+1\rangle$(0.4040)\tabularnewline
\hline 
 &  &  &  &  & $\vert H-1\rightarrow L;H\text{\textrightarrow}L+16\rangle$(0.2540)\tabularnewline
\hline 
 &  &  &  &  & $\vert H\text{\textrightarrow}L+6\rangle$(0.2391)\tabularnewline
\hline 
IV & 2.14 & $^{2}A_{1}$ & 0.6381 & perpendicular & $\vert H-1\rightarrow H;H\text{\textrightarrow}L+2\rangle$(0.4266)\tabularnewline
\hline 
 &  &  &  &  & $\vert H-1\rightarrow H;H\text{\textrightarrow}L+8\rangle$(0.3772)\tabularnewline
\hline 
 &  &  &  &  & $\vert H-1\rightarrow H;H\text{\textrightarrow}L+1\rangle$(0.3019)\tabularnewline
\hline 
 &  &  &  &  & $\vert H\text{\textrightarrow}L+6\rangle$(0.2752)\tabularnewline
\hline 
V & 2.44 & $^{2}B_{1}$ & 0.5725 & in-plane & $\vert H-1\text{\ensuremath{\rightarrow}}L+6\rangle$(0.4432)\tabularnewline
\hline 
 &  &  &  &  & $\vert H-1\rightarrow H;H-1\rightarrow L+16\rangle$(0.3314)\tabularnewline
\hline 
 &  &  &  &  & $\vert H-1\rightarrow H;H\text{-1\ensuremath{\rightarrow}}L+2\rangle$(0.3175)\tabularnewline
\hline 
 &  &  &  &  & $\vert H\rightarrow L+2\rangle$(0.2603)\tabularnewline
\hline 
\end{tabular}\label{tab:na5-bipyramid}
\end{table}

\begin{table}
\caption{Excitation energies, $E$, and many-particle wave functions of excited
states corresponding to the peaks in the linear absorption spectrum
of Na$_{6}$ $C_{5v}$ structure (\emph{cf}. Fig. 16(a), in main text),
along with the oscillator strengths $f$ of the transitions. Given
the \textit{$^{1}A_{1}$} symmetry of the ground state, optically
active excited states are of doubly degenerate $^{1}E_{1}$ symmetry
with polarization along the basal plane of the pyramid, and of $^{1}A_{1}$
symmetry, with the polarization perpendicular to it. We refer to these
polarizations as ``in-plane'', and ``perpendicular'', respectively.
Because of the $C_{5v}$ symmetry, the cluster has doubly-degenerate
HOMOs indicated as $H_{1}$ and $H_{2}$. Rest of the information
is same as in previous tables. }

\begin{tabular}{|c|c|c|c|c|c|}
\hline 
Peak	 & $E$ (eV) & Symmetry & $f$ & Polarization & Wave Function\tabularnewline
\hline 
\hline 
I & 1.19  & $^{1}E_{1}$ & 0.0097 & in-plane & \selectlanguage{american}%
$\left|H_{2}\rightarrow L\right\rangle $$(0.7419)$\selectlanguage{english}%
\tabularnewline
\hline 
 &  &  &  &  & \selectlanguage{american}%
$\left|H_{2}\rightarrow L+5\right\rangle $$(0.3170)$\selectlanguage{english}%
\tabularnewline
\hline 
 &  &  &  &  & \selectlanguage{american}%
$\left|H_{2}\rightarrow L+2\right\rangle $$(0.2526)$\selectlanguage{english}%
\tabularnewline
\hline 
 &  &  &  &  & \selectlanguage{american}%
$\left|H_{2}\rightarrow L+12\right\rangle $$(0.1383)$\selectlanguage{english}%
\tabularnewline
\hline 
 & 1.19 & $^{1}E_{1}$ & 0.0097 & in-plane & \selectlanguage{american}%
$\left|H_{1}\rightarrow L\right\rangle $$(0.7420)$\selectlanguage{english}%
\tabularnewline
\hline 
 &  &  &  &  & \selectlanguage{american}%
$\left|H_{1}\rightarrow L+5\right\rangle $$(0.3170)$\selectlanguage{english}%
\tabularnewline
\hline 
 &  &  &  &  & \selectlanguage{american}%
$\left|H_{1}\rightarrow L+2\right\rangle $$(0.2526)$\selectlanguage{english}%
\tabularnewline
\hline 
 &  &  &  &  & \selectlanguage{american}%
$\left|H_{1}\rightarrow L+12\right\rangle $$(0.1383)$\selectlanguage{english}%
\tabularnewline
\hline 
II & 2.150  & $^{1}E_{1}$ & 1.4732 & in-plane & \selectlanguage{american}%
$\left|H_{1}\rightarrow(L+1)_{2}\right\rangle $$(0.4811)$\selectlanguage{english}%
\tabularnewline
\hline 
 &  &  &  &  & \selectlanguage{american}%
$\left|H_{2}\rightarrow(L+1)_{1}\right\rangle $$(0.4808)$\selectlanguage{english}%
\tabularnewline
\hline 
 &  &  &  &  & \selectlanguage{american}%
$\left|H_{1}\rightarrow L\right\rangle $$(0.3100)$\selectlanguage{english}%
\tabularnewline
\hline 
 &  &  &  &  & \selectlanguage{american}%
$\left|H_{1}\rightarrow(L+2)\right\rangle $$(0.3028)$\selectlanguage{english}%
\tabularnewline
\hline 
 & 2.150  & $^{1}E_{1}$ & 1.4732 & in-plane & \selectlanguage{american}%
$\left|H_{1}\rightarrow(L+1)_{1}\right\rangle $$(0.4811)$\selectlanguage{english}%
\tabularnewline
\hline 
 &  &  &  &  & \selectlanguage{american}%
$\left|H_{2}\rightarrow(L+1)_{2}\right\rangle $$(0.4808)$\selectlanguage{english}%
\tabularnewline
\hline 
 &  &  &  &  & \selectlanguage{american}%
$\left|H_{2}\rightarrow L\right\rangle $$(0.3100)$\selectlanguage{english}%
\tabularnewline
\hline 
 &  &  &  &  & \selectlanguage{american}%
$\left|H_{2}\rightarrow(L+2)\right\rangle $$(0.3028)$\selectlanguage{english}%
\tabularnewline
\hline 
III & 2.527 & $^{1}E_{1}$ & 0.2422 & in-plane & \selectlanguage{american}%
$\left|H_{1}\rightarrow(L+4)_{3}\right\rangle $$(0.4774)$\selectlanguage{english}%
\tabularnewline
\hline 
 &  &  &  &  & \selectlanguage{american}%
$\left|H_{1}\rightarrow L+17\right\rangle $$(0.4643)$\selectlanguage{english}%
\tabularnewline
\hline 
 &  &  &  &  & \selectlanguage{american}%
$\left|H_{1}\rightarrow(L+4)_{1}\right\rangle $$(0.3869)$\selectlanguage{english}%
\tabularnewline
\hline 
 &  &  &  &  & \selectlanguage{american}%
$\left|H_{1}\rightarrow(L+1)_{2}\right\rangle $$(0.1903)$\selectlanguage{english}%
\tabularnewline
\hline 
 & 2.527 & $^{1}E_{1}$ & 0.2422 & in-plane & \selectlanguage{american}%
$\left|H_{2}\rightarrow(L+4)_{3}\right\rangle $$(0.4774)$\selectlanguage{english}%
\tabularnewline
\hline 
 &  &  &  &  & \selectlanguage{american}%
$\left|H_{2}\rightarrow L+17\right\rangle $$(0.4643)$\selectlanguage{english}%
\tabularnewline
\hline 
 &  &  &  &  & \selectlanguage{american}%
$\left|H_{2}\rightarrow(L+4)_{1}\right\rangle $$(0.3869)$\selectlanguage{english}%
\tabularnewline
\hline 
 &  &  &  &  & \selectlanguage{american}%
$\left|H_{1}\rightarrow(L+1)_{1}\right\rangle $$(0.1903)$\selectlanguage{english}%
\tabularnewline
\hline 
IV & 2.786  & $^{1}A_{1}$ & 0.4687 & perpendicular & \selectlanguage{american}%
$\left|H_{1}\rightarrow(L+3)_{1}\right\rangle $$(0.4135)$\selectlanguage{english}%
\tabularnewline
\hline 
 &  &  &  &  & \selectlanguage{american}%
$\left|H_{2}\rightarrow(L+3)_{3}\right\rangle $$(0.4133)$\selectlanguage{english}%
\tabularnewline
\hline 
 &  &  &  &  & \selectlanguage{american}%
$\left|H_{1}\rightarrow(L+4)_{2}\right\rangle $$(0.3280)$\selectlanguage{english}%
\tabularnewline
\hline 
 &  &  &  &  & \selectlanguage{american}%
$\left|H_{2}\rightarrow(L+4)_{1}\right\rangle $$(0.2549)$\selectlanguage{english}%
\tabularnewline
\hline 
\end{tabular}\label{tab:na6-c5v}
\end{table}

\begin{table}
\caption{Excitation energies, $E$, and many-particle wave functions of excited
states corresponding to the peaks in the linear absorption spectrum
of Na$_{6}$ $D_{3h}$- planar structure (\emph{cf}. Fig. 16(a), in
main text), along with the oscillator strength $f$ of the transitions.
Given the \textit{$^{1}A_{1}^{'}$} symmetry of the ground state,
optically active excited states are of doubly degenerate $^{1}E^{'}$
symmetry with polarization in plane of the cluster, and of $^{1}A_{2}^{''}$
symmetry, with polarization perpendicular to it. We refer to these
polarizations as ``in-plane'', and ``perpendicular'', respectively.
Because of the $D_{3h}$ symmetry, the cluster has doubly-degenerate
HOMOs indicated as $H_{1}$ and $H_{2}$. Rest of the information
is same as in previous tables. }

\begin{tabular}{|c|c|c|c|c|c|c|}
\hline 
Peak & State & $E$ (eV) & Symmetry & Osci. Strength($f$) & Polarization & Wave Function\tabularnewline
\hline 
\hline 
I &  & 1.36 & \selectlanguage{american}%
$^{1}E'$\selectlanguage{english}%
 & 0.1326  & in-plane & \selectlanguage{american}%
$\left|H_{1}\rightarrow L\right\rangle $$(0.5247)$\selectlanguage{english}%
\tabularnewline
\hline 
 &  &  &  &  &  & \selectlanguage{american}%
$\left|H_{2}\rightarrow L\right\rangle $$(0.3580)$\selectlanguage{english}%
\tabularnewline
\hline 
 &  &  &  &  &  & \selectlanguage{american}%
$\left|H_{1}\rightarrow L+4\right\rangle $$(0.2071)$\selectlanguage{english}%
\tabularnewline
\hline 
 &  &  &  &  &  & \selectlanguage{american}%
$\left|H_{1}\rightarrow(L+1)_{1}\right\rangle $$(0.1855)$\selectlanguage{english}%
\tabularnewline
\hline 
 &  & 1.36 & \selectlanguage{american}%
$^{1}E'$\selectlanguage{english}%
 & 0.1326 &  & \selectlanguage{american}%
$\left|H_{2}\rightarrow L\right\rangle $$(0.5247)$\selectlanguage{english}%
\tabularnewline
\hline 
 &  &  &  &  &  & \selectlanguage{american}%
$\left|H_{1}\rightarrow L\right\rangle $$(0.3580)$\selectlanguage{english}%
\tabularnewline
\hline 
 &  &  &  &  &  & \selectlanguage{american}%
$\left|H_{2}\rightarrow L+4\right\rangle $$(0.2071)$\selectlanguage{english}%
\tabularnewline
\hline 
 &  &  &  &  &  & \selectlanguage{american}%
$\left|H_{2}\rightarrow(L+1)_{1}\right\rangle $$(0.1855)$\selectlanguage{english}%
\tabularnewline
\hline 
II &  & 1.94 & \selectlanguage{american}%
$^{1}E'$\selectlanguage{english}%
 & 2.6326 & in-plane & \selectlanguage{american}%
$\left|H_{1}\rightarrow L\right\rangle $$(0.3499)$\selectlanguage{english}%
\tabularnewline
\hline 
 &  &  &  &  &  & \selectlanguage{american}%
$\left|H_{2}\rightarrow L\right\rangle $$(0.3055)$\selectlanguage{english}%
\tabularnewline
\hline 
 &  &  &  &  &  & \selectlanguage{american}%
$\left|H_{2}\rightarrow(L+1)_{2}\right\rangle $$(0.2989)$\selectlanguage{english}%
\tabularnewline
\hline 
 &  &  &  &  &  & \selectlanguage{american}%
$\left|H_{1}\rightarrow(L+1)_{1}\right\rangle $$(0.2989)$\selectlanguage{english}%
\tabularnewline
\hline 
 &  & 1.94 & \selectlanguage{american}%
$^{1}E'$\selectlanguage{english}%
 & 2.6325  &  & \selectlanguage{american}%
$\left|H_{2}\rightarrow L\right\rangle $$(0.3499)$\selectlanguage{english}%
\tabularnewline
\hline 
 &  &  &  &  &  & \selectlanguage{american}%
$\left|H_{1}\rightarrow L\right\rangle $$(0.3055)$\selectlanguage{english}%
\tabularnewline
\hline 
 &  &  &  &  &  & \selectlanguage{american}%
$\left|H_{2}\rightarrow(L+1)_{1}\right\rangle $$(0.2989)$\selectlanguage{english}%
\tabularnewline
\hline 
 &  &  &  &  &  & \selectlanguage{american}%
$\left|H_{1}\rightarrow(L+1)_{2}\right\rangle $$(0.2989)$\selectlanguage{english}%
\tabularnewline
\hline 
III &  & 2.49 & \selectlanguage{american}%
$^{1}E'$\selectlanguage{english}%
 & 0.0748 & in-plane & \selectlanguage{american}%
$\left|H_{1}\rightarrow(L+3)_{2}\right\rangle $$(0.2471)$\selectlanguage{english}%
\tabularnewline
\hline 
 &  &  &  &  &  & \selectlanguage{american}%
$\left|H_{2}\rightarrow(L+3)_{1}\right\rangle $$(0.2471)$\selectlanguage{english}%
\tabularnewline
\hline 
 &  &  &  &  &  & \selectlanguage{american}%
$\left|H_{2}\rightarrow(L+3)_{2}\right\rangle $$(0.2028)$\selectlanguage{english}%
\tabularnewline
\hline 
 &  &  &  &  &  & \selectlanguage{american}%
$\left|H_{1}\rightarrow(L+3)_{1}\right\rangle $$(0.2028)$\selectlanguage{english}%
\tabularnewline
\hline 
 &  & 2.49 & \selectlanguage{american}%
$^{1}E'$\selectlanguage{english}%
 & 0.0748 & in-plane & \selectlanguage{american}%
$\left|H_{2}\rightarrow(L+3)_{2}\right\rangle $$(0.2471)$\selectlanguage{english}%
\tabularnewline
\hline 
 &  &  &  &  &  & \selectlanguage{american}%
$\left|H_{1}\rightarrow(L+3)_{1}\right\rangle $$(0.2471)$\selectlanguage{english}%
\tabularnewline
\hline 
 &  &  &  &  &  & \selectlanguage{american}%
$\left|H_{1}\rightarrow(L+3)_{2}\right\rangle $$(0.2028)$\selectlanguage{english}%
\tabularnewline
\hline 
 &  &  &  &  &  & \selectlanguage{american}%
$\left|H_{2}\rightarrow(L+3)_{1}\right\rangle $$(0.2028)$\selectlanguage{english}%
\tabularnewline
\hline 
IV &  & 2.81 & \textcolor{black}{$^{1}A_{2}^{''}$} & 0.8056 & perpendicular & \selectlanguage{american}%
$\left|H_{2}\rightarrow(L+5)_{2}\right\rangle $$(0.4686)$\selectlanguage{english}%
\tabularnewline
\hline 
 &  &  &  &  &  & \selectlanguage{american}%
$\left|H_{1}\rightarrow(L+5)_{1}\right\rangle $$(0.4685)$\selectlanguage{english}%
\tabularnewline
\hline 
 &  &  &  &  &  & \selectlanguage{american}%
$\left|H-1\rightarrow L+2\right\rangle $$(0.3190)$\selectlanguage{english}%
\tabularnewline
\hline 
 &  &  &  &  &  & \selectlanguage{american}%
$\left|H_{2}\rightarrow(L+17)_{2}\right\rangle $$(0.1671)$\selectlanguage{english}%
\tabularnewline
\hline 
\end{tabular}\label{tab:na6-d3h}
\end{table}

\begin{table}
\caption{Excitation energies, $E$, and many-particle wave functions of excited
states corresponding to the peaks in the linear absorption spectrum
of Na$_{6}$, $D_{4h}$ structure (\emph{cf}. Fig. 16(b), in main
text), along with the oscillator strength $f$ of the transitions.
Because of the $D_{4h}$ symmetry, the cluster has doubly-degenerate
HOMOs indicated as $H_{1}$ and $H_{2}$. Given the bipyramidal structure
of the cluster (\emph{cf}. Fig. 6(p) of the main text) and \textit{$^{1}A_{1g}$}
symmetry of the ground state, optically active excited states are
of doubly degenerate $^{1}E_{u}$ symmetry with polarization along
the common base of the pyramids, and of $^{1}A_{2u}$ symmetry, with
polarization perpendicular to the base. We refer to these polarizations
as ``in-plane'', and ``perpendicular'', respectively. Rest of
the information is same as in previous tables. }

\begin{tabular}{|c|c|c|c|c|c|}
\hline 
Peak	 & $E$ (eV) & Symmetry & $f$ & Polarization & Wave Function\tabularnewline
\hline 
\hline 
I & 1.647 & $^{1}E_{u}$ & 0.2786 & in-plane & \selectlanguage{american}%
$\left|H_{2}\rightarrow(L+1)\right\rangle $$(0.4269)$\selectlanguage{english}%
\tabularnewline
\hline 
 &  &  &  &  & \selectlanguage{american}%
$\left|H_{2}\rightarrow(L+2)\right\rangle $$(0.3583)$\selectlanguage{english}%
\tabularnewline
\hline 
 &  &  &  &  & \selectlanguage{american}%
$\left|H_{1}\rightarrow(L+6)\right\rangle $$(0.3087)$\selectlanguage{english}%
\tabularnewline
\hline 
 &  &  &  &  & \selectlanguage{american}%
$\left|H_{2}\rightarrow(L+9)\right\rangle $$(0.2703)$\selectlanguage{english}%
\tabularnewline
\hline 
 & 1.647  & $^{1}E_{u}$ & 0.2786 & in-plane & \selectlanguage{american}%
$\left|H_{1}\rightarrow(L+1)\right\rangle $$(0.4269)$\selectlanguage{english}%
\tabularnewline
\hline 
 &  &  &  &  & \selectlanguage{american}%
$\left|H_{1}\rightarrow(L+2)\right\rangle $$(0.3583)$\selectlanguage{english}%
\tabularnewline
\hline 
 &  &  &  &  & \selectlanguage{american}%
$\left|H_{2}\rightarrow(L+6)\right\rangle $$(0.3087)$\selectlanguage{english}%
\tabularnewline
\hline 
 &  &  &  &  & \selectlanguage{american}%
$\left|H_{1}\rightarrow(L+9)\right\rangle $$(0.2703)$\selectlanguage{english}%
\tabularnewline
\hline 
II & 1.90 & $^{1}E_{u}$ & 0.0421 & in-plane & \selectlanguage{american}%
$\left|H_{2}\rightarrow(L+5)\right\rangle $$(0.4200)$\selectlanguage{english}%
\tabularnewline
\hline 
 &  &  &  &  & \selectlanguage{american}%
$\left|H_{2}\rightarrow(L+15)\right\rangle $$(0.3930)$\selectlanguage{english}%
\tabularnewline
\hline 
 &  &  &  &  & \selectlanguage{american}%
$\left|H_{1}\rightarrow(L+5)\right\rangle $$(0.2992)$\selectlanguage{english}%
\tabularnewline
\hline 
 &  &  &  &  & \selectlanguage{american}%
$\left|H_{1}\rightarrow(L+15)\right\rangle $$(0.2799)$\selectlanguage{english}%
\tabularnewline
\hline 
 & 1.90 & $^{1}E_{u}$ & 0.0421 & in-plane & \selectlanguage{american}%
$\left|H_{1}\rightarrow(L+5)\right\rangle $$(0.4200)$\selectlanguage{english}%
\tabularnewline
\hline 
 &  &  &  &  & \selectlanguage{american}%
$\left|H_{1}\rightarrow(L+15)\right\rangle $$(0.3930)$\selectlanguage{english}%
\tabularnewline
\hline 
 &  &  &  &  & \selectlanguage{american}%
$\left|H_{2}\rightarrow(L+5)\right\rangle $$(0.2992)$\selectlanguage{english}%
\tabularnewline
\hline 
 &  &  &  &  & \selectlanguage{american}%
$\left|H_{2}\rightarrow(L+15)\right\rangle $$(0.2799)$\selectlanguage{english}%
\tabularnewline
\hline 
III & 2.336 & $^{1}E_{u}$ & 1.3035 & in-plane & \selectlanguage{american}%
$\left|H_{1}\rightarrow(L+6)\right\rangle $$(0.5922)$\selectlanguage{english}%
\tabularnewline
\hline 
 &  &  &  &  & \selectlanguage{american}%
$\left|H_{2}\rightarrow(L+2)\right\rangle $$(0.4008)$\selectlanguage{english}%
\tabularnewline
\hline 
 &  &  &  &  & \selectlanguage{american}%
$\left|H_{2}\rightarrow(L+5)\right\rangle $$(0.2602)$\selectlanguage{english}%
\tabularnewline
\hline 
 &  &  &  &  & \selectlanguage{american}%
$\left|H_{1}\rightarrow(L+16)\right\rangle $$(0.2542)$\selectlanguage{english}%
\tabularnewline
\hline 
 & 2.336 & $^{1}E_{u}$ & 1.3035 & in-plane & \selectlanguage{american}%
$\left|H_{2}\rightarrow(L+6)\right\rangle $$(0.5922)$\selectlanguage{english}%
\tabularnewline
\hline 
 &  &  &  &  & \selectlanguage{american}%
$\left|H_{1}\rightarrow(L+2)\right\rangle $$(0.4008)$\selectlanguage{english}%
\tabularnewline
\hline 
 &  &  &  &  & \selectlanguage{american}%
$\left|H_{1}\rightarrow(L+5)\right\rangle $$(0.2602)$\selectlanguage{english}%
\tabularnewline
\hline 
 &  &  &  &  & \selectlanguage{american}%
$\left|H_{2}\rightarrow(L+16)\right\rangle $$(0.2542)$\selectlanguage{english}%
\tabularnewline
\hline 
IV & 2.71 & $^{1}A_{2u}$ & 0.7969 & perpendicular & \selectlanguage{american}%
$\left|H_{2}\rightarrow(L+7)_{1}\right\rangle $$(0.5801)$\selectlanguage{english}%
\tabularnewline
\hline 
 &  &  &  &  & \selectlanguage{american}%
$\left|H_{1}\rightarrow(L+7)_{2}\right\rangle $$(0.5801)$\selectlanguage{english}%
\tabularnewline
\hline 
 &  &  &  &  & \selectlanguage{american}%
$\left|H-1\rightarrow L\right\rangle $$(0.2014)$\selectlanguage{english}%
\tabularnewline
\hline 
 &  &  &  &  & $\vert H_{1}\rightarrow L;H_{2}\text{\ensuremath{\rightarrow}}(L+6)\rangle$$(0.1636)$\tabularnewline
\hline 
\end{tabular}\label{tab:na6-d4h}
\end{table}